\documentclass[longauth]{aa}
\usepackage[utf8]{inputenc}
\usepackage[varg]{txfonts}
\usepackage[breaklinks, colorlinks, citecolor=blue, linkcolor=blue]{hyperref}
\usepackage[usenames, dvipsnames]{color}
\usepackage{amsmath}
\usepackage{graphicx}
\usepackage[english]{babel}
\usepackage{siunitx}
\usepackage{float}
\usepackage{url}
\usepackage{longtable}
\usepackage{fancyhdr}
\usepackage{natbib}
\usepackage[normalem]{ulem}
\usepackage[version=4]{mhchem}

\makeatletter

\def\instrefs#1{{\def\scsep{\def\scsep{,}}\@for\w:=#1\do{\scsep\ref{inst:\w}}}}
\renewcommand{\inst}[1]{\unskip$^{\instrefs{#1}}$}

\let\orgautoref\autoref
\renewcommand{\autoref}
        {\def\equationautorefname{Eq.}%
         \def\figureautorefname{Fig.}%
         \def\sectionautorefname{Sect.}%
         \def\subsectionautorefname{Sect.}%
         \def\subsubsectionautorefname{Sect.}%
         \orgautoref}

\renewcommand*\aa@pageof{, page \thepage{} of \pageref*{LastPage}} 

\makeatother
\begin{document}

\title{A planetary system with two transiting mini-Neptunes near the radius valley transition around the bright M dwarf TOI-776\thanks{Based on observations made with ESO Telescopes at the La Silla Observatory under programs ID 1102.C-0923 and 60.A-9709.}}


\titlerunning{A multi-planetary system around TOI-776}

\author{
R.~Luque\inst{iac,ull}  
\and L.\,M.~Serrano\inst{torino}
\and K.~Molaverdikhani\inst{mpia,lsw}
\and M.\,C.~Nixon\inst{ucambridge}
\and J.\,H.~Livingston\inst{utokyo}
\and E.\,W.~Guenther\inst{tls}
\and E.~Pall\'e\inst{iac,ull} 
\and N.~Madhusudhan\inst{ucambridge}
\and G.~Nowak\inst{iac,ull}
\and J.~Korth\inst{koln}
\and W.\,D.~Cochran\inst{utexas} 
\and T.~Hirano\inst{depstokyo}
\and P.~Chaturvedi\inst{tls}
\and E.~Goffo\inst{torino}
\and S.~Albrecht\inst{aarhus}
\and O.~Barragán\inst{oxford}
\and C~Brice\~{n}o\inst{ctio} 
\and J.~Cabrera\inst{dlr}
\and D.~Charbonneau\inst{cfa} 
\and R.~Cloutier\inst{cfa}
\and K.\,A.~Collins\inst{cfa}
\and K.\,I.~Collins\inst{georgemason}
\and K.\,D.~Colón\inst{gsfcexo}
\and I.\,J.\,M.~Crossfield\inst{kansas}
\and Sz.~Csizmadia\inst{dlr}
\and F.~Dai\inst{caltech}
\and H.\,J.~Deeg\inst{iac,ull}
\and M.~Esposito\inst{tls}
\and M.~Fridlund\inst{leiden,chalm}
\and D.~Gandolfi\inst{torino}
\and I.~Georgieva\inst{chalm}
\and A.~Glidden\inst{MIT1,kavli}
\and R.\,F.~Goeke\inst{MIT1}
\and S.~Grziwa\inst{koln}
\and A.\,P.~Hatzes\inst{tls}
\and C.\,E.~Henze\inst{Ames}
\and S.\,B.~Howell\inst{Ames}
\and J.~Irwin\inst{cfa}
\and J.\,M.~Jenkins\inst{Ames} 
\and E.\,L.\,N.~Jensen\inst{swarthmore}
\and P.~K\'abath\inst{chec}
\and R.\,C.~Kidwell~Jr.\inst{stsci} 
\and J.\,F.~Kielkopf\inst{louisville}
\and E.~Knudstrup\inst{aarhus}
\and K.\,W.\,F.~Lam\inst{tub}
\and D.\,W.~Latham\inst{cfa}
\and J.\,J.~Lissauer\inst{Ames}
\and A.\,W.~Mann\inst{northcarolina}
\and E.\,C.~Matthews\inst{kavli}
\and I.~Mireles\inst{kavli}
\and N.~Narita\inst{komaba,jst,abc,iac}
\and M.~Paegert\inst{cfa}
\and C.\,M.~Persson\inst{chalm}
\and S.~Redfield\inst{wes}
\and G.\,R.~Ricker\inst{kavli}
\and F.~Rodler\inst{eso}
\and J.\,E.~Schlieder\inst{gsfcexo}
\and N.\,J.~Scott\inst{Ames}
\and S.~Seager\inst{kavli,MIT1,MIT2}
\and J.~Šubjak\inst{chec}
\and T.\,G.~Tan\inst{pest}
\and E.\,B.~Ting\inst{Ames}
\and R.~Vanderspek\inst{kavli}
\and V.~Van~Eylen\inst{dacp}
\and J.\,N.~Winn\inst{princeton}
\and C.~Ziegler\inst{dunlap}
}

\institute{
\label{inst:iac}Instituto de Astrof\'isica de Canarias, 38205 La Laguna, Tenerife, Spain
\email{rluque@iac.es}
\and 
\label{inst:ull}Departamento de Astrof\'isica, Universidad de La Laguna, 38206 La Laguna, Tenerife, Spain
\and
\label{inst:torino}Dipartimento di Fisica, Universit\`a degli Studi di Torino, via Pietro Giuria 1, I-10125, Torino, Italy
\and 
\label{inst:mpia}Max-Planck-Institut f\"ur Astronomie, K\"onigstuhl 17, 69117 Heidelberg, Germany
\and 
\label{inst:lsw}Landessternwarte, Zentrum für Astronomie der Universität Heidelberg, Königstuhl 12, 69117 Heidelberg, Germany
\and
\label{inst:ucambridge}Institute of Astronomy, University of Cambridge, Madingley Road, Cambridge CB3 0HA, UK
\and
\label{inst:utokyo}Department of Astronomy, University of Tokyo, 7-3-1 Hongo, Bunkyo-ky, Tokyo 113-0033, Japan
\and 
\label{inst:tls}Th\"uringer Landessternwarte Tautenburg, Sternwarte 5, 07778 Tautenburg, Germany
\and
\label{inst:koln}Rheinisches Institut f\"ur Umweltforschung an der Universit\"at zu K\"oln, Aachener Strasse 209, 50931 K\"oln, Germany
\and
\label{inst:utexas}Center for Planetary Systems Habitability and McDonald Observatory, The University of Texas at Austin, Austin, TX 78730, USA
\and
\label{inst:depstokyo}Department  of  Earth  and  Planetary  Sciences,  Tokyo  Institute  of Technology, 2-12-1 Ookayama, Meguro-ku, Tokyo 152-8551, Japan
\and
\label{inst:aarhus}Stellar Astrophysics Centre, Department of Physics and Astronomy, Aarhus University, Ny Munkegade 120, DK-8000 Aarhus C, Denmark
\and 
\label{inst:oxford}Sub-department of Astrophysics, Department of Physics, University of Oxford, Oxford, OX1 3RH, UK
\and
\label{inst:ctio}Cerro Tololo Inter-American Observatory/NSF’s NOIRLab, Casilla 603, La Serena, Chile
\and
\label{inst:dlr}Deutsches Zentrum f\"ur Luft- und Raumfahrt, Institut f\"ur Planetenforschung, 12489 Berlin, Rutherfordstrasse 2., Germany 
\and
\label{inst:cfa}Center for Astrophysics \textbar \ Harvard \& Smithsonian, 60 Garden Street, Cambridge, MA 02138, USA
\and
\label{inst:georgemason}George Mason University, 4400 University Drive, Fairfax, VA, 22030 USA
\and
\label{inst:gsfcexo}Exoplanets and Stellar Astrophysics Laboratory, Mail Code 667, NASA Goddard Space Flight Center, 8800 Greenbelt Rd., Greenbelt MD, 20771, USA
\and
\label{inst:kansas}Department of Physics and Astronomy, University of Kansas, Lawrence, KS, USA
\and
\label{inst:caltech}Division of Geological and Planetary Sciences, California Institute of Technology, 1200 East California Blvd, Pasadena, CA, 91125, USA
\and
\label{inst:leiden}Leiden Observatory, Leiden University, 2333CA Leiden, The Netherlands
\and
\label{inst:chalm}Department of Space, Earth and Environment, Chalmers University of Technology, Onsala Space Observatory, 439 92 Onsala, Sweden
\and
\label{inst:MIT1}Department of Earth, Atmospheric and Planetary Sciences, Massachusetts Institute of Technology, Cambridge, MA 02139, USA
\and
\label{inst:kavli}Department of Physics and Kavli Institute for Astrophysics and Space Research, Massachusetts Institute of Technology, Cambridge, MA 02139, USA
\and
\label{inst:Ames}Space Science \& Astrobiology Division, NASA Ames Research Center, Moffett Field, CA 94035, USA
\and
\label{inst:swarthmore}Dept.\ of Physics \& Astronomy, Swarthmore College, Swarthmore PA 19081, USA
\and
\label{inst:chec}Astronomical Institute, Czech Academy of Sciences, Fri\v{c}ova 298, 25165, Ond\v{r}ejov, Czech Republic
\and
\label{inst:stsci}Space Telescope Science Institute, Baltimore, MD, USA
\and
\label{inst:louisville}Department of Physics and Astronomy, University of Louisville, Louisville, KY 40292, USA
\and
\label{inst:tub}Center for Astronomy and Astrophysics, Technical University Berlin, Hardenbergstr. 36, 10623 Berlin, Germany
\and
\label{inst:northcarolina}Department of Physics and Astronomy, The University of North Carolina at Chapel Hill, Chapel Hill, NC 27599-3255, USA
\and
\label{inst:komaba}Komaba Institute for Science, The University of Tokyo, 3-8-1 Komaba, Meguro, Tokyo 153-8902, Japan
\and
\label{inst:jst}JST, PRESTO, 3-8-1 Komaba, Meguro, Tokyo 153-8902, Japan
\and
\label{inst:abc}Astrobiology Center, 2-21-1 Osawa, Mitaka, Tokyo 181-8588, Japan
\and
\label{inst:wes}Astronomy Department and Van Vleck Observatory, Wesleyan University, Middletown, CT 06459, USA
\and
\label{inst:eso}European Southern Observatory (ESO), Alonso de C\'ordova 3107, Vitacura, Casilla 19001, Santiago de Chile
\and
\label{inst:MIT2}Department of Aeronautics and Astronautics, Massachusetts Institute of Technology, 77 Massachusetts Avenue, Cambridge, MA 02139, USA
\and
\label{inst:pest}Perth Exoplanet Survey Telescope, Perth, Western Australia
\and
\label{inst:dacp}Mullard Space Science Laboratory, University College London, Holmbury St. Mary, Dorking, Surrey, RH5 6NT, UK
\and
\label{inst:princeton}Department of Astrophysical Sciences, Princeton University, 4 Ivy Lane, Princeton, NJ 08544, USA 
\and
\label{inst:dunlap}Dunlap Institute for Astronomy and Astrophysics, University of Toronto, 50 St. George Street, Toronto, Ontario M5S 3H4, Canada
}

\date{Received 17.09.2020 / Accepted 30.11.2020}

\abstract{We report the discovery and characterization of two transiting planets around the bright M1\,V star LP~961-53 (TOI-776, $J\,=\,8.5\,\mathrm{mag}$,$M\,=\,0.54\pm0.03\,M_\odot$) detected during Sector 10 observations of the Transiting Exoplanet Survey Satellite (\textit{TESS}). Combining the \textit{TESS} photometry with HARPS radial velocities, as well as ground-based follow-up transit observations from MEarth and LCOGT telescopes, we measured for the inner planet, TOI-776~b, a period of $P_b=8.25\,\mathrm{d}$,
a radius of $R_b=1.85\pm0.13\,R_\oplus$, and a mass of $M_b=4.0\pm0.9\,M_\oplus$; and for the outer planet, TOI-776~c, a period of $P_c=15.66\,\mathrm{d}$, 
a radius of $R_c=2.02\pm0.14\,R_\oplus$, and a mass of $M_c=5.3\pm1.8\,M_\oplus$. The Doppler data shows one additional signal, with a period of $\sim 34\,\mathrm{d}$, associated with the rotational period of the star. The analysis of fifteen years of ground-based photometric monitoring data and the inspection of different spectral line indicators confirm this assumption. The bulk densities of TOI-776~b and c allow for a wide range of possible interior and atmospheric compositions. However, both planets have retained a significant atmosphere, with slightly different envelope mass fractions. Thanks to their location near the radius gap for M dwarfs, we can start to explore the mechanism(s) responsible for the radius valley emergence around low-mass stars as compared to solar-like stars. While a larger sample of well-characterized planets in this parameter space is still needed to draw firm conclusions, we tentatively estimate that the stellar mass below which thermally-driven mass loss is no longer the main formation pathway for sculpting the radius valley is between 0.63 and 0.54\,$M_\odot$. Due to the brightness of the star, the TOI-776 system is also an excellent target for the James Webb Space Telescope, providing a remarkable laboratory to break the degeneracy in planetary interior models and to test formation and evolution theories of small planets around low-mass stars.

}


\maketitle

\section{Introduction} \label{sec:intro}

Exoplanets with masses between those of Earth and Uranus are characterized by a broad range of measured bulk densities \citep[e.g.,][]{Hatzes2015ApJ...810L..25H}. A low density suggests the presence of an extended H/He-envelope around a solid core. On the contrary, if the density is high, the exoplanet is considered to be fully rocky or enriched in light elements (e.g., water, methane, ammonia). In a nutshell, the absence of an envelope might be the result of two opposite scenarios: the planet is born without it or the planet loses it over time. In the first case, the planet forms in a gas-poor inner proto-planetary disk without a thick H/He-envelope \citep[e.g.,][]{Lee2014ApJ...797...95L,Lee2016ApJ...817...90L}. For the second case different mechanisms have been proposed in the last years, such as slow atmospheric escape powered by the planetary core's primordial energy reservoir from formation \citep{Ginzburg2018MNRAS.476..759G,Gupta2019MNRAS.487...24G,Gupta2020MNRAS.493..792G}, impact erosion by planetesimals \citep{Shuvalov2009M&PS...44.1095S,Schlichting2015Icar..247...81S,Wyatt2020MNRAS.491..782W}, or erosion processes driven by the stellar X-ray+EUV (XUV)-radiation \citep[e.g.,][]{Murray-Clay2009ApJ...693...23M,Lammer2012EP&S...64..179L,Owen2012MNRAS.425.2931O,2013ApJ...775..105O,2013AsBio..13.1030K,2014A&A...562A.116K,Lopez2014ApJ...792....1L,2014ApJ...795...65J,2016ApJ...831..180C,2017A&A...604A..19O,2018ApJ...853..163J,Lopez2018MNRAS.479.5303L,Wu2019ApJ...874...91W,Mordasini2020A&A...638A..52M}.


For the latter, the erosion rate becomes faster if the planetary surface gravity decreases and the amount of XUV-radiation that the planet receives increases. In addition, the intensity of XUV-radiation depends on the orbital semi-major axis and on the stellar activity level. The XUV-radiation is particularly high at young ages and then it declines as a result of age, mass and stellar rotation \citep{Walter1988AJ.....96..297W,Briceno1997AJ....113..740B,Tu2015A&A...577L...3T}. A star that begins its life rapidly rotating will suffer a more rapid decline in rotation than a star that was initially a slow rotator. Thus, for a several Gyr old star, understanding its original activity level is challenging. The presence, or absence, of a hydrogen-rich envelope in a system containing just one planet can thus be equally explained by assuming that the host star was either a slow or rapid rotator when it was young. Systems containing more than one planet are necessary to test the theory of atmospheric erosion, because the origin of all the planets of a system should be explained with a unique evolutionary history of the host's XUV-radiation \citep{Owen2020MNRAS.491.5287O}. 

On the other hand, the amount of XUV-radiation also depends on the stellar type. The XUV luminosities of young G- and M-stars are similar to each other. The average X-ray luminosity of G-stars is $10^{29}$\,erg\,s$^{-1}$, while in the case of M dwarfs, the 50\,Myr stars in $\alpha$-Per, for example, have luminosities of $10^{28}$\,erg\,s$^{-1}$ \citep{France2016ApJ...820...89F}. The main difference is that M dwarfs remain in the high activity phase for up to 2\,Gyr \citep{Johnstone2015A&A...577A..28J}, a much longer time compared to the 300\,Myr of G-stars \citep{Gudel2004A&A...416..713G}. This makes M dwarfs preferred targets to study planetary systems that have experienced significant stellar XUV irradiation. Another advantage of M dwarfs is their small size, which makes it easier to detect smaller transiting planets. The paucity of close-in planets around mid-K to mid-M dwarfs between approximately 1.4 and $1.7\,R_\oplus$ \citep{Cloutier2020AJ....160...22C}, known as the radius valley, marks the transition between rocky planets and sub-Neptunes orbiting low-mass stars. As such, M dwarfs multi-planetary systems which include sub-Neptunes and/or rocky planets represent an ideal benchmark for testing the theory of atmospheric erosion.
 
Gas-poor formation provides an alternative to explain the absence of H/He envelopes in some low-mass planets, since the erosion scenario presents some issues. For instance, if a close-in $10\,M_\oplus$ rocky planet forms while there is still a gaseous disk, its mass is high enough to undergo runaway accretion and become a Jupiter-type planet. The detection of close-in Jupiter-mass planets, at least in A-stars, shows that it is hard to reconstruct a mechanism which transforms a Jupiter into a rocky super-Earth since any working physical process should be able to completely strip off the H/He atmosphere. On the other hand, stars hosting hot Jupiters have high metallicities, while rocky planets are equally distributed between metal-poor and metal-rich stars \citep{Winn2017AJ....154...60W}. Thus, there are two alternative scenarios within gas-poor formation models that could explain the existence of rocky super-Earths. Either the dust-to-gas ratio of the inner disk is 20 times higher than solar, or the gas accretion is delayed until just before the disk disperses \citep{Lee2014ApJ...797...95L,Lee2016ApJ...817...90L}. 

\citet{Lopez2018MNRAS.479.5303L} proposed a statistical test that could allow us to understand the most likely formation history for super-Earths. If a high percentage of rocky planets are the evaporated cores of sub-Neptunes, the transition radius from rocky to sub-Neptune planets should decrease for longer orbital periods. On the contrary, if the gas-poor formation scenario is correct, the transition radius should increase with orbital period. Another methodology to test the formation theory of super-Earths requires studying the position of the radius valley for stars with different masses, thus of different stellar types. If the photo-evaporation scenario is correct, the radius valley shifts towards planets of smaller radii for stars of lower mass. If, on the contrary, the gas-poor formation scenario is at work, the valley position is not affected by the stellar mass \citep{CloutierMenou2020AJ....159..211C}. However, since the radius valley represents the range of radii in which the transition between rocky planets and sub-Neptunes occurs, it is necessary to accurately determine the mass and radius of the planets to calculate the mass-fraction of their envelope and unveil their nature. Therefore, the ideal test to understand which model is more realistic between the gas-poor formation and the photo-evaporation consists of measuring the masses and radii of the planets close to, or inside, the radius valley, preferably in a multi-planetary system around low-mass stars. In this way, we can also constrain these models in a much better way than through the radius distribution alone. 

As of today, there is a limited number of known multi-planetary systems which orbit M-dwarfs ($3000\,\mathrm{K} < T_\mathrm{eff} < 4000\,\mathrm{K}$; as a proxy of M0\,V to M5\,V, \citealt{Cifuentes2020A&A...642A.115C}) and respect the condition ($M_p < 10\,M_\oplus$) required to test the two mentioned formation theories. 
There are only two systems with three transiting planets with measured dynamical masses, Kepler-138 \citep{Almenara2018MNRAS.478..460A} and L~98-59 \citep{Cloutier2019A&A...629A.111C},
and four systems with two transiting planets: LHS~1140 \citep{Lillo-Box2020A&A...642A.121L}, LTT~3780 \citep{Nowak2020A&A...642A.173N,Cloutier2020AJ....160...22C}, K2-146 \citep{Lam2020AJ....159..120L,Hamann2019AJ....158..133H}, and Kepler-26 \citep{Jontof-Hutter2016ApJ...820...39J}. This paucity of systems is inadequate for understanding the formation and evolution of planetary systems around M dwarfs. The discovery of each new system is thus important, especially if the host star is bright and the planets are close to the radius valley. 

In this paper, we present the discovery of two transiting planets orbiting an M1\,V star. The inner one has a period of 8.2\,d and a radius of $\sim 1.8\,R_\oplus$; thus, it is close to the radius valley. The outer planet has a period of 15.7\,d and a radius of $2.0\,R_\oplus$, in the sub-Neptune regime. By measuring their masses, we explore whether these new planets are characterized by extended H/He envelopes. Since they orbit a relatively bright, nearby M dwarf, these new objects represent ideal targets for follow-up atmospheric studies.

\section{\textit{TESS} photometry} \label{sec:tess}

\begin{figure}[ht!]
    \centering
    \includegraphics[width=0.99\hsize]{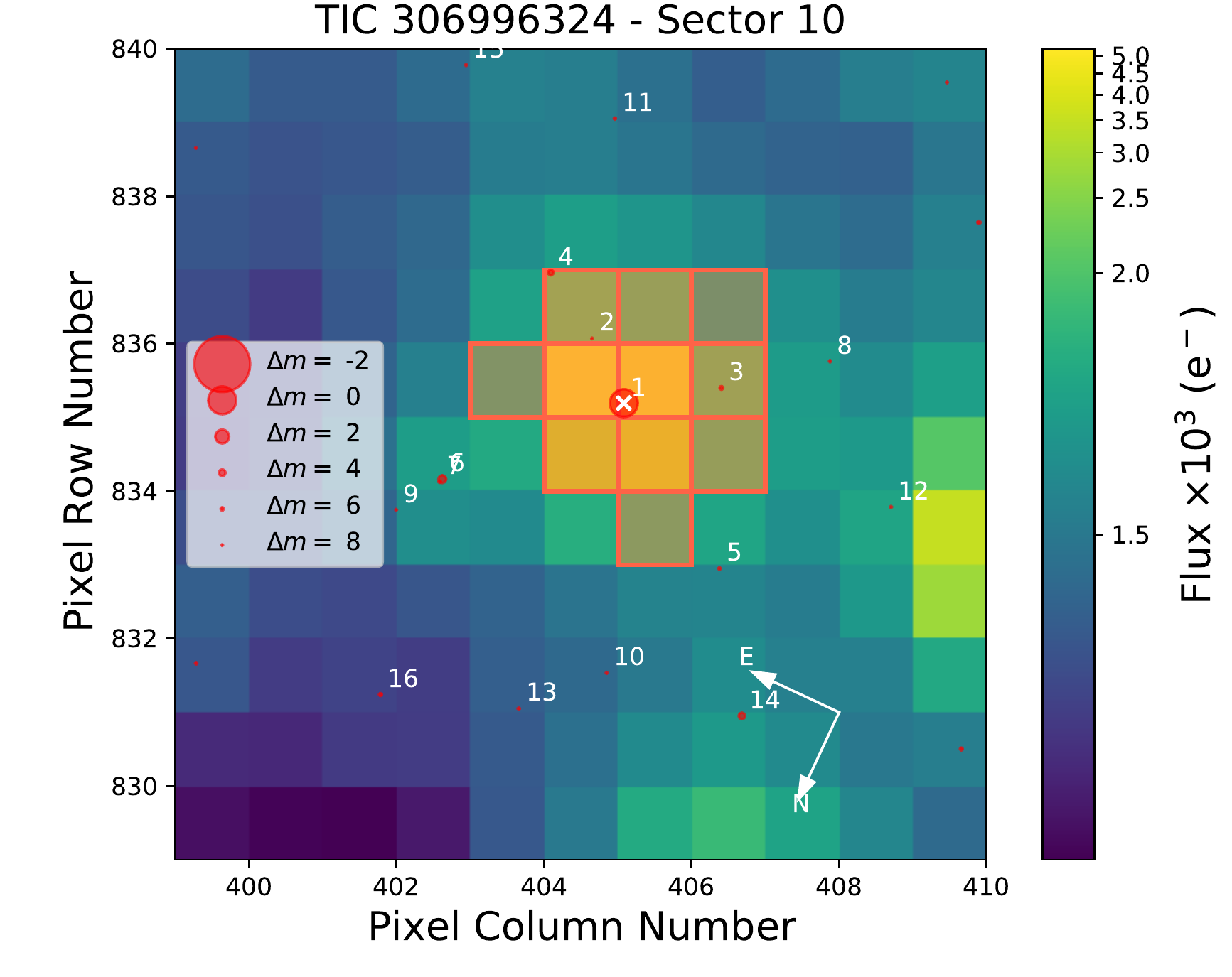}
    \caption{\textit{TESS} target pixel file image of LP~961-53 in Sector~10 (created with \texttt{tpfplotter}, \citealt{aller20}). The electron counts are color-coded. The red bordered pixels are used in the simple aperture photometry. The size of the red circles indicates the \textit{TESS} magnitudes of all nearby stars and LP~961-53 (label \#1 with the "$\times$"). Positions are corrected for proper motions between \textit{Gaia} DR2 epoch (2015.5) and \textit{TESS} Sector~10 epoch (2019.2). The \textit{TESS} pixel scale is approximately 21\arcsec. }
    \label{fig:tpfgaia}
\end{figure}

\begin{figure*}[ht!]
    \centering
    \includegraphics[width=\hsize]{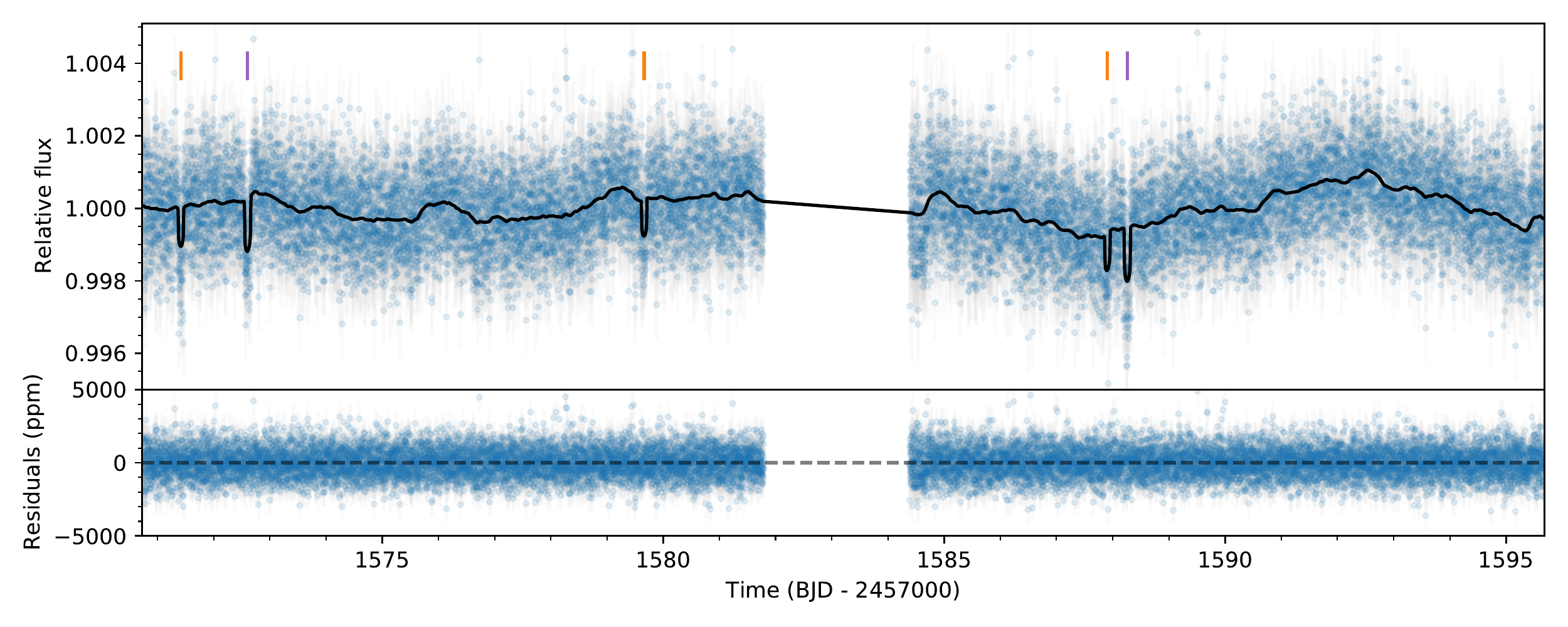}
    \caption{\textit{TESS} PDC-corrected SAP transit photometry from SPOC pipeline with the best-fit \texttt{juliet} model (black line; see Sect.~\ref{subsubsec:photonly} for details on the modeling). Purple and orange ticks above the light curve mark the transits of the candidates TOI–776.01 (purple) and TOI–732.02 (orange).}
    \label{fig:tess_lc}
\end{figure*}

LP~961-53 (TIC~306996324) was observed with \textit{TESS} in Sector~10 (Camera \#2, CCD \#4) from March 26, 2019 until April 22, 2019, with 2-min cadence exposures and it will be observed again in Sector~37 from April 2 to 28, 2021. Data collection was paused for 0.98\,d during perigee passage, while downloading data. The Science Processing Operations Center \citep[SPOC;][]{SPOC} at the NASA Ames Research Center made the data available at the Mikulski Archive for Space Telescopes (MAST)\footnote{\url{https://mast.stsci.edu}} on June 1, 2019. SPOC provided for this target simple aperture photometry (SAP) and systematics-corrected photometry, a procedure consisting of an adaptation of the Kepler Presearch Data Conditioning algorithm \citep[PDC,][]{Smith2012PASP..124.1000S, Stumpe2012PASP..124..985S, Stumpe2014PASP..126..100S} to \textit{TESS}. Figure~\ref{fig:tpfgaia} shows the \textit{TESS} pixels included in the computation of the SAP and PDC-corrected SAP data. For the remainder of this work we make use of the latter photometric data, shown in Fig.~\ref{fig:tess_lc}.

On June 11, 2019, two transiting candidates orbiting LP~961-53 were announced in the \textit{TESS} data public website\footnote{\url{https://tev.mit.edu/data/}} under the \textit{TESS} Object of Interest (TOI) number 776. TOI-776.01 is a planet candidate with a period of 15.65\,d, a transit depth of $1484\pm127\,\mathrm{ppm}$, and an estimated planet radius of $2.2\pm0.6\,R_\oplus$; while TOI-776.02 is a planet candidate with a period of 8.24\,d, a transit depth of $1063\pm104\,\mathrm{ppm}$, and an estimated planet radius of $1.8\pm1.4\,R_\oplus$. Both candidates passed all the tests from the TCE (Threshold Crossing Event) Data Validation Report \citep[DVR;][]{Twicken2018PASP..130f4502T, Li2019PASP..131b4506L}: even-odd transits comparison, eclipsing binary (EB) discrimination tests, ghost diagnostic tests to help rule out scattered light, or background EB, among others. However, the vetting team at the \textit{TESS} Science Office proposed the possibility that TOI-776.01 could be an EB, where the secondary transit is the primary transit of TOI-776.02 candidate. The ground-based follow-up observations discussed in the next Section refuted this scenario and confirmed the two announced candidates as bona-fide planets.

\section{Ground-based observations} \label{sec:obs}

\subsection{Transit follow-up}

\begin{table*}[t!]
\caption{TESS Follow-up Program transit observations.}
\label{table:followup-obs}
\centering
\begin{tabular}{llcccccc}
\noalign{\smallskip}
\hline\hline
\noalign{\smallskip}
Observatory     & Date  & Filter    & Exposure  & Total & Aperture  & Pixel scale   & FOV \\
                & [UTC] &           & [s]       & [h]   & [m]       & [arcsec]      & [arcmin]\\
 \hline\\[-1.5mm]
{\it  TOI-776.01 = TOI-776~c }\\
 \cline{1-1}\\[-2mm]
MEarth-South, CTIO, Chile & Jul 1, 2019 & RG715 & 10 & 4.2 & $7 \times 0.4$ & 0.84 & $29 \times 29$ \\
LCOGT, CTIO, Chile  & Jul 1, 2019 & $i'$ & 20 & 3.4 & 1.0 & 0.39 & $26.5\times26.5$ \\[2mm]
{\it  TOI-776.02 = TOI-776~b} \\
\cline{1-1}\\[-2mm]
LCOGT, SAAO, South Africa  & Feb 29, 2020 & $z_s$ & 45 & 4.7 & 1.0 & 0.39 & $26.5 \times 26.5$ \\
LCOGT, SSO, Australia  & Mar 17, 2020 & $z_s$ & 45 & 4.1 & 1.0 & 0.39 & $26.5\times26.5$ \\
PEST, Australia  & May 22, 2020 & $R_C$ & 60 & 3.6 & 0.3 & 1.23 & $31\times21$ \\

\hline
\end{tabular}
\end{table*}

\begin{figure*}
    \centering
    \includegraphics[width=0.45\hsize]{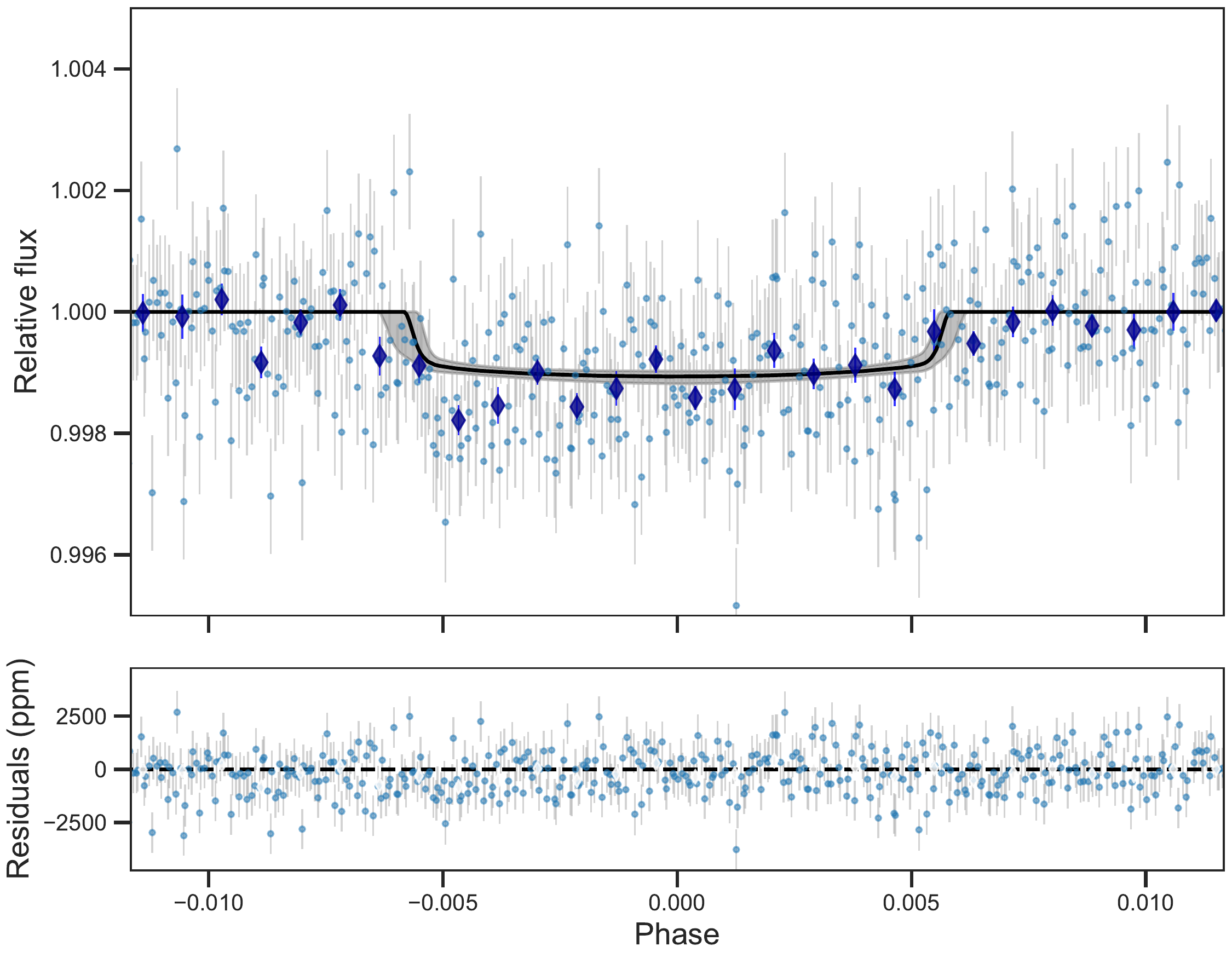}
    \includegraphics[width=0.45\hsize]{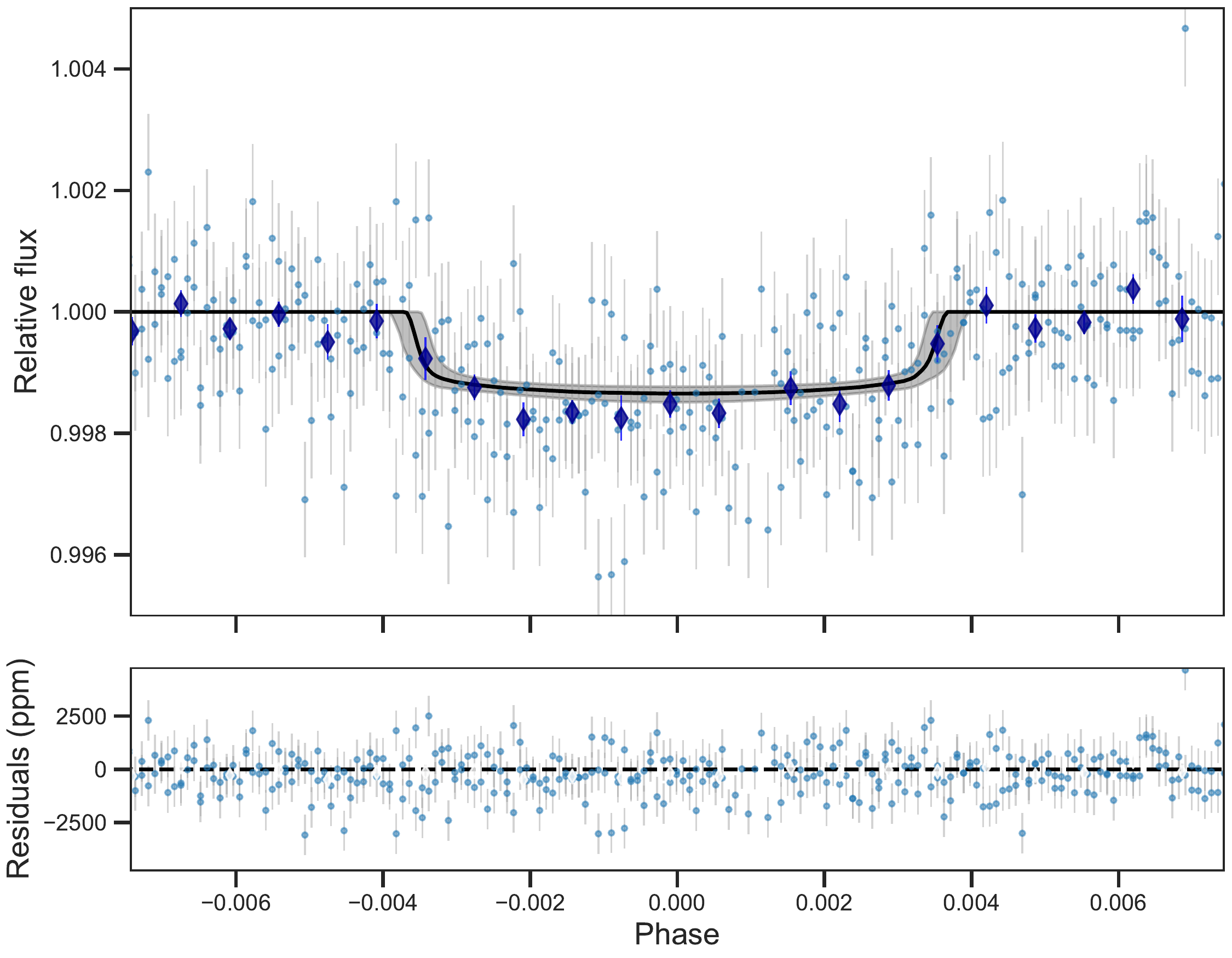}\\
    \includegraphics[width=0.45\hsize]{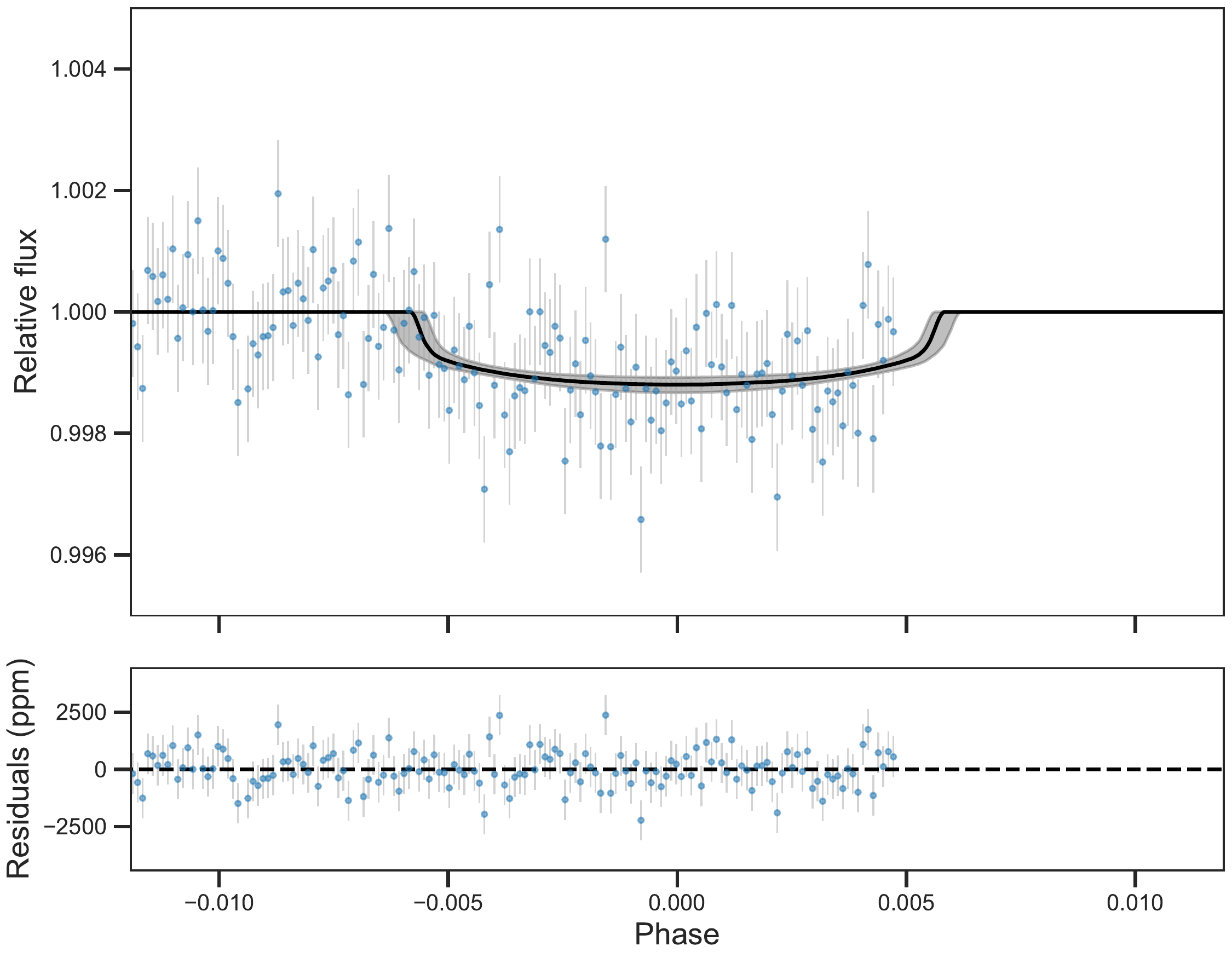}
    \includegraphics[width=0.45\hsize]{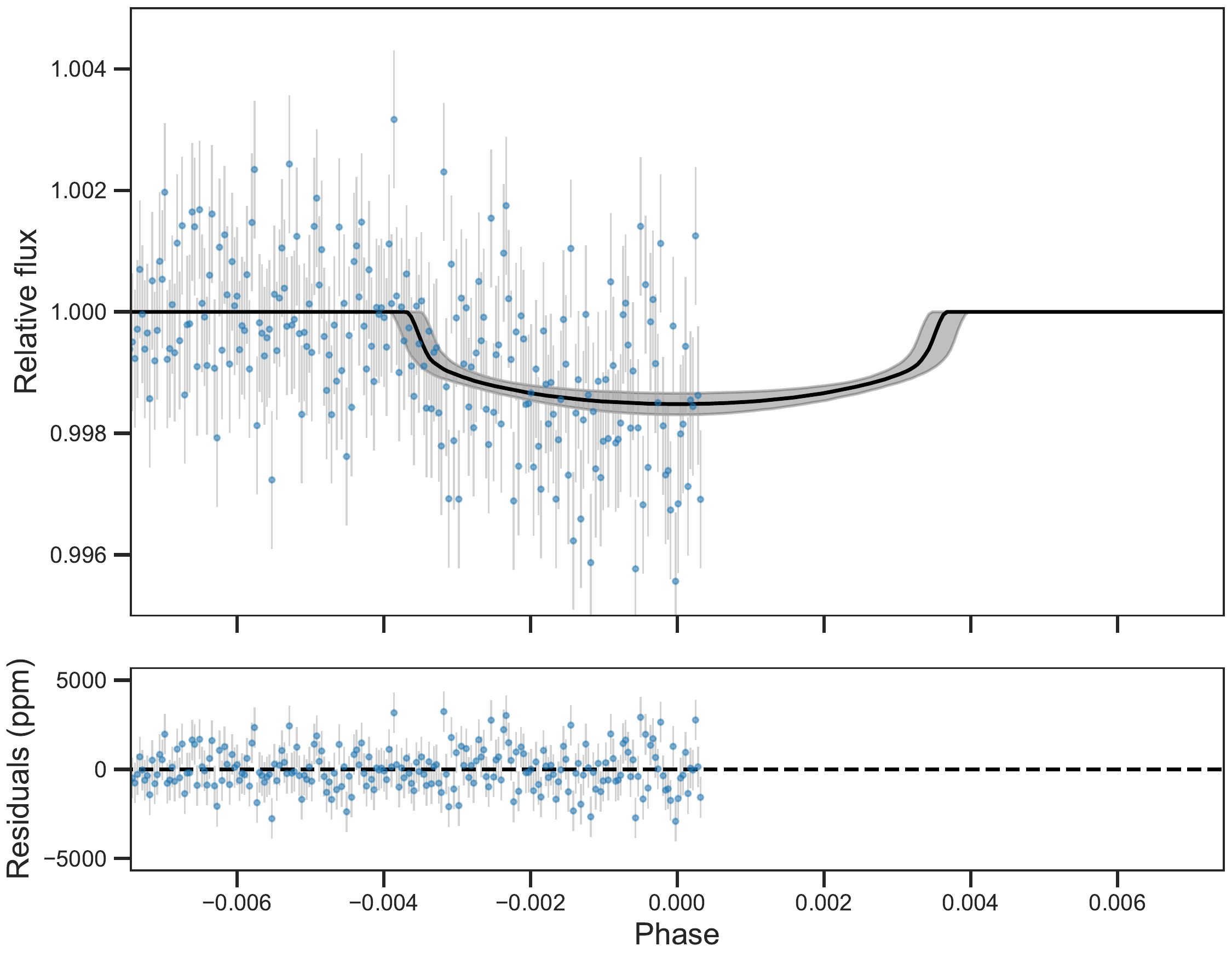}\\
    \includegraphics[width=0.45\hsize]{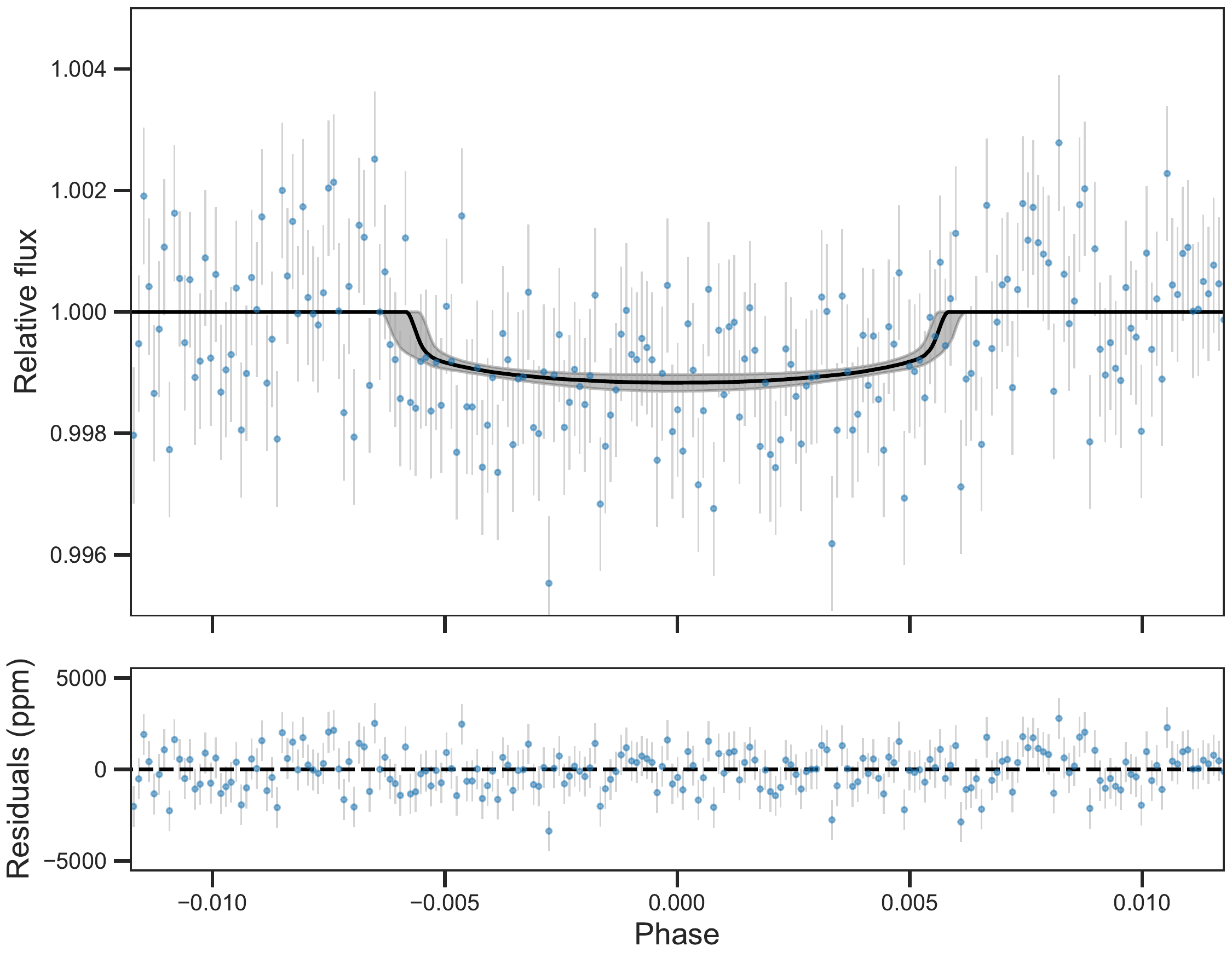}
    \includegraphics[width=0.45\hsize]{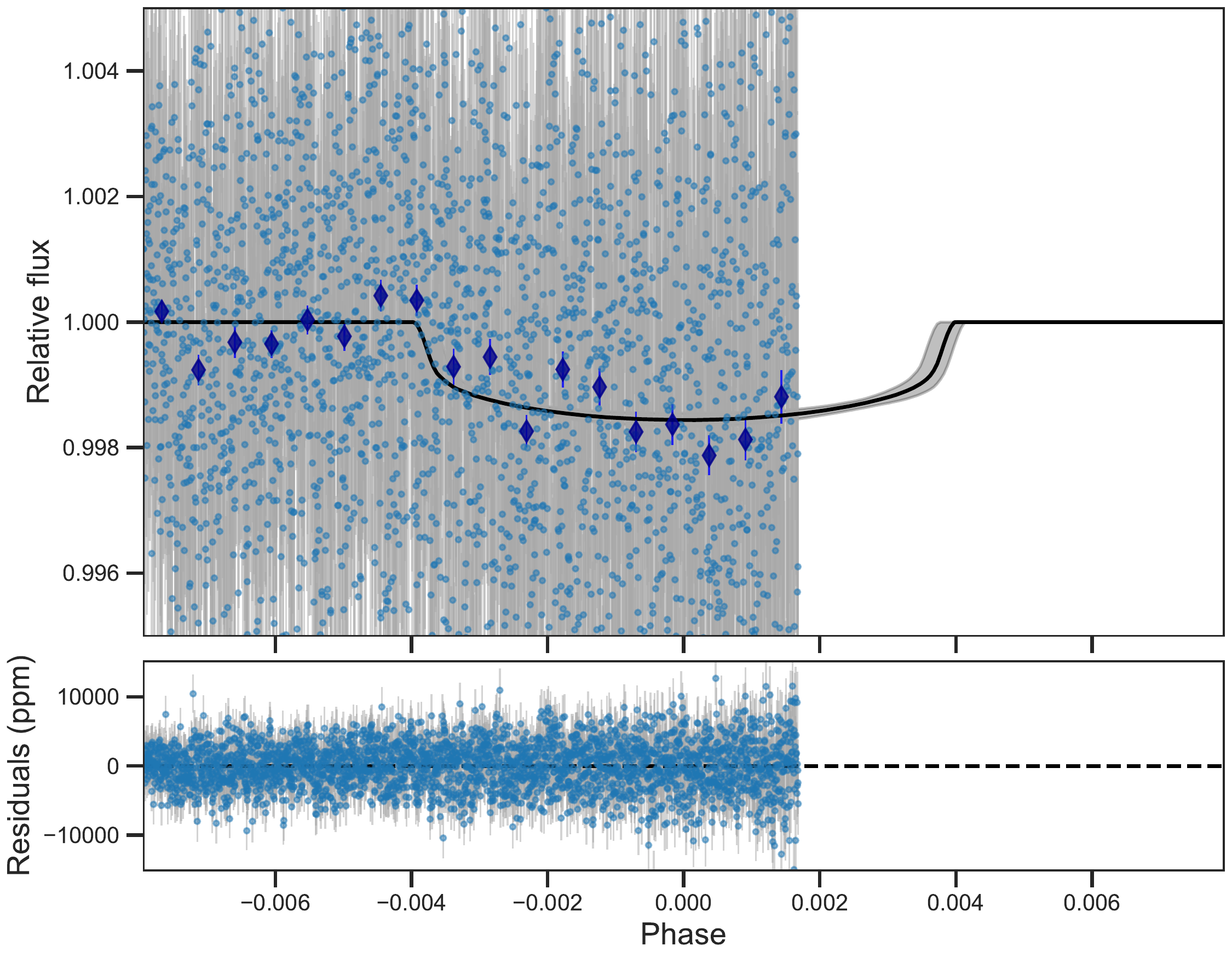}
    \caption{Phase-folded light curves of TOI-776~b and c. First column: transits of TOI-776~b observed with \textit{TESS} (top) in Sector 10, LCO-SAAO (middle) on Feb 29, 2020, and LCO-SSO (bottom) on May 22, 2020. Second column: transits of TOI-776~c observed with \textit{TESS} (top) in Sector 10, and LCO-CTIO (middle) and MEarth-South (bottom) on July 2, 2020. \textit{TESS} and MEarth-South photometry binned every 10\,min are marked with blue diamonds to improve visualization. In all panels, the black lines and shaded areas indicate the detrended best fit model from Sect.~\ref{subsubsec:joint} and its 1$\sigma$ confidence interval. Below each panel are represented the residuals after the subtraction of the median best fit model. }
    \label{fig:all_lc}
\end{figure*}

We observed the TOI-776 candidates as part of the {\it TESS} Follow-up Observing Program (TFOP)\footnote{\url{https://tess.mit.edu/followup}}. The goals of these ground-based photometric follow-up observations were to verify that the transits observed by \textit{TESS} are on target, and to refine the transit ephemeris and depth measurements. We used the {\tt TESS Transit Finder}, a customized version of the {\tt Tapir} software package \citep{Jensen:2013}, to schedule photometric time-series follow-up observations. We observed two transits of TOI-776.01 and three transits of TOI-776.02, as summarized in Table \ref{table:followup-obs} and discussed further below.

\subsubsection{MEarth-South}
A single transit of TOI-776.01 was observed with the 40\,cm MEarth-South telescope array \citep{2015csss...18..767I} at Cerro Tololo Inter-American Observatory (CTIO), Chile on June 1, 2019.  Seven telescopes observed continuously from evening twilight until the target star set below airmass 2, using an exposure time of 10\,s with all telescopes in focus. The target star was west of the meridian throughout the observation to avoid meridian flips.

Data were reduced following the standard procedures in \citet{2007MNRAS.375.1449I} and \citet{2012AJ....144..145B} with a photometric extraction aperture radius of $r = 6$\,pix (5\arcsec\ on sky given the pixel scale of 0\farcs84\,pix$^{-1}$). The light curve is shown in Fig.~\ref{fig:all_lc}, lower right. Due to the large variation in airmass and relatively red target star compared to the available field comparison stars, we found the light curve exhibited a small amount of residual second-order (color-dependent) atmospheric extinction, so the transit model was fitted including an extinction term (linear decorrelation against airmass).

\subsubsection{LCOGT} 

One transit of TOI-776.01 and two transits of TOI-776.02 were observed with the 1.0\,m telescopes in the Las Cumbres Observatory (LCOGT) telescope network \citep{Brown:2013}. The $4096\times4096$\,pix LCOGT SINISTRO cameras have an image scale of 0\farcs389\,pix$^{-1}$, resulting in a $26\arcmin\times26\arcmin$ field of view. The images were calibrated using the standard LCOGT {\tt BANZAI} pipeline, and photometric data were extracted with {\tt AstroImageJ} \citep{Collins:2017}. 

An ingress of TOI-776.01 was observed from the LCOGT node at CTIO on July 1, 2019 in the $i'$ filter, simultaneous with the MEarth-South observations mentioned above (Fig.~\ref{fig:all_lc}, middle right). Transits of TOI-776.02 were observed from the LCOGT nodes at the South African Astronomical Observatory (SAAO) on February 29, 2020  (Fig.~\ref{fig:all_lc}, middle left) and from the Siding Spring Observatory (SSO) on March 17, 2020,  (Fig.~\ref{fig:all_lc}, lower left). Both observations were made in the $z_s$ filter, with the telescopes defocused. 

\subsubsection{PEST}
A full transit of TOI-776.02 was observed with the 30\,cm Perth Exoplanet Survey Telescope\footnote{\url{http://pestobservatory.com/}} (PEST) on May 22, 2020. These data have a scatter that is too large to reliably detect the transit. For this reason, we did not include them in the global fit.

\subsection{Long-term photometric monitoring} \label{subsec:phot}

We compiled ground-based, long baseline photometric series from automated surveys. The following public surveys observed TOI-776: the All-Sky Automated Survey for Supernovae \citep[ASAS-SN;][]{Kochanek2017PASP..129j4502K}, All-Sky Automated Survey \citep[ASAS;][]{Pojmanski2002AcA....52..397P}, Northern Sky Variability Survey \citep[NSVS;][]{Wozniak2004AJ....127.2436W}, and the Catalina surveys \citep{Drake2014ApJS..213....9D}. The telescope location, instrument configurations, and photometric bands of each public survey were summarized in Table~1 of \citet{DiezAlonso2019A&A...621A.126D}. All together, the measurements span a period of 15\,yr.

Additionally, TOI-776 is a candidate of the Super-Wide Angle Search for Planets  \citep[SuperWASP;][]{Pollacco2006PASP..118.1407P}. SuperWASP acquired more than 11\,000 photometric observations, using a broad-band optical filter spanning three consecutive seasons from May to July 2006, January to June 2007, and January to June 2008. In order to detect long-term photometric modulations associated with the stellar rotation, we binned the data into one day intervals, resulting in 201 epochs.

\subsection{High-spatial resolution imaging} \label{subsec:hri}

The large pixel size of \textit{TESS} increase the possibility of contamination by nearby sources that are not detected in the seeing-limited photometry or in \textit{Gaia} DR2. Close companions can dilute the transit depth and thus alter the measured planet radius, or lead to false positives if the companion is itself an EB \citep[e.g.][]{Ciardi2015}. We thus searched for companions by collecting adaptive optics (AO) and speckle images of TOI-776 using 4 and 8\,m class telescopes, providing robust limits on the presence of companions and the level of photometric dilution.

\subsubsection{Adaptive optics imaging} \label{subsubsec:ao}

\begin{figure}
    \centering
    \includegraphics[width=\columnwidth]{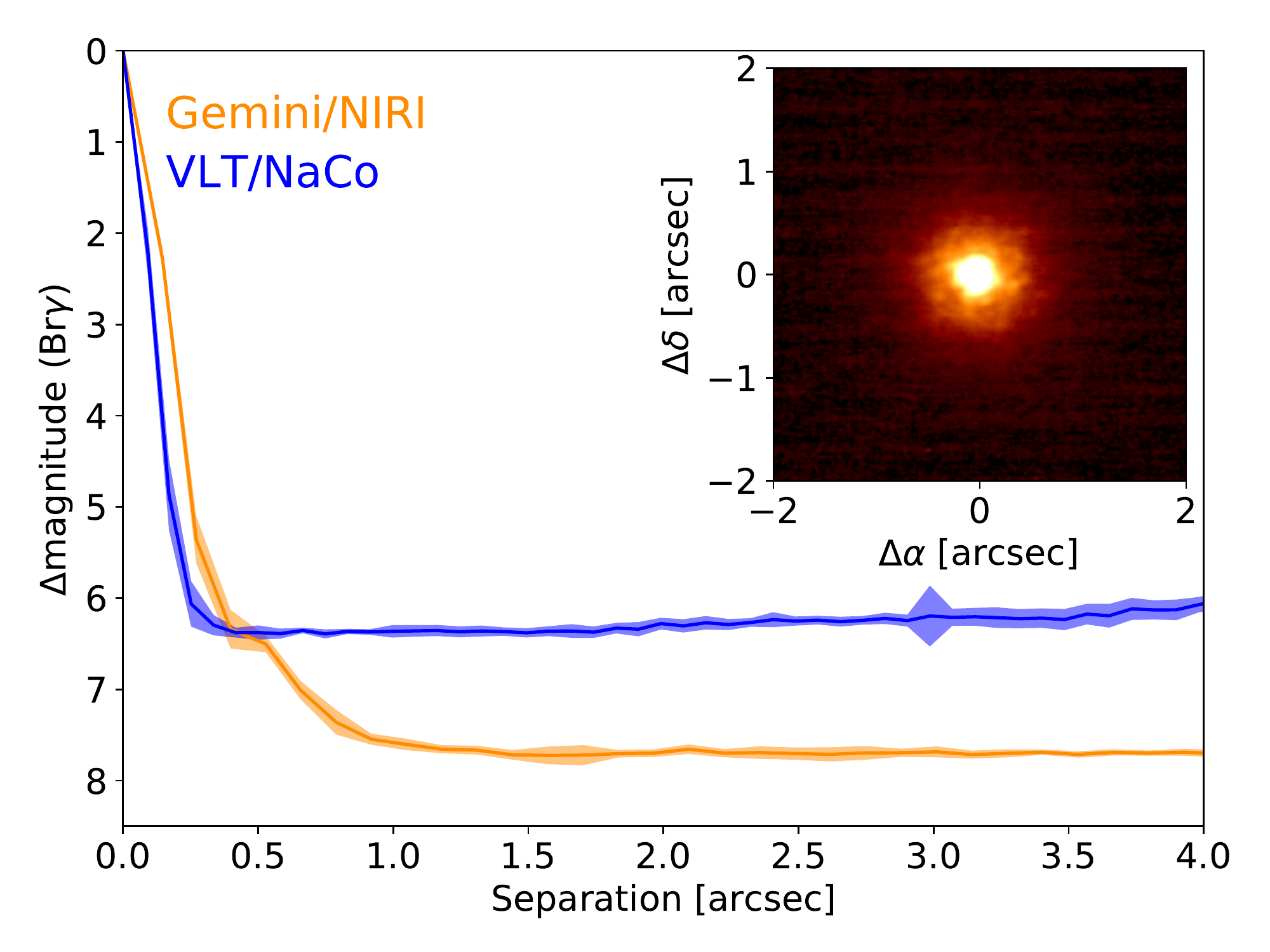}
    \caption{Contrast curves from NIRI (orange) and NaCo (blue), and the central 4\arcsec$\times$4\arcsec of the NIRI image (inset). We rule out companions 6\,mag fainter than TOI-776 beyond 250\,mas, and 7.5\,mag fainter beyond 900\,mas. The NaCo observations have a slightly tighter inner working angle, while the NIRI observations reach a deeper sensitivity beyond 0.5\arcsec.}
    \label{fig:ao}
\end{figure}

\paragraph{Gemini/NIRI}
On June 15, 2019, TOI-776 was observed using the adaptive optics near-infrared imager (NIRI) mounted on the 8.1\,m Gemini North telescope at Mauna Kea, Hawai'i. We collected a total of 9$\times$1.4\,s images in the Br$\gamma$ filter centered on \SI{2.166}{\micro\metre}. We dithered the telescope between exposures, so that the sky background can be constructed from the science frames themselves. After removing bad pixels, flat-fielding, and subtracting the sky background, we aligned the stellar position between frames and co-added the images. The sensitivity of our observations was calculated as a function of radius by injecting fake companions, and scaling their brightness such that they could be detected at 5$\sigma$. The contrast curve and image are shown in Fig.~\ref{fig:ao}. Only the central 4\arcsec$\times$4\arcsec are shown, but no companions are seen anywhere in the field, which has a field of view $\sim$13\arcsec$\times$13\arcsec.

\paragraph{VLT/NaCo}
On July 4, 2019, TOI-776 was observed in Br$\gamma$ using the NAOS-CONICA AO instrument (NaCo), mounted at the Nasmyth A port of the 8\,m UT1 Very Large Telescope (VLT) in Paranal, Chile. We collected a total of 9$\times$10\,s Br$\gamma$ images. Data were reduced and analyzed using the same procedures as described above for the NIRI data, and no companions were found in the reduced image. The NaCo contrast curve is shown in Fig.~\ref{fig:ao}.

\subsubsection{Speckle imaging} \label{subsubsec:speckle}

\begin{figure}
    \centering
    \includegraphics[width=\columnwidth]{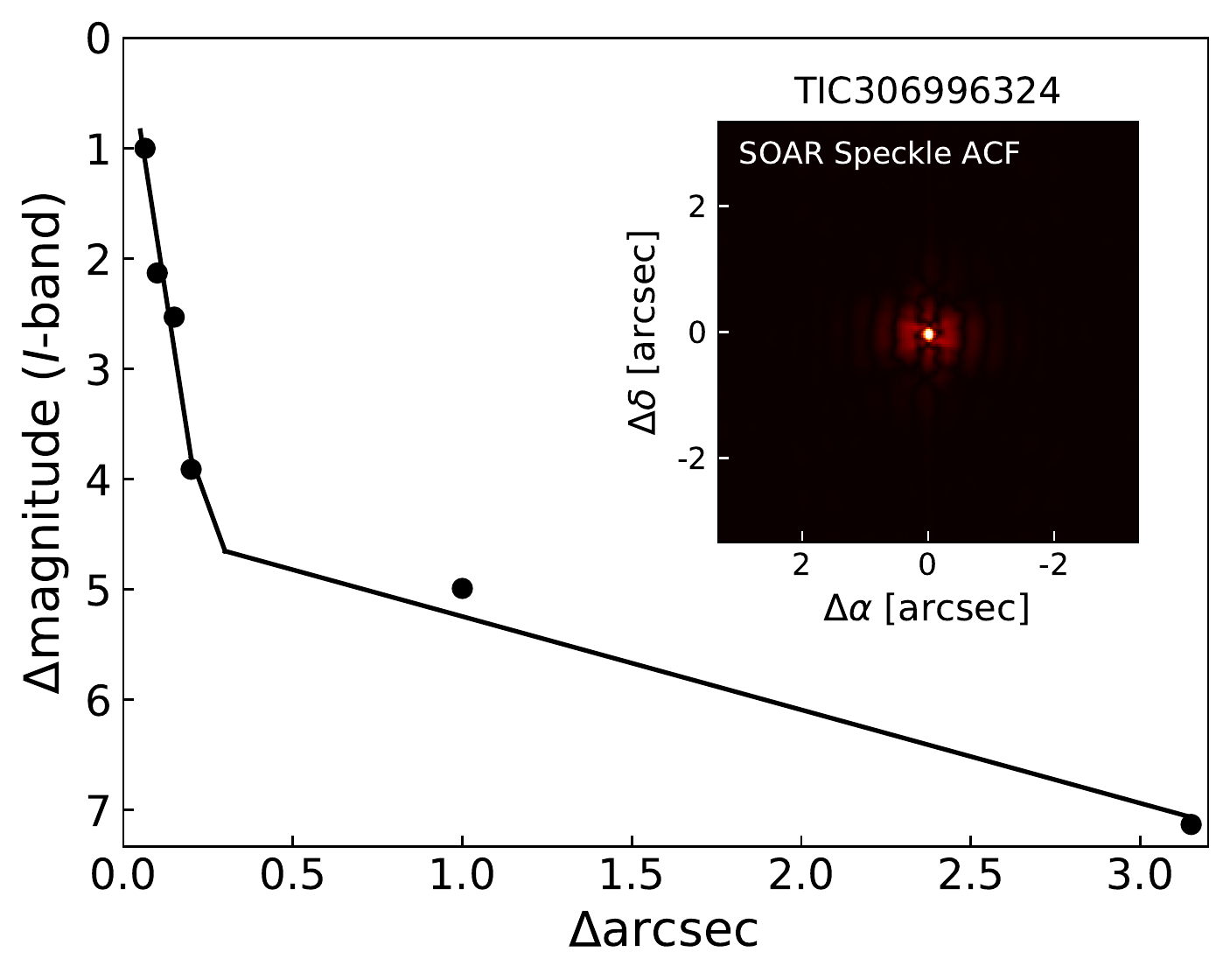}
    \includegraphics[width=\columnwidth]{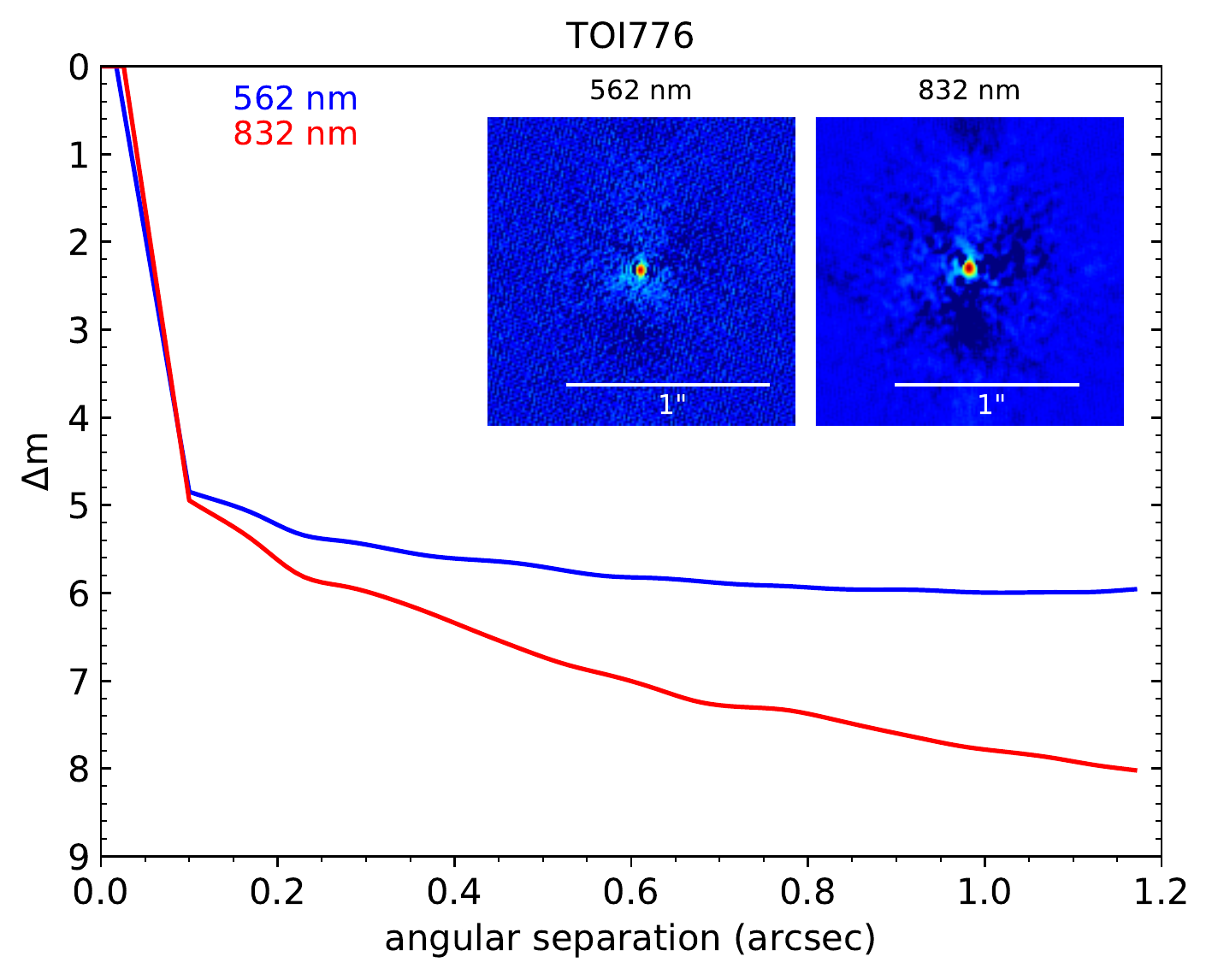}
    \caption{\textit{Top}: SOAR contrast curve and 6\arcsec$\times$6\arcsec reconstructed image (inset). \textit{Bottom}: Gemini/Zorro contrast curves and 1.2\arcsec$\times$1.2\arcsec reconstructed images (inset).}
    \label{fig:speckle}
\end{figure}

\paragraph{SOAR/HRCam}
On December 12, 2019, TOI-776 was observed in $I$ band with a pixel scale of 0.01575\arcsec\,pix$^{-1}$ using the HRCam imager, mounted on the 4.1\,m Southern Astrophysical Research (SOAR) telescope at Cerro Tololo Inter-American Observatory, Chile. The data were acquired and reduced following the procedures described in \citet{Tokovinin2018} and \citet{Ziegler2020}. The resulting reconstructed image achieved a contrast of $\Delta\mathrm{mag}=7.1$ at a separation of 3\arcsec (see top panel of Fig.~\ref{fig:speckle}).

\paragraph{Gemini/Zorro}
On March 15, 2020, TOI-776 was observed using the Zorro speckle imager \citep{Scott2019}, mounted on the 8.1\,m Gemini South telescope in Cerro Pachón, Chile. Zorro uses high speed electron-multiplying CCDs (EMCCDs) to simultaneously acquire data in two bands centered at 562\,nm and 832\,nm. The data were collected and reduced following the procedures described in \citet{Howell2011}. The resulting reconstructed image achieved a contrast of $\Delta\mathrm{mag}=7.8$ at a separation of 1\arcsec in the 832\,nm band (see bottom panel of Fig.~\ref{fig:speckle}). We note that at the distance of TOI-776, our Zorro speckle images cover a spatial range of 0.46 to 32\,au around the star with contrasts between 5 to 8\,mag.

\subsection{Radial velocity observations}

We obtained 29 high-resolution (R$\approx$115000) spectra of TOI-776 using the High Accuracy Radial velocity Planet Searcher (HARPS) spectrograph mounted at the ESO 3.6\,m telescope of La Silla Observatory, Chile \citep{HARPS}. The observations were carried out as part of our large observing program 1102.C-0923 (PI: Gandolfi) starting on February 5 and until March 23, 2020, when ESO observatories stopped the operations due to the COVID-19 pandemic. One spectrum was acquired under the program 60.A-9709. We used the second fiber of the instrument to monitor the sky background and we reduced the data with the HARPS data reduction software \citep[DRS;][]{2007A&A...468.1115L}. To compute precise radial velocities and spectral diagnostics, we applied on the reduced data the codes \texttt{serval} \citep{SERVAL} and \texttt{TERRA} \citep{TERRA}. Both programs employ a template-matching algorithm that is better suited to derive precise radial velocities for M dwarfs, if compared to the cross-correlation function (CCF) technique implemented in the DRS. In the CCF technique, the line lists of M dwarfs used to define the binary mask are incomplete and they thus produce a CCF which is often a poor match for cool stars. The RVs have a median internal uncertainty of $1.5\,\mathrm{m\,s^{-1}}$ ($1.5\,\mathrm{m\,s^{-1}}$) and a root mean square of $5.2\,\mathrm{m\,s^{-1}}$ ($3.5\,\mathrm{m\,s^{-1}}$) around the mean value for the \texttt{serval} (\texttt{TERRA}) extractions, respectively. We report in Tables~\ref{tab:rvs-serval} and \ref{tab:rvs-terra} of the Appendix the HARPS measurements, the extracted RVs and the associated uncertainties, Na\,{\sc i}\,D, Na\,{\sc ii}\,D, and H$\alpha$ line indices from both programs together with the chromatic index (CRX) and differential line width (dLW) computed by \texttt{serval}, and the Mount Wilson S-index computed by \texttt{TERRA}.

\section{Stellar properties} \label{sec:star}

\begin{table}
\centering
\small
\caption{Stellar parameters of TOI-776.} \label{tab:star}
\begin{tabular}{lcr}
\hline\hline
\noalign{\smallskip}
Parameter                               & Value                 & Reference \\ 
\hline
\noalign{\smallskip}
\multicolumn{3}{c}{Name and identifiers}\\
\noalign{\smallskip}
Name                            & LP~961-53                     & {\citet{1974IAUS...61..169L}}      \\
TOI                             & 776                           & {\it TESS} Science Office      \\  
TIC                             & 306996324                     & {\citet{2018AJ....156..102S}}      \\  
\noalign{\smallskip}
\multicolumn{3}{c}{Coordinates and spectral type}\\
\noalign{\smallskip}
$\alpha$                                & 11:54:18.39       & {\it Gaia} DR2     \\
$\delta$                                & $-$37:33:09.8     & {\it Gaia} DR2     \\
SpT                                     & M1\,V             & {\citet{Gaidos2014MNRAS.443.2561G}}             \\
\noalign{\smallskip}
\multicolumn{3}{c}{Magnitudes}\\
\noalign{\smallskip}
$V$ [mag]                               & $11.54\pm0.04$        & UCAC4       \\
$g$ [mag]                               & $12.35\pm0.12$        & UCAC4       \\
$G$ [mag]                               & $10.7409\pm0.0005$    & {\it Gaia} DR2       \\
$r$ [mag]                               & $10.92\pm0.03$        & UCAC4       \\
$i$ [mag]                               & $10.05\pm0.09$        & UCAC4       \\
$J$ [mag]                               & $8.483\pm0.018$       & 2MASS       \\
$H$ [mag]                               & $7.877\pm0.040$       & 2MASS       \\
$K_s$ [mag]                             & $7.615\pm0.020$       & 2MASS       \\
\noalign{\smallskip}
\multicolumn{3}{c}{Parallax and kinematics}\\
\noalign{\smallskip}
$\pi$ [mas]                             & $36.78\pm0.04$       & {\it Gaia} DR2             \\
$d$ [pc]                                & $27.19\pm0.03 $      & {\it Gaia} DR2             \\
$\mu_{\alpha}\cos\delta$ [$\mathrm{mas\,yr^{-1}}$]  & $+251.112 \pm 0.051$ & {\it Gaia} DR2          \\
$\mu_{\delta}$ [$\mathrm{mas\,yr^{-1}}$]            & $-145.059 \pm 0.083$ & {\it Gaia} DR2          \\
$V_r$ [$\mathrm{km\,s^{-1}}]$           & 49.34$\pm$0.22    & {\it Gaia} DR2    \\
$U$ [$\mathrm{km\,s^{-1}}]$             & 60.71$\pm$0.08    & This work\tablefootmark{a}      \\
$V$ [$\mathrm{km\,s^{-1}}]$             &$-$28.27$\pm$0.17  & This work\tablefootmark{a}      \\
$W$ [$\mathrm{km\,s^{-1}}]$             & 18.73$\pm$0.09    & This work\tablefootmark{a}      \\
\noalign{\smallskip}
\multicolumn{3}{c}{Photospheric parameters}\\
\noalign{\smallskip}
$T_{\mathrm{eff}}$ [K]                      & $3709 \pm 70$             & This work  \\
                                            & $3766 \pm 100$            & \citet{Gaidos2014MNRAS.443.2561G}   \\
$\log g$                                    & $4.727 \pm 0.025$         & This work   \\
{[Fe/H]}                                    & $-0.20 \pm 0.12$          & This work   \\
\noalign{\smallskip}
\multicolumn{3}{c}{Physical parameters}\\
\noalign{\smallskip}
$R$ [$R_{\odot}$]                       & $0.538^{+0.024}_{-0.024}$     & This work   \\ 
                                        & $0.53\pm0.05$                 & \citet{Gaidos2014MNRAS.443.2561G}   \\ 
$L$ [$L_\odot$]                         & $0.049\pm0.002$               & This work \\
                                        & $0.050\pm0.013$               & \citet{Gaidos2014MNRAS.443.2561G} \\
$M$ [$M_{\odot}$]                       & $0.544^{+0.028}_{-0.028}$     & This work   \\ 
                                        & $0.56\pm0.07$                 & \citet{Gaidos2014MNRAS.443.2561G}   \\ 
Age [Gyr]                               & $7.8^{+3.9}_{-6.3}$           & This work   \\ 
\noalign{\smallskip}
\hline
\end{tabular}
\tablebib{
    {\it Gaia} DR2: \citet{GaiaDR2};
    UCAC4: \citet{UCAC4};
    2MASS: \citet{2MASS}.
}
\tablefoot{
\tablefoottext{a}{Computed in the local standard of rest.}
}
\end{table}

\subsection{Stellar parameters}

TOI-776 belongs to the Catalog Of Nearby Cool Host-Stars for Habitable ExopLanets and Life (CONCH-SHELL) compiled by \citet{Gaidos2014MNRAS.443.2561G}. For an all-sky sample of approximately 3000 M- or late K-type stars, the authors provide spectroscopically determined values of the spectral type, effective temperature and metallicity, which combined with empirical relations for cool stars, allow to estimate stellar radius, luminosity and mass. In particular, they measure that TOI-776 is a relatively inactive M1\,V dwarf star with the stellar properties shown in Table~\ref{tab:star}.

We carried out an independent analysis to improve the photospheric and fundamental parameters of TOI-776. We used \texttt{SpecMatch-Emp} \citep{specmatch} to empirically estimate the effective temperature, metallicity, and stellar radius by comparing the co-added HARPS high-resolution spectrum with a spectroscopic library of well-characterized stars. The results of this analysis are in agreement with the values of \citet{Gaidos2014MNRAS.443.2561G} within the errors. Then, we derived the stellar radius and luminosity combining \textit{Gaia} $G$, $G_{\rm BP}$, $G_{\rm RP}$ photometry and 2MASS $J$, $H$, $K_s$ magnitudes with the spectroscopic parameters from the \texttt{SpecMatch-Emp} analysis and the \textit{Gaia} parallax. We corrected the Gaia $G$ photometry for the magnitude dependent offset using Eq.~3 from \citet{CasagrandeVandenberg2018MNRAS.479L.102C}, and adopted a minimum uncertainty of 0.01\,mag for the \textit{Gaia} magnitudes to account for additional systematic uncertainties. We added 0.06\,mas to the nominal \textit{Gaia} parallax to account for the systematic offset found by \citet{StassunTorres2018ApJ...862...61S,Riess2018ApJ...861..126R,Zinn2019ApJ...878..136Z}. Our best estimate of the stellar radius is consistent with the value from \citet{Gaidos2014MNRAS.443.2561G} and in agreement with each of the radius estimates obtained independently using only one of the magnitudes. Finally, we computed the mass using the mass-radius relations for M dwarfs from \citet{Schweitzer2019A&A...625A..68S}.

We also applied the methods of \citet{2006MNRAS.367.1329R} to \textit{Gaia} DR2 astrometry for TOI-776 to compute galactic $U$, $V$, $W$ velocities in the local standard of rest and the probabilities of kinematic membership in galactic stellar populations. We found that TOI-776 has a probability of 96.3\% of belonging to the thin disk population, which is in excellent agreement with the galactic population probabilities for this star in the recent catalog of \citet{2020MNRAS.491.4365C}. Additionally, using the code \texttt{isochrones} \citep{isochrones}, we estimated the age of TOI-776 to be loosely constrained between 2 to 10\,Gyr. From the metallicity, age and kinematics given in Table~\ref{tab:star}, we can conclude that TOI-776 is a relatively old member of the galactic thin disk population.

\subsection{Stellar rotation period} \label{subsec:rotation}


To determine the rotational period of the star, we used the publicly available photometric data for TOI-776. Using \texttt{juliet} (see more details about the algorithm in Sect.~\ref{subsec:fit}) we modeled the ASAS-SN, ASAS, NSVS, Catalina, and daily binned SuperWASP data with Gaussian processes (GPs). In particular, we adopted the quasi-periodic GP kernel introduced in \cite{celerite} of the form 
\begin{equation*}
k_{i,j}(\tau) = \frac{B}{2+C}e^{-\tau/L}\left[\cos \left(\frac{2\pi \tau}{P_\textnormal{rot}}\right) + (1+C)\right] \quad ,
\end{equation*}
where $\tau = |t_{i} - t_{j}|$ is the time-lag, $B$ and $C$ define the amplitude of the GP, $L$ is a timescale for the amplitude-modulation of the GP, and $P_\textnormal{rot}$ is the rotational period of the modulations. As in \citet{Luque2019A&A...628A..39L}, we considered each of the five data-sets to have different values of $B$ and $C$, in order to account for the possibility that different bands could have different GP amplitudes, while we imposed the timescale of the modulation and the rotational period as common parameters for all the data sets. In addition, we fitted for an extra jitter term for each photometric time series. We considered wide uninformative priors for the jitter, $B$, $C$, $L$, and a uniform rotation period prior between 10 and 100\,d. 

\begin{figure}
    \centering
    \includegraphics[width=\hsize]{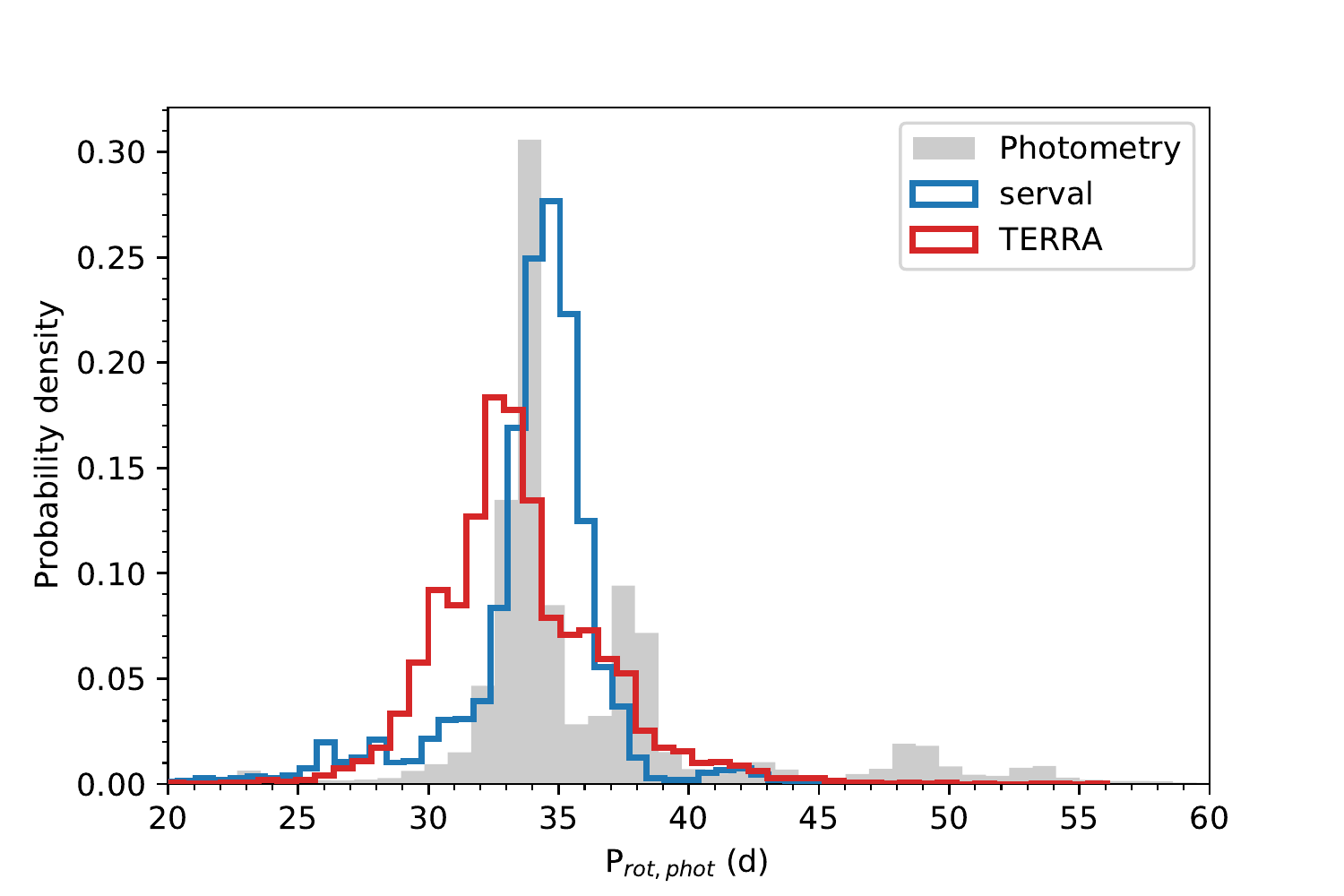}
    \caption{Probability density of the samples of the $P_\textnormal{rot}$ parameter from the GP fit of the ground-based, long-term photometric monitoring (grey) from Sect.~\ref{subsec:rotation} and of the period of the additional sinusoidal signal from the RV fit from Sect.~\ref{subsubsec:rvonly} using \texttt{serval} (blue) or \texttt{TERRA} (red) reductions.  }
    \label{fig:prot-samples}
\end{figure}

Figure~\ref{fig:prot-samples} shows the posterior samples of the GP hyperparameter $P_\textnormal{rot}$ after fitting all the long-term monitoring ground-based photometry. The distribution is bimodal with peaks at $33\pm1\,\mathrm{d}$ and $38\pm1\,\mathrm{d}$, where the samples from the first peak have the highest likelihood. From this we can estimate that the stellar rotation of TOI-776 is between 30 to 40\,d over the course of 15\,yr. The 38\,d peak may be an alias of the true 33\,d rotation due to 1\,yr window function in the photometry. Alternatively, the bimodal distribution of the $P_\textnormal{rot}$ can be explained as a consequence of the stellar differential rotation coupled with the activity cycle \citep{Rudiger2014A&A...572L...7R,Kuker2019A&A...622A..40K}.
For early M dwarfs with rotational periods similar to TOI-776, the expected dynamo cycle time is between 3 to 6\,yr \citep{Kuker2019A&A...622A..40K}, thus detectable in our data. Additionally, assuming that this star is a solar-like rotator, the rotational velocity of the star decreases as the latitude increases. The two peaks correspond to two different groups of activity features, a bigger one, closer to the equator, which generates the first peak of the posterior distribution, and a smaller one, at a higher latitude, which produces the second peak. The opposite situation, with an anti-solar like rotator, is less likely, considering that TOI-776 is an adult star, still belonging to the main sequence. 

\section{Analysis} \label{sec:fit}

\begin{figure}
    \centering
    \includegraphics[width=\hsize]{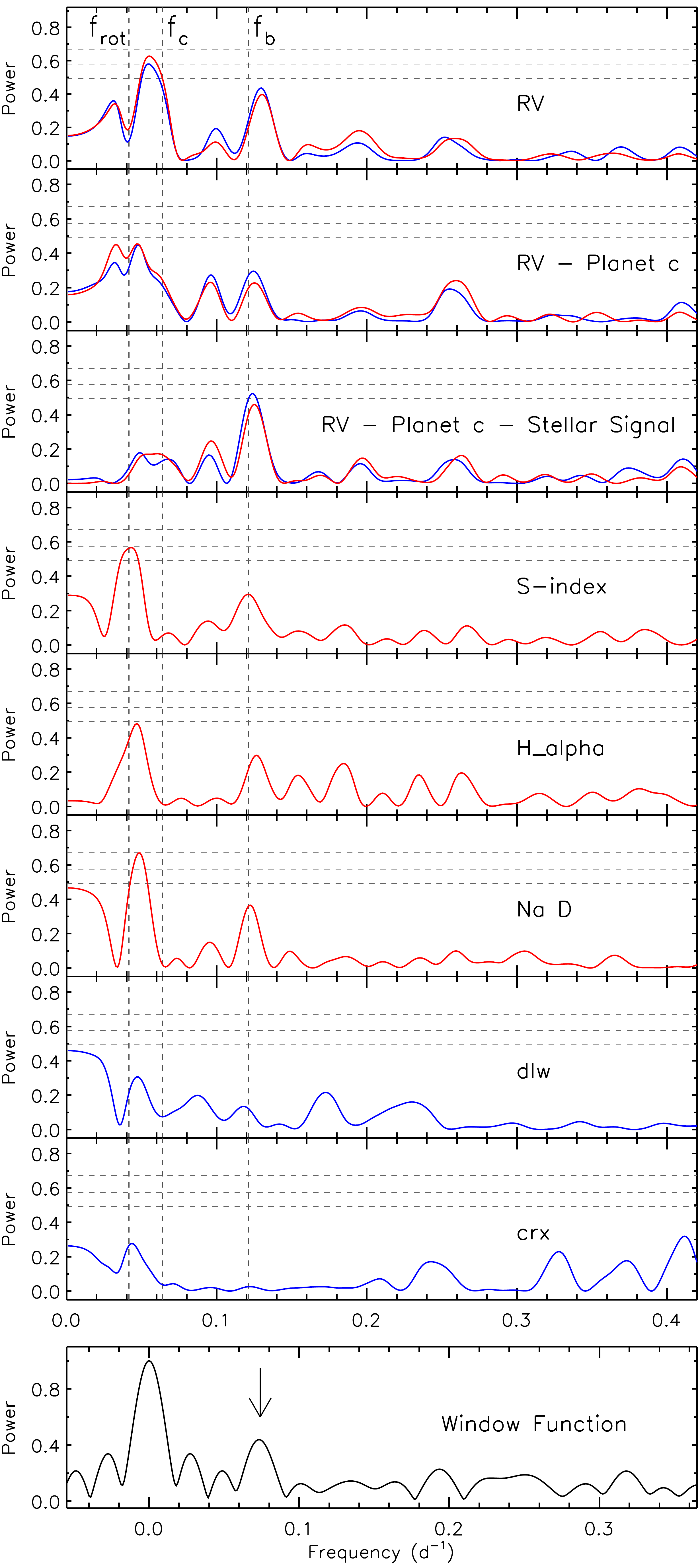}
    \caption{Generalized Lomb-Scargle periodograms of the HARPS RVs and spectral activity indicators from \texttt{serval} (blue) and \texttt{TERRA} (red). The horizontal dashed lines mark, from bottom to top, the 5\%, 1\%, and 0.1\% FAP levels, respectively. The vertical dashed lines mark the orbital frequencies of the two transiting planets (f$_\mathrm{b}$\,=\,0.121\,d$^{-1}$ and f$_\mathrm{c}$\,=\,0.064\,d$^{-1}$) and of the stellar signal at $\sim$0.03\,d$^{-1}$. \emph{Upper panel}: HARPS RVs. \emph{Second panel}: RV residuals following the subtraction of the signal of TOI-776~c. \emph{Third panel}: RV residuals following the subtraction of the reflex motion of TOI-776~c and of the activity-induced stellar signal. \emph{Fourth panel}: S-index. \emph{Fifth panel}: H$\alpha$ line. \emph{Sixth panel}: Na D lines. \emph{Seventh panel}: differential line width (dLW). \emph{Eight panel}: chromatic index (CRX). \emph{Bottom panel}: Window function. The arrow in the bottom panel indicates the peak at 0.07\,d$
   ^{-1}$ referred in the discussion of Sect.~\ref{subsec:gls}.}
    \label{fig:gls_rv}
\end{figure}

\subsection{Frequency analysis of the HARPS data} \label{subsec:gls}

We performed a frequency analysis of the HARPS \texttt{serval}/\texttt{TERRA} extracted measurements to search for the Doppler reflex motion induced by the two transiting planets discovered in the \texttt{TESS} light curve and to unveil the presence of additional signals associated with the star and/or other orbiting planets. 

Figure~\ref{fig:gls_rv} shows the generalized Lomb Scargle \citep[GLS;][]{Zechmeister2009A&A...505..859Z} periodograms of the HARPS RVs and activity indicators extracted with \texttt{serval} (blue lines) and with \texttt{TERRA} (red lines). The horizontal dashed lines mark the GLS powers corresponding to the 0.1, 1, and 5\% false alarm probability\footnote{Following the bootstrap method described, e.g., in \citet{Murdoch1993} and \citet{Hatzes2016}, we estimated the FAP by computing the GLS periodogram of 10$^6$ time series obtained by randomly shuffling the measurements and their uncertainties, while keeping the time-stamps fixed.} (FAP). The vertical dashed lines mark the orbital frequencies of the two transiting planets detected in the \textit{TESS} light curve (f$_\mathrm{b}$\,=\,0.121\,d$^{-1}$ and f$_\mathrm{c}$\,=\,0.064\,d$^{-1}$) and the stellar signal at $\sim$0.03\,d$^{-1}$ (see below).

The upper panel of Fig.~\ref{fig:gls_rv} displays the GLS periodogram of the HARPS RVs in the frequency range 0\,-\,0.42\,d$^{-1}$. The highest peak is found at 0.055\,d$^{-1}$ (FAP\,$\approx$\,0.3\,\%), which is close to the orbital frequency of TOI-776~c (f$_\mathrm{c}$\,=\,0.064\,d$^{-1}$). Taking into account our frequency resolution\footnote{The frequency resolution is defined as the inverse of the time-baseline. The baseline of our HARPS observations is about 47 days, corresponding to a frequency resolution of about 1/47=0.021\,d$^{-1}$.} of 0.021\,d$^{-1}$, the two frequencies are indistinguishable. This suggests that the highest peak seen in the periodogram of the HARPS RVs is the stellar reflex motion induced by the outer transiting planet TOI-776~c. The second highest peak is found at 0.129\,d$^{-1}$ (Fig.~\ref{fig:gls_rv}, upper panel), which is close to the orbital frequency of TOI-776~b. However, this signal is an alias of the signal at 0.055\,d$^{-1}$. The periodogram of the window function indeed shows a peak at 0.074\,d$^{-1}$ (highlighted with an arrow in the bottom panel of Fig.~\ref{fig:gls_rv}), which is equal to the frequency spacing between the two highest peaks seen in the periodogram of the HARPS RVs.

We used the code \texttt{pyaneti} \citep{2019MNRAS.482.1017B} (Sect.~\ref{subsubsec:joint}) to subtract the Doppler signal of TOI-776~c from the HARPS RVs. We assumed a circular model (see also Sect.~\ref{subsubsec:rvonly}), fixing period and time of first transit to the \textit{TESS} ephemeris, while allowing for the systemic velocity and RV semi-amplitude to vary. The periodogram of the RV residuals shows a broad peak centered around $\sim$0.04\,d$^{-1}$ with a FAP of about 10\,\%. Although the Doppler signal is not significant, the GLS periodograms of the CRX, dLW, H$\alpha$, and S-index activity indicators show also peaks at $\sim$0.04\,d$^{-1}$, suggesting that this signal is caused by the presence of active regions appearing and disappearing from the visible stellar disk as the star rotates around its axis. It is worth noting that the peak at 0.130\,d$^{-1}$ is not observed in the GLS periodogram of the RV residuals, corroborating the interpretation that this peak is an alias of the dominant frequency detected in the periodogram of the HARPS data.

We removed the Doppler reflex motion of TOI-776~c and the activity-induced RV signal by jointly modeling the HARPS measurements with a circular Keplerian orbit and a sine curve. For TOI-776~c we followed the same procedure described in the previous paragraph. For the stellar signal we fitted for the phase, amplitude, and frequency. The latter was allowed to vary within a wide uniform prior centered around 0.04\,d$^{-1}$. The GLS periodogram of the RV residuals displays a peak at 0.125\,d$^{-1}$ (FAP\,$\approx$\,11\,\%), which is very close to the frequency of the inner transiting planet TOI-776~b (f$_\mathrm{b}$\,=\,0.121\,d$^{-1}$). We note that the activity indicators show also peaks close to the orbital frequency of TOI-776~b. Yet, those peaks are separated by 0.074\,d$^{-1}$ from the stellar signal at $\sim$0.04\,d$^{-1}$. As such they are very likely aliases of the latter.

\subsection{Modeling results} \label{subsec:fit}

\begin{table}
    \centering
    \small
    \caption{Model comparison of RV-only fits with \texttt{juliet}. The prior label $\mathcal{N}$ represents a normal distribution. The final model used for the joint fit is marked in boldface (see Sect.~\ref{subsubsec:rvonly} for details about the selection of the final model). }  \label{tab:models}
    \begin{tabular}{lllcc}
        \hline
        \hline
        \noalign{\smallskip}
        Model & Prior $P_{\rm planet}$ & GP  & $\ln Z_\mathrm{\tt serval}$   & $\ln Z_\mathrm{\tt TERRA}$  \\
        \noalign{\smallskip}
        \hline
        \noalign{\smallskip}
0pl         & \dots                                     & \dots                     & $-$84.3 & $-$85.4     \\[0.1cm]
2pl         & $\mathcal{N}_\mathrm{b}(8.24,0.05^2)$     & \dots                     & $-$81.3 & $-$81.8     \\
            & $\mathcal{N}_\mathrm{c}(15.65,0.05^2)$    &                           & &         \\[0.1cm]
2pl+GP1   & $\mathcal{N}_\mathrm{b}(8.24,0.05^2)$       & EXP\tablefootmark{a}      & $-$80.7 & $-$81.7      \\
            & $\mathcal{N}_\mathrm{c}(15.65,0.05^2)$    &                           & &          \\[0.1cm]
2pl+GP2   & $\mathcal{N}_\mathrm{b}(8.24,0.05^2)$       & ESS\tablefootmark{b}      & $-$80.6 & $-$81.8     \\
            & $\mathcal{N}_\mathrm{c}(15.65,0.05^2)$    &                           & &         \\[0.1cm]
{\bf 2pl+sinusoid}     & $\mathcal{N}_\mathrm{b}(8.24,0.05^2)$     & \dots          & {\bf -78.9} & $-$79.4     \\
            & $\mathcal{N}_\mathrm{c}(15.65,0.05^2)$    &                           & &         \\
            & $\mathcal{N}_\mathrm{d}(35.0,10.0^2)$       &                         & &         \\[0.1cm]
        \noalign{\smallskip}
        \hline
    \end{tabular}
    \tablefoot{
        \tablefoottext{a}{Simple exponential kernel (EXP) of the form $k_{i,j} = \sigma^2_\mathrm{GP,RV} \exp\left(-|t_i - t_j|/T_\mathrm{GP,RV}\right)$}.
        \tablefoottext{b}{Exponential-sine-squared kernel (ESS) of the form $k_{i,j} = \sigma^2_\mathrm{GP,RV} \exp\left(- \alpha_\mathrm{GP,RV} (t_i - t_j)^2 - \Gamma_\mathrm{GP,RV} \sin^2 \left[\frac{\pi |t_i - t_j|}{P_{\rm rot;GP,RV}}\right]\right)$ with a uniform prior in $P_{\rm rot;GP,RV}$ ranging from 5 to 50\,d.} 
    }
\end{table}

\begin{table}[t]
    \centering
    \caption{Median and the 68\% credibility intervals of the posterior distributions for each fitted parameter of the final joint model obtained for the TOI-776 system using \texttt{juliet}. Priors and descriptions for each parameter can be found in Table~\ref{tab:priors}.}
    \label{tab:posteriors}
    \begin{tabular}{l@{\hspace{-3mm}}c@{\hspace{3mm}}c@{\hspace{3mm}}} 
        \hline
        \hline
        \noalign{\smallskip}
        Parameter & TOI-776~b & TOI-776~c  \\
        \noalign{\smallskip}
        \hline
        \noalign{\smallskip}
        \multicolumn{3}{c}{\it Stellar parameters} \\[0.1cm]
        \noalign{\smallskip}
        $\rho_\star$ ($\mathrm{kg\,m\,^{-3}}$)       & \multicolumn{2}{c}{$6024^{+650}_{-640}$} \\[0.1 cm]
        \noalign{\smallskip}
        \multicolumn{3}{c}{\it Planet parameters} \\[0.1cm]
        \noalign{\smallskip}
        $P$ (d)                                & $8.24661^{+0.00005}_{-0.00004}$ & $15.6653^{+0.0004}_{-0.0003}$  \\[0.1 cm]
        $t_0$\tablefootmark{(a)}                  & $8571.4167^{+0.0010}_{-0.0011}$ & $8572.5999^{+0.0018}_{-0.0016}$  \\[0.1 cm]
        $r_1$                                & $0.43^{+0.10}_{-0.07}$ & $0.51^{+0.08}_{-0.07}$  \\[0.1 cm]
        $r_2$                                & $0.0316^{+0.0008}_{-0.0011}$ & $0.03437^{+0.0009}_{-0.0008}$  \\[0.1 cm]
        $e$\tablefootmark{(b)}              & $0.06^{+0.03}_{-0.02}~(<0.18)$  & $0.04^{+0.02}_{-0.01}~(<0.18)$  \\[0.1 cm]
        $\omega$                                & $-67^{+117}_{-73}$ & $-11^{+55}_{-79}$  \\[0.1 cm]
        $K$ ($\mathrm{m\,s^{-1}}$)           & $1.88^{+0.40}_{-0.44}$ & $2.05^{+0.67}_{-0.68}$  \\[0.1 cm]
        \noalign{\smallskip}
        \multicolumn{3}{c}{\it Photometry parameters} \\[0.1cm]
        \noalign{\smallskip}
        $\sigma_{\textnormal{TESS}}$ (ppm)     & \multicolumn{2}{c}{$1.1^{+8.0}_{-1.0}$} \\[0.1 cm] 
        $q_{1,\textnormal{TESS}}$                & \multicolumn{2}{c}{$0.26^{+0.29}_{-0.17}$} \\[0.1 cm]
        $q_{2,\textnormal{TESS}}$                & \multicolumn{2}{c}{$0.36^{+0.21}_{-0.20}$} \\[0.1 cm]
        $\sigma_{\textnormal{LCO-CTIO}}$ (ppm)     & \multicolumn{2}{c}{$1000^{+36}_{-34}$} \\[0.1 cm] 
        $M_{\textnormal{LCO-CTIO}}$ (ppm)            & \multicolumn{2}{c}{$890^{+395}_{-468}$} \\[0.1 cm]
        $\theta_{\textnormal{LCO-CTIO}}$           & \multicolumn{2}{c}{$0.0011^{+0.0003}_{-0.0004}$} \\[0.1 cm]
        $q_{1,\textnormal{LCO-CTIO}}$                & \multicolumn{2}{c}{$0.76^{+0.13}_{-0.14}$} \\[0.1 cm]
        $\sigma_{\textnormal{LCO-SAAO}}$ (ppm)     & \multicolumn{2}{c}{$471^{+60}_{-60}$} \\[0.1 cm] 
        $M_{\textnormal{LCO-SAAO}}$ (ppm)            & \multicolumn{2}{c}{$-810^{+250}_{-260}$} \\[0.1 cm]
        $\theta_{\textnormal{LCO-SAAO}}$           & \multicolumn{2}{c}{$-0.0004\pm0.0002$} \\[0.1 cm]
        $q_{1,\textnormal{LCO-SAAO}}$                & \multicolumn{2}{c}{$0.73^{+0.14}_{-0.15}$} \\[0.1 cm]
        $\sigma_{\textnormal{LCO-SSO}}$ (ppm)     & \multicolumn{2}{c}{$885^{+45}_{-40}$} \\[0.1 cm] 
        $M_{\textnormal{LCO-SSO}}$ (ppm)            & \multicolumn{2}{c}{$-4057^{+386}_{-413}$} \\[0.1 cm]
        $\theta_{\textnormal{LCO-SSO}}$           & \multicolumn{2}{c}{$-0.0031\pm0.0003$} \\[0.1 cm]
        $q_{1,\textnormal{LCO-SSO}}$                & \multicolumn{2}{c}{$0.46^{+0.16}_{-0.18}$} \\[0.1 cm]
        $\sigma_{\textnormal{MEarth}}$ (ppm)     & \multicolumn{2}{c}{$1734^{+52}_{-47}$} \\[0.1 cm] 
        $M_{\textnormal{MEarth}}$ (ppm)            & \multicolumn{2}{c}{$-2^{+47}_{-44}$} \\[0.1 cm]
        $q_{1,\textnormal{MEarth}}$                & \multicolumn{2}{c}{$0.73^{+0.13}_{-0.14}$} \\[0.1 cm]
        \noalign{\smallskip}
        \multicolumn{3}{c}{\it RV parameters} \\[0.1cm]
        \noalign{\smallskip}
        $\mu_{\textnormal{HARPS}}$ ($\mathrm{m\,s^{-1}}$)            & \multicolumn{2}{c}{$4.33^{+0.51}_{-0.58}$} \\[0.1 cm]
        $\sigma_{\textnormal{HARPS}}$ ($\mathrm{m\,s^{-1}}$)       & \multicolumn{2}{c}{$1.66^{+0.35}_{-0.30}$} \\[0.1 cm]
        \noalign{\smallskip}
        \multicolumn{3}{c}{\it GP hyperparameters and additional sinusoid} \\
        \noalign{\smallskip}
        $\sigma_\mathrm{GP,TESS}$ (ppm) & \multicolumn{2}{c}{$0.17^{+0.06}_{-0.04}$} \\[0.1 cm]
        $T_\mathrm{GP,TESS}$ (d)        & \multicolumn{2}{c}{$0.56^{+0.19}_{-0.15}$} \\[0.1 cm]
        $K$ ($\mathrm{m\,s^{-1}}$)      & \multicolumn{2}{c}{$2.71^{+0.53}_{-0.60}$}  \\[0.1 cm]
        $t_0$\tablefootmark{(b)}        & \multicolumn{2}{c}{$8607.0^{+11.6}_{-12.1}$}  \\[0.1 cm]
        $P$ (d)                         & \multicolumn{2}{c}{$34.4^{+1.4}_{-2.0}$}  \\[0.1 cm]

        \noalign{\smallskip}
        \hline
    \end{tabular}
    \tablefoot{
        \tablefoottext{a}{Units are BJD - 2450000.}
        \tablefoottext{b}{3$\sigma$ upper limit in parenthesis.}

    }
\end{table}

\begin{table}[t]
    \centering
    \caption{Derived planetary parameters obtained for the TOI-776 system using the posterior values from Table~\ref{tab:posteriors} and stellar parameters from Table~\ref{tab:star}.}
    \label{tab:derivedparams}
    \begin{tabular}{lcc} 
        \hline
        \hline
        \noalign{\smallskip}
        Parameter\tablefootmark{(a)} & TOI-776~b & TOI-776~c  \\
        \noalign{\smallskip}
        \hline
        \noalign{\smallskip}
        \multicolumn{3}{c}{\it Derived transit parameters} \\[0.1cm]
        \noalign{\smallskip}
        $p = R_{\rm p}/R_\star$            & $0.0316^{+0.0008}_{-0.0011}$ & $0.0344^{+0.0009}_{-0.0008}$  \\[0.1 cm]
        $b = (a/R_\star)\cos i_{\rm p}$    & $0.25^{+0.10}_{-0.14}$       & $0.27^{+0.12}_{-0.11}$ \\[0.1 cm]
        $a/R_\star$                  & $27.87^{+0.97}_{-1.02}$      & $42.75^{+1.49}_{-1.57}$  \\[0.1 cm]
        $i_p$ (deg)              & $89.65^{+0.22}_{-0.37}$      & $89.51^{+0.25}_{-0.21}$  \\[0.1 cm]
        $t_T$ (h)                & $2.41^{+0.11}_{-0.10}$       & $2.99^{+0.16}_{-0.13}$  \\[0.1 cm]
        \noalign{\smallskip}
        \multicolumn{3}{c}{\it Derived physical parameters} \\[0.1cm]
        \noalign{\smallskip}
        $M_{\rm p}$ ($M_\oplus$)    & $4.0\pm0.9$   & $5.3\pm1.8$ \\[0.1 cm]
        $R_{\rm p}$ ($R_\oplus$)        & $1.85\pm0.13$   & $2.02\pm0.14$  \\[0.1 cm]
        $\rho_{\rm p}$ (g cm$^{-3}$)             & $3.4^{+1.1}_{-0.9}$      & $3.5^{+1.4}_{-1.3}$  \\[0.1 cm]
        $g_{\rm p}$ (m s$^{-2}$)                 & $11.2^{+3.1}_{-2.8}$     & $12.8^{+4.9}_{-4.4}$  \\[0.1 cm]
        $a_{\rm p}$ (au)                         & $0.0652\pm0.0015$ & $0.1000\pm0.0024$ \\[0.1 cm]
        $T_\textnormal{eq}$ (K)\tablefootmark{(b)}          & $514\pm17$   & $415\pm14$ \\[0.1 cm]
        $S$ ($S_\oplus$)            & $11.5\pm0.6$  & $4.9\pm0.2$  \\[0.1 cm]
        \noalign{\smallskip}
        \hline
    \end{tabular}
    \tablefoot{
      \tablefoottext{a}{Error bars denote the $68\%$ posterior credibility intervals.}
      \tablefoottext{b}{Equilibrium temperatures were calculated assuming zero Bond albedo and uniform surface temperatures across the entire planet.}
      }
\end{table}

In this section, we use \texttt{juliet} \citep{juliet} to model the photometric and Doppler data, both separately and jointly. The algorithm is built on several publicly available tools which model transits \citep[\texttt{batman},][]{batman}, RVs \citep[\texttt{radvel},][]{radvel}, and GPs (\texttt{george}, \citealt{Ambikasaran2015ITPAM..38..252A}; \texttt{celerite}, \citealt{celerite}). 


\subsubsection{Photometry} \label{subsubsec:photonly}

First, to constrain the properties of the transiting planets and use them for further analyses, we modeled the \textit{TESS}, LCO, and MEarth photometry with \texttt{juliet}. We adopted a quadratic limb darkening law for \textit{TESS}, since \citet{espinoza_jordan_limb:2015} showed it was appropriate as well for space-based missions. The limb darkening parameters were then parametrized with a uniform sampling prior ($q_1, q_2$), introduced by \citet{Kipping13}. For LCO and MEarth transits, we used a more simple linear limb darkening law, because the lower data precision with respect to\textit{TESS}  prevents us from adopting a more complex law. Additionally, we followed the parametrization introduced in \cite{Espinoza18}. In particular, for each transiting planet, rather than fitting for the planet-to-star radius ratio $p=R_p/R_*$ and the impact parameter of the orbit $b$, we sampled from the uniform priors assigned to two parameters, $r_1$ and $r_2$, which are connected to $p$ and $b$ with the equations (1)-(4) in \cite{Espinoza18}. $r_1$ and $r_2$ were shown in \citet{Espinoza18} to guarantee a full exploration of the physically plausible values in the ($p, b$) plane. We assumed as well circular orbits and fixed the \textit{TESS} dilution factor to $1$, based on our analysis from Sect.~\ref{subsec:hri}. Finally, we added in quadrature a jitter term $\sigma$ to the {\it TESS}, LCO, and MEarth photometric uncertainties. The details of the priors and the description for each parameter are presented in Table~\ref{tab:priors} of the Appendix.

To account for the time-correlated noise in the light curve in Fig.~\ref{fig:tess_lc}, even using the PDC-corrected SAP, we modeled the \textit{TESS} photometry with the exponential GP kernel
\begin{equation*}
    k_{i,j} = \sigma^2_\mathrm{GP,TESS} \exp\left(- |t_i - t_j|/T_\mathrm{GP,TESS}\right)
\end{equation*}
where $T_\mathrm{GP,TESS}$ is a characteristic timescale and $\sigma_\mathrm{GP,TESS}$ is the amplitude of this GP modulation. For the LCO photometry, on the other hand, we used a linear model to detrend the data from airmass correlations.

Our photometry-only analysis increases significantly the precision of the planet parameters with respect to the \texttt{TESS} DVR. The uncertainties in the period decreases by two orders of magnitudes which eases up future ground- and space-based follow-up efforts. The radii of the planets are determined to a precision better than 5\%. 
Finally, we searched for an additional planets in the system by modeling a three-planet fit with the same priors as in Table~\ref{tab:priors} for the transiting planets, and varying the period and mid-transit time of the third hypothetical planet. Our result significantly exclude the presence of any additional transits in the light curve ($\Delta \ln Z = \ln Z_{\rm 2pl} - \ln Z_{\rm 3pl} > 7$).

\subsubsection{RV} \label{subsubsec:rvonly}

Even though the results of the RVs extraction slightly change whether we use \texttt{serval} or \texttt{TERRA}, the GLS analysis in both cases show the evidence of a stellar signal together with the RV trends associated with the transiting planets. To adequately describe the data, we considered several RV-only models and carried out a model comparison scheme as in \citet{Luque2019A&A...628A..39L}. We used \texttt{juliet}, a code which efficiently computes the Bayesian log-evidence of each tested model and explores the parameter space using the importance nested sampling included in MultiNest \citep{multinest} via the \texttt{PyMultiNest} package \citep{pymultinest}. As discussed in \citet{Nelson2020AJ....159...73N}, this method outperforms other samplers in choosing robustly the best model for those with 3 or less planets. We considered a model to be moderately favored over another if the difference in its Bayesian log-evidence $\Delta \ln Z$ is greater than two, while strongly favored if it is greater than five \citep{2008ConPh..49...71T}. If $\Delta \ln Z \lesssim 2$, then the models are indistinguishable. In this case, the model with fewer degrees of freedom would be chosen. 

Due to the sampling and the scarce number of RV measurements, if we model the eccentricity with a wide uninformative prior we derive nonphysically high eccentricities for both planets that would make the system unstable in less than a hundred orbits. The eccentricity of systems with multiple transiting planets is low but not necessarily zero \citep{VanEylen2015ApJ...808..126V,Xie2016PNAS..11311431X,Hadden2017AJ....154....5H}. Therefore, instead of assuming circular orbits, we place a prior on the orbital eccentricity of a Beta distribution with $\alpha = 1.52$ and $\beta = 29$ following \citet{vanEylen19}.   
Table~\ref{tab:models} summarizes the results of our analysis on both \texttt{serval}- or \texttt{TERRA}-extracted RVs. As seen in Table~\ref{tab:models}, including the two transiting planets in the model is favored against the fiducial model (0pl). On the other hand, we tested different types of two planet models. First, we considered just the two transiting planets (2pl), without accounting for additional noise sources. Then, we accounted for the stellar noise, modeling it in three different ways: with an exponential GP kernel (2pl+GP1), with an exponential sine-squared GP kernel (2pl+GP2) and with a simple sinusoid (2pl+sinusoid). All the tested two-planet models are statistically indistinguishable, with their Bayesian log-evidences within $\Delta \ln Z < 2$. 

However, for both \texttt{serval} and \texttt{TERRA}-extracted RVs, the nominal best model accounts for two circular orbits and an additional sinusoidal curve, whose period is equal to the stellar period of rotation we estimated through the long-term ground-based photometric data. For this test we imposed a normal prior on $P_{\mathrm{rot}}$, with a wide standard deviation ($10$). We additionally tried wide, uninformative priors for the period of the sinusoidal signal and we retrieved the same posterior distributions and log-evidences (Fig.~\ref{fig:prot-samples}) as for the test with a gaussian prior. With the RV analysis, we estimated a stellar period of rotation $P_{\mathrm{rot}}=34.4^{+1.4}_{-2.0}\,\mathrm{d}$, consistent with the rotational period estimated from the ground-based long-term photometry in Section \ref{subsec:rotation}. Additionally, all models presented in Table~\ref{tab:models} derive the same RV semi-amplitude for TOI-776~b and TOI-776~c, well within their 1$\sigma$ uncertainties. This proves the robustness of the mass determination for the transiting planets, independently of the stellar noise distribution.

Leveraging the prior information on the stellar rotation from photometry discussed in Sect.~\ref{subsec:rotation} with the presence of a significant periodicity in the RV residuals of a two-planet model (Fig.~\ref{fig:gls_rv}b), we decided to choose the 2pl+sinusoid as our final model for the joint fit. With respect to the RV extraction, we preferred to use the \texttt{serval} extracted RVs in the final joint fit due to their nominal highest log-evidence and lower jitter compared to \texttt{TERRA}.


\subsubsection{Joint fit} \label{subsubsec:joint}

\begin{figure*}[t]
    \centering
    \includegraphics[width=\hsize]{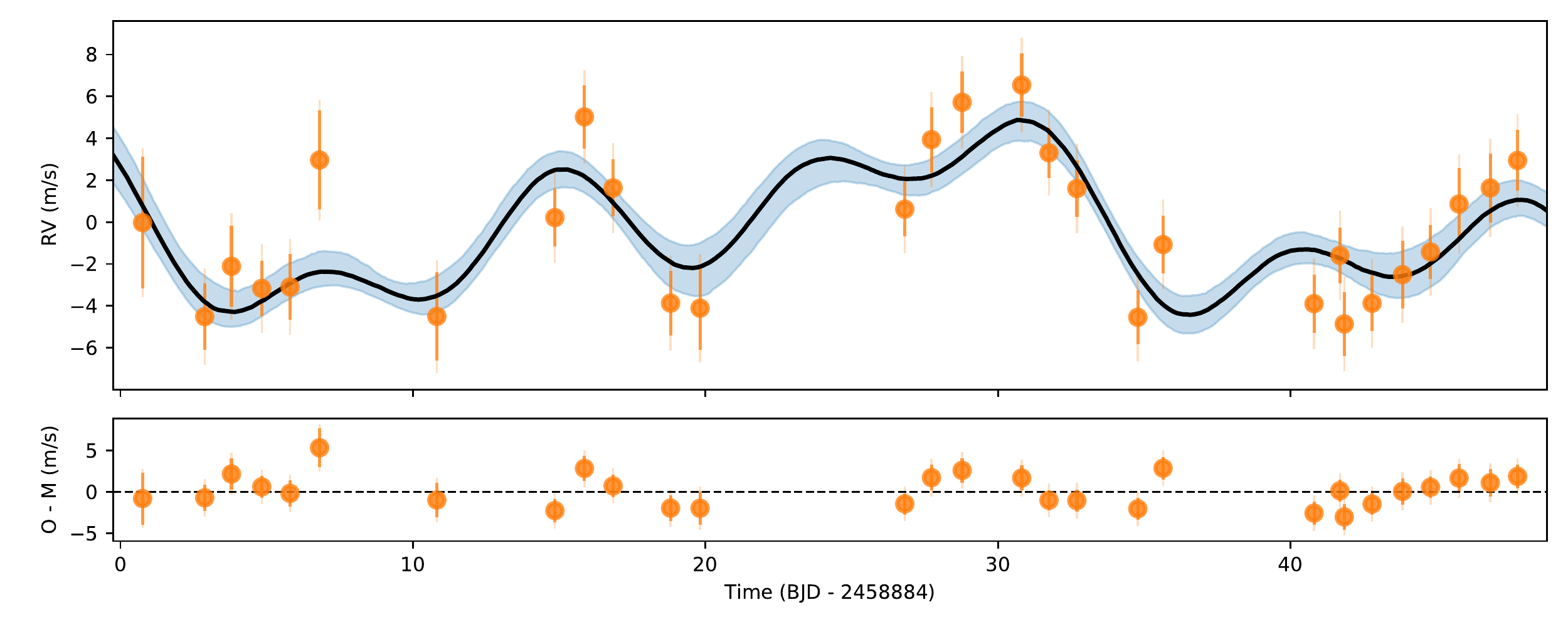}\\
    \includegraphics[width=\hsize]{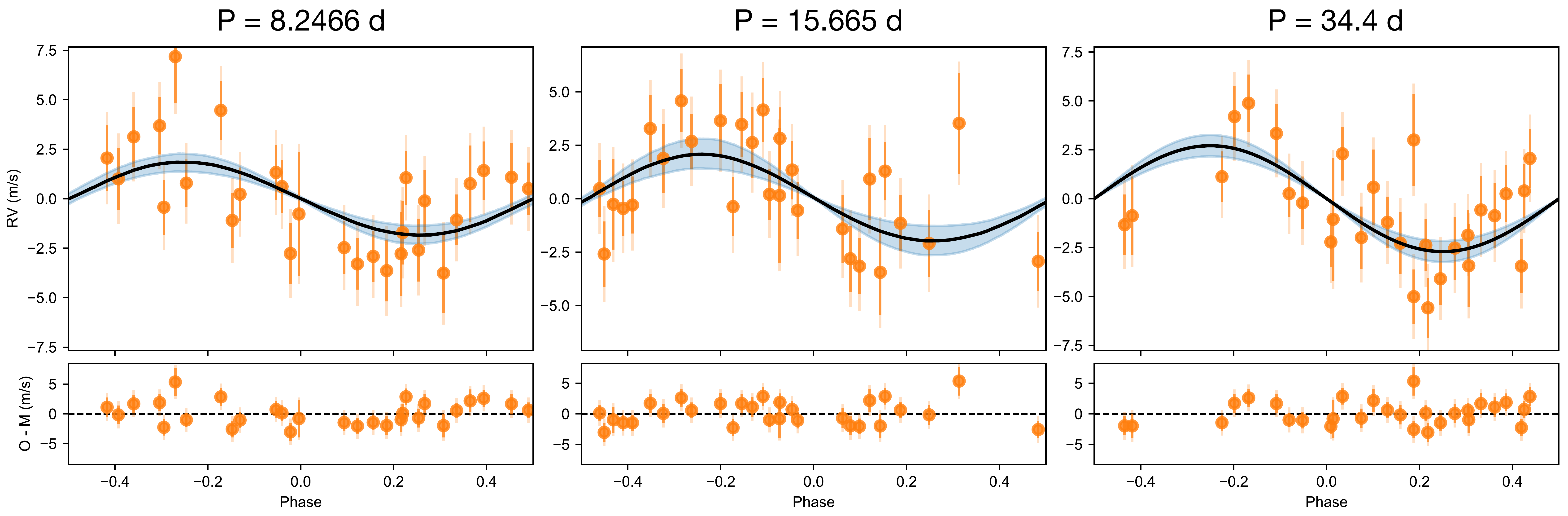}
    \caption{Top panel: time series of the HARPS \texttt{serval} RVs and the best model discussed in Sect.~\ref{subsubsec:rvonly} and the residuals from the fit below. The blue shaded area corresponds to the 1$\sigma$ confidence interval of the model. Bottom panel: RVs phase-folded to the period (shown above each panel) of the two confirmed planets (TOI-776~b, left; TOI-776~c, center) and the additional sinusoid associated with the stellar variability. In both panels the error bars of the RV data have the extra jitter term added in quadrature and plotted in lighter orange for its visualization.} 
    \label{fig:jointfit}
\end{figure*}

We performed a joint fit using \texttt{juliet} of the \textit{TESS}, LCO, and MEarth photometry and HARPS \texttt{serval} extracted RVs, using the 2pl+sinusoid model we selected after the RV-only analysis in Sect.~\ref{subsubsec:rvonly}. Table~\ref{tab:priors} and \ref{tab:posteriors} shows the priors and posteriors of all the fitted parameters, respectively. Figure~\ref{fig:corner_main} shows a corner plot of the orbital parameters of planets b and c. The data, residuals, and joint fit best model are shown in Figs.~\ref{fig:all_lc}~and~\ref{fig:jointfit} for the photometry and the RVs, respectively. Table~\ref{tab:derivedparams} lists the transit and physical parameters, derived using the stellar parameters in Table~\ref{tab:star}. 

As a sanity check, we performed an independent joint analysis of the transit photometry and Doppler measurements using the code \texttt{pyaneti} \citep{2019MNRAS.482.1017B}, which estimates the parameters of planetary systems in a Bayesian framework, combined with an MCMC sampling. We imposed uniform priors for all the fitted parameters. Following \citet{Winn10}, we sampled for the mean stellar density $\rho_\star$ and recovered the scaled semi-major axis ($R_{\rm p}/R_\star$) for each planet using Kepler’s third law. We found that the modeling of the transit light curves provides a mean stellar density of $\rho_\star$\,=\,5203$^{+1782}_{-1228}$\,kg\,m$^{-3}$, which agrees with the density of 4834$^{+651}_{-559}$\,kg\,m$^{-3}$ derived from the stellar mass and radius presented in Sect.~\ref{sec:star}. As for the remaining parameters, the analysis provides consistent parameter estimates with those derived with \texttt{juliet}, corroborating our results.

\section{Results and Discussion} \label{sec:discussion}

The TOI-776 system consists of two transiting planets. The inner planet, TOI-776~b, has a period of 8.25\,d, 
a radius of $1.85\pm0.13\,R_\oplus$, a mass of $4.0\pm0.9\,M_\oplus$, and a bulk density of $3.4^{+1.1}_{-0.9}\,\mathrm{g\,cm^{-3}}$. The outer planet, TOI-776~c, has a period of 15.66\,d, 
a radius of $2.02\pm0.14\,R_\oplus$, a mass of $5.3\pm1.8\,M_\oplus$, and a bulk density of $3.5^{+1.4}_{-1.3}\,\mathrm{g\,cm^{-3}}$. The RV data show only one additional signal with 
a semi-amplitude of $\sim 2.7\,\mathrm{m\,s^{-1}}$ and a period of 34\,d associated with the stellar rotation, as suggested by our analyses of the photometry and spectral line indicators.

\subsection{System architecture} \label{subsec:architecture}

\begin{figure*}
    \centering
    \includegraphics[width=\hsize]{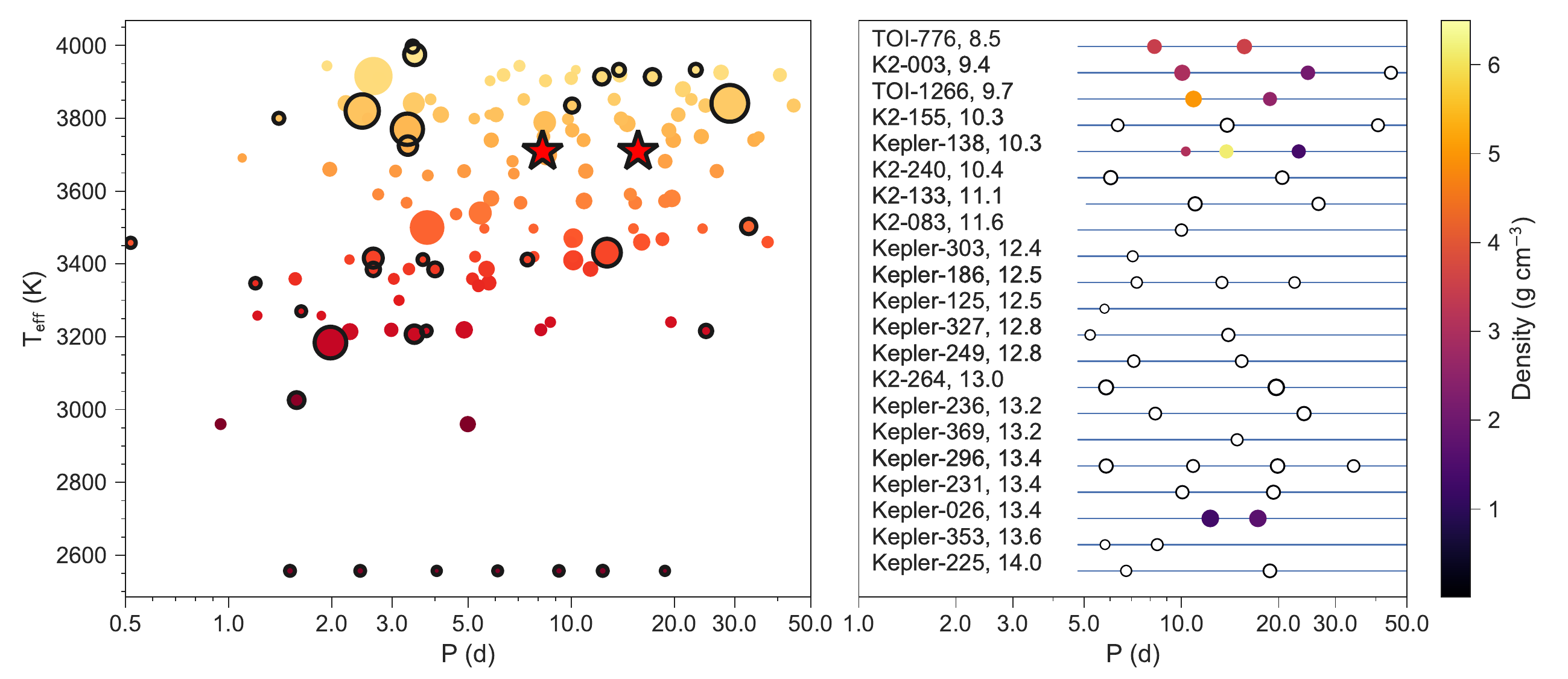}
    \caption{{\it Left}: Confirmed transiting planets from the TEPCat database \citep{Southworth2011MNRAS.417.2166S} around M dwarfs as a function of period. Black circled points indicate planets with a mass determination better than 30\%. Circles are color-coded by the host effective temperature and their sizes are proportional to the planet radius. The red stars mark the two planets in the TOI-776 system. {\it Right}: Transiting multi-planetary systems around early-type M dwarfs ($3500\,\mathrm{K} < T_\mathrm{eff} < 4000\,\mathrm{K}$) with similar architectures to the TOI-776 system. Sizes are proportional to the planet radius and color indicates their bulk density. Planets with no mass determination are marked in white. Next to the system's name is indicated the brightness of the host star in $J$ band.}
    \label{fig:multis}
\end{figure*}

While the occurrence rate of planets around early M dwarfs ($3500\,\mathrm{K} < T_\mathrm{eff} < 4000\,\mathrm{K}$) has been investigated in detail with \textit{Kepler} and \textit{K2} samples \citep[see e.g.,][]{Dress13,Dress15,Montet2015ApJ...809...25M,Hirano2018AJ....155..127H}, the number of currently known planets transiting low-mass stars is still much smaller with respect to those discovered around solar-type stars. While none of these surveys were optimized for M dwarfs, we expect more statistically significant results from the \textit{TESS} mission for these stars. Figure~\ref{fig:multis} shows the confirmed transiting planets around M dwarfs as a function of the orbital period and the effective temperature of the host star. However, very few of these systems have precise determinations of the planetary masses (i.e densities), eccentricities and orbital architectures that would be required to link the statistical properties of this population with planet formation and evolution models in the low stellar mass regime.

There are several validated transiting multi-planetary systems orbiting early M dwarfs similar to TOI-776 in terms of planetary architecture. 
Kepler-225, Kepler-236 and Kepler-231 are two-planet transiting systems composed of super-Earth and/or mini-Neptune sized companions with similar periods and semi-major axes, all validated by \citet{Rowe2014ApJ...784...45R}. However, these systems are on average 5\,mag fainter than TOI-776 and the planets do not have a mass determination nor precise stellar parameters.
Similarly, K2-240 \citep{DiezAlonso2018MNRAS.480L...1D} has two transiting super-Earths with periods of 6 and 20.5\,d, although they do not have mass determination and orbit an active star that is 2\,mag fainter with a clear photometric rotational period of 10.8\,d. 
The two outermost planets of the four-planet system K2-133 have periods and sizes similar to TOI-776~b and c, but the star is at the faint-end for RV follow-up and does not exhibit transit timing variations (TTVs).

If compared to systems with mass determination, TOI-776 shows some similarities with Kepler-26 \citep{Steffen2012MNRAS.421.2342S}, Kepler-138 \citep{Rowe2014ApJ...784...45R}, TOI-1266 \citep{Demory2020A&A...642A..49D}, and K2-3 \citep{Montet2015ApJ...809...25M,Crossfield2015ApJ...804...10C}. 
Kepler-26~b and c have periods of 12.3 and 17.2\,d, respectively, and bulk densities compatible with those of sub-Neptunes determined from TTVs. However, the system has two more planets without mass determination, an inner Earth-sized planet and an outer mini-Neptune sized planet. 
Kepler-138 is a very interesting system of three small planets, whose densities were estimated through photodynamical modeling \citep{Almenara2018MNRAS.478..460A}. The most similar to the TOI-776 planets in terms of orbital period, Kepler-138~b (10.3\,d) and c (13.8\,d), are very different in composition, the former being a Mars analogue and the latter a prototypical rocky super-Earth. The third, outermost planet seems to have retained a substantial volatile-rich envelope.
TOI-1266 is the system that resembles TOI-776 the most. The two planets of the system have tentative dynamical masses determined from TTVs, although RVs are likely to become available in the future. The planets straddle the radius valley and, interestingly, the innermost is larger and more massive than the outer one.
K2-3, the brightest of all four systems, has three small transiting planets and only the two inner ones (with periods of 10 and 24.6\,d) have a mass determination using HARPS-N, HARPS, HIRES and PFS RVs \citep{Almenara2015A&A...581L...7A,Damasso2018A&A...615A..69D,Kosiarek2019AJ....157...97K}, only an upper limit is measured for the third (with a period of 44.5\,d). The planets have a similar composition, compatible to that of water-worlds or water-poor planets with gaseous envelopes, however the poor bulk density estimations of planets c and d impede further conclusions. 
The right panel of Fig.~\ref{fig:multis} shows all of the aforementioned systems, color-coded by bulk density and with the $J$-band magnitude of their host stars indicated.

Therefore, we conclude that, although multi-planetary systems of super-Earths and/or sub-Neptunes are common around early-type M dwarfs, only TOI-776 has all of its planets well characterized, bulk density uncertainties better than 30\%, precise stellar parameters and a host star bright enough for atmospheric follow-up observations with current and planned facilities.

\subsection{Dynamics and TTV analysis}

We investigated possible TTVs through a 3-body simulation, using the Python Tool for Transit Variations \citep[{\tt PyTTV};][]{Korth2020}. We simulated the estimated TTVs and RVs using the stellar and planetary parameters reported in Table~\ref{tab:star}, \ref{tab:posteriors} and \ref{tab:derivedparams} and found an expected TTV signal with a period of $\sim$\,150\,d and a maximum amplitude of $\sim$\;2\,min for the inner planet. Thus, the time span of the photometric observations, their cadence and signal-to-noise would prevent a detection of TTVs with the currently available data. 

Additionally, we carried out a set of dynamical simulations to study the long-term stability of the system. We used the parameters in Table~\ref{tab:posteriors} and \ref{tab:derivedparams} and randomly drew 1000 samples from the posterior distributions as initial parameters for the dynamical simulations. We integrated each parameter set for $10^{6}$ orbits of the inner planet, using the tool \texttt{REBOUND} \citep{Rein2012} with the standard IAS15 integrator \citep{Rein2015}. We also explored the stability using the MEGNO criteria as implemented in \texttt{REBOUND}. In the cases of close encounters between the bodies or one body ejection, the system would be flagged as unstable for the specific set of parameters. We found that the systems is dynamically stable over the entire integration time and for the whole parameter posterior space. 


\subsection{Planetary composition and interior structure} \label{subsec:interior}

\begin{figure*}
    \centering
    \includegraphics[width=0.49\hsize]{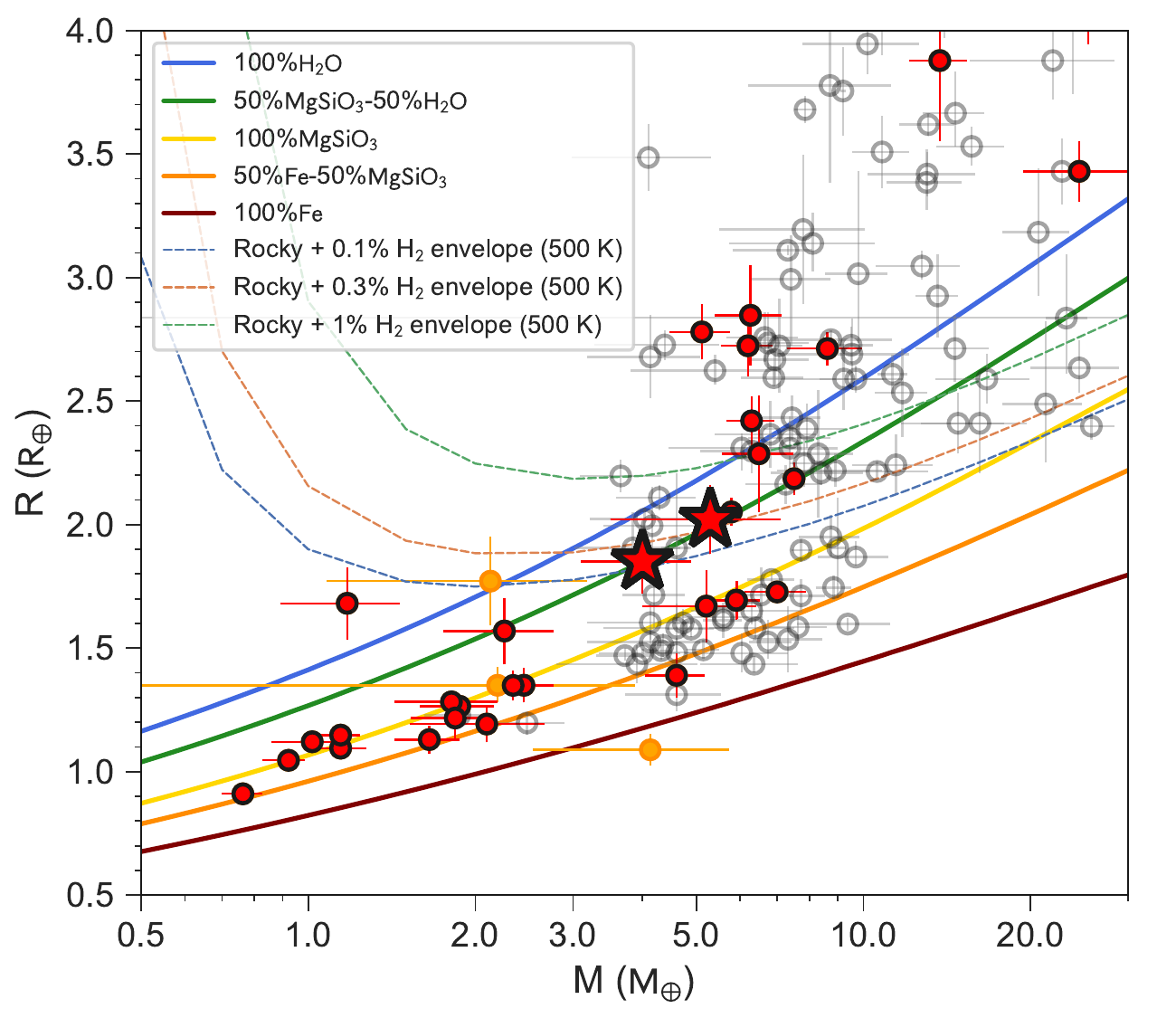}
    \includegraphics[width=0.49\hsize]{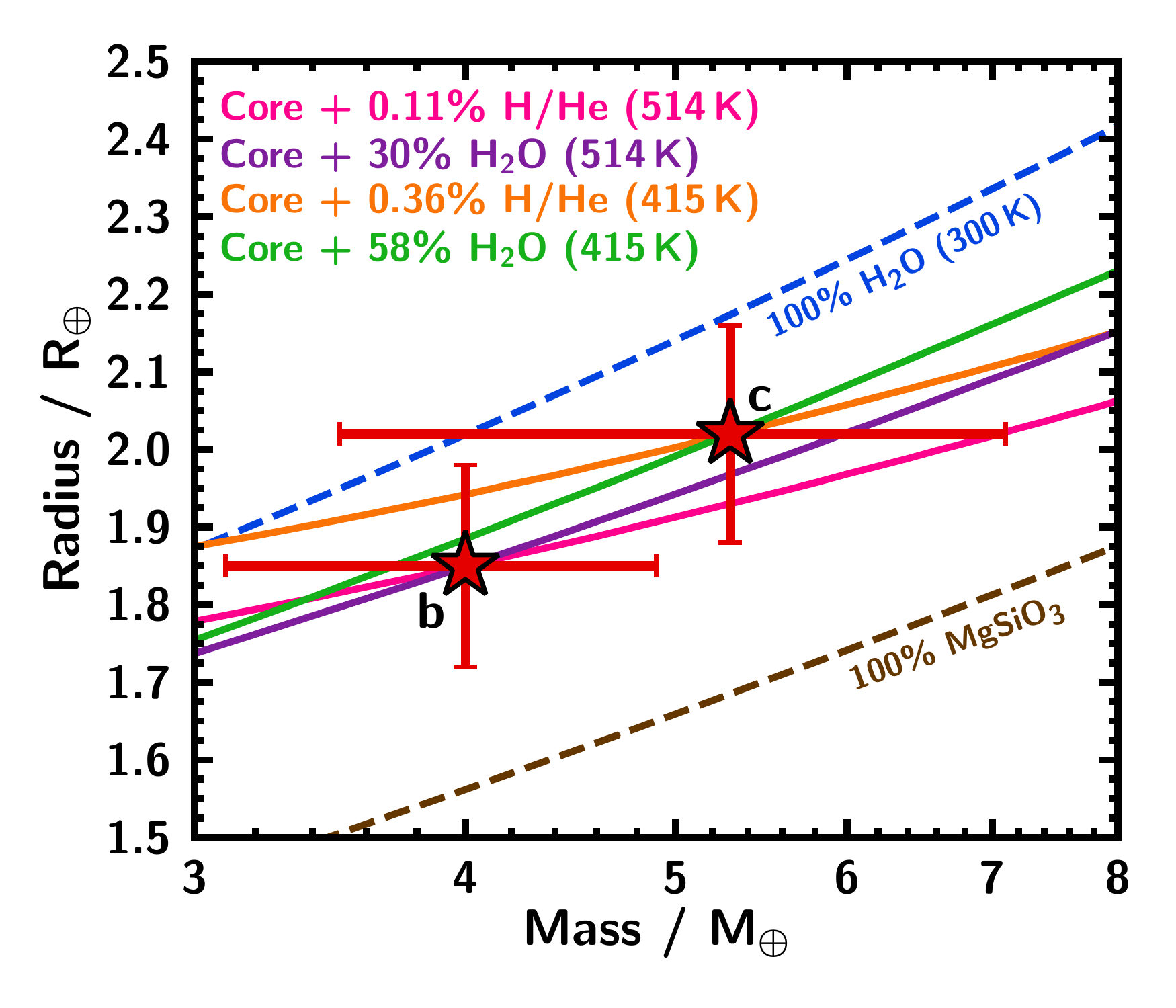}
    \caption{Mass-radius diagrams in Earth units. In the {\it left} panel, open circles are transiting planets around F-, G-, and K-type stars with mass and radius measurement better than 30\,\% from the TEPCat database of well-characterized planets \citep{Southworth2011MNRAS.417.2166S}, red circles are planets around M dwarfs with mass and radius measurement, orange filled circles are planets around M dwarfs with mass determinations worse than 30\,\%, and the red stars are TOI-776~b and c which have masses determined with accuracies of 23\,\% and 34\,\%, respectively. In the {\it left} panel, the color lines are the theoretical $R$-$M$ models of \citet{Zeng2016ApJ...819..127Z} and \citet{Zeng2019PNAS..116.9723Z}. In the {\it right} panel, the solid pink and purple lines show the models from Sect.~\ref{subsec:interior} that are consistent with the mass and radius of TOI-776~b, and the orange and green lines show compositions consistent with the mass and radius of TOI-776~c, assuming an Earth-like core (1/3 iron, 2/3 silicates). }
    \label{fig:massradius}
\end{figure*}

\begin{figure}
    \centering
    \includegraphics[width=\hsize]{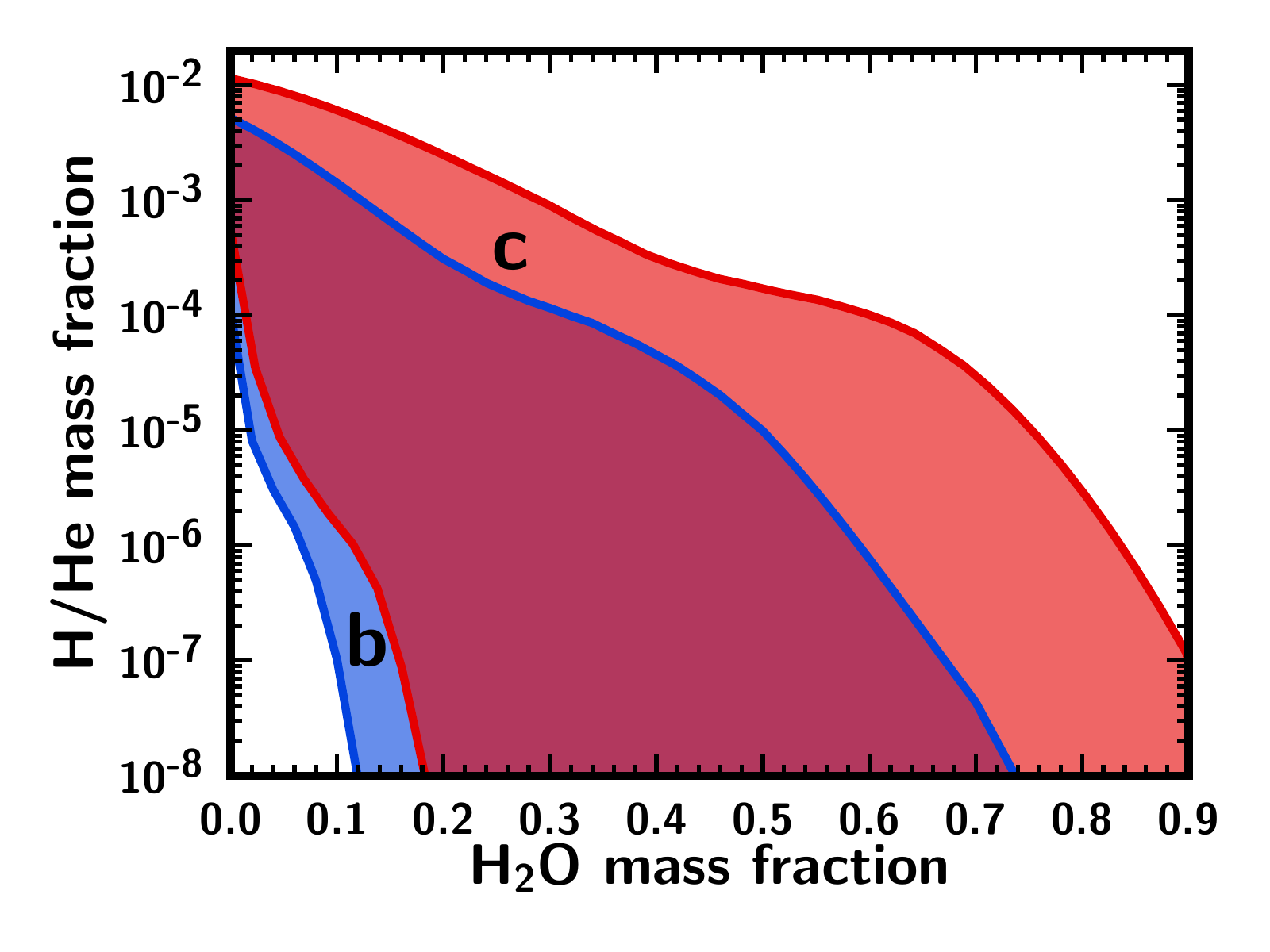}
    \caption{H/He vs. H$_2$O mass fractions for the best-fitting interior compositions ($\leq1\sigma$) permitted by the masses and radii of TOI-776~b and c, assuming an Earth-like core, for two different pressure-temperature profiles with radiative-convective boundaries at 1 and 100 bar. The blue shaded region indicates possible compositions for TOI-776~b, and the red shaded region shows compositions for TOI-776~c. The darker red shaded area between the two corresponds to the range of possible compositions that could explain both planets. For TOI-776~b, the H$_2$O mass fraction is constrained to be $\leq 73\%$ and the H/He mass fraction is $\leq 0.52\%$. For TOI-776~c the upper limit for H/He is $1.2\%$.  A purely H$_2$O planet would be consistent with this mass and radius, but we only show H$_2$O mass fractions up to $90\%$.}
    \label{fig:h2o_hhe}
\end{figure}

Figure~\ref{fig:massradius} shows the location of the TOI-776 system in a mass-radius diagram. Both planets occupy a scarcely populated region, characterized by a lack of planets around M dwarfs and with precise bulk density measurements. A comparison with the theoretical models by \citet{Zeng2016ApJ...819..127Z}, reported in the left panel of Fig.~\ref{fig:massradius}, shows that TOI-776~b and c are consistent with mixtures of silicates and water in a 50-50 proportion. We adopted the three-layer models from \citet{2013PASP..125..227Z} and \citet{Zeng2016ApJ...819..127Z} to infer the interior structure of the planets. However, given the mass and radius input, the solution of the model is degenerate. As a consequence, the same mass-radius pair can lead to a broad range of combinations of iron, silicate and water mass fractions. On the other hand, when we applied the latest models by \citet{Zeng2019PNAS..116.9723Z}, assuming a 1\,mbar surface pressure level and an equilibrium temperature of 500\,K (from Table~\ref{tab:derivedparams}), we found that an Earth-like rocky core with a 0.1\% and a 0.3\% molecular hydrogen atmosphere is consistent with the bulk densities of TOI-776~b and c, respectively. Nonetheless, it is clear that both of the planets in the system have an internal composition ranging from water worlds to rocky planets that have retained a significant atmosphere. 
    
For a better understanding of the nature of the two exoplanets, we performed a more detailed modeling of their interior compositions, using their masses, radii and surface temperatures. Our model considers a canonical four-layer structure consisting of a two-component iron and silicate core, a layer of H$_2$O and a H/He envelope. We assume that the core is Earth-like in composition (a third of iron, two-thirds of silicates by mass), meaning the core, water and H/He envelope mass fractions ($x_{\rm core}$, $x_{\rm H_2O}$, $x_{\rm H/He}$) are free parameters which sum to unity. The model solves the planetary structure equations of mass continuity and hydrostatic equilibrium assuming spherical symmetry. Further detail regarding the internal structure model can be found in \citet{Madhu2020} and Nixon~\&~Madhusudhan (submitted).

The equation of state (EOS) prescriptions for the iron and silicate layers are adopted from \citet{Seager2007}, who used a Vinet EOS of the $\epsilon$ phase of Fe \citep{Vinet1989,Anderson2001} and a Birch-Murnaghan EOS of MgSiO$_3$ perovskite \citep{Birch1952,Karki2000}. Thermal effects in these layers are ignored, since they have a small effect on the planetary radius \citep{Howe2014}. However, thermal effects in the outer envelope can alter the mass-radius relation significantly \citep{Thomas2016}. For this reason the model uses a temperature-dependent EOS for the outer H$_2$O and H/He layers. For H$_2$O, we used a patchwork EOS in order to cover all possible phases of H$_2$O that might be present in the interior, compiled from \citet{Salpeter1967,Fei1993,Wagner2002,Feistel2006,Seager2007,French2009,Klotz2017}, and \citet{Journaux2020}. For H/He we use the EOS in \citet{Chabrier2019}, which assumes a solar helium fraction ($Y=0.275$). The temperature profile in the envelope is isothermal from the surface down to some radiative-convective boundary, where it becomes adiabatic. The pressure at the radiative-convective boundary $P_{\rm rc}$ is a free parameter in the model. For this study, we considered values of $P_{\rm rc}$ ranging from 1--100\,bar.

We explore the parameter space of possible compositions in ($x_{\rm core}$, $x_{\rm H_2O}$, $x_{\rm H/He}$) space. For each composition, we consider a range of masses that agree with the observed mass of the planet to within 1$\sigma$. For a given mass $\hat{M}$, the model radius $\hat{R}$ is computed and the $\chi^2$ statistic is calculated:
\begin{equation}
    \chi^2 = \frac{(M_p-\hat{M})^2}{\sigma_M^2} + \frac{(R_p-\hat{R})^2}{\sigma_R^2},
\end{equation}
where ($\sigma_M,\sigma_R$) are the observed uncertainties on the mass and radius of each planet.

The bulk densities of TOI-776~b and c ($3.4^{+1.1}_{-0.9}\,\mathrm{g\,cm^{-3}}$ and $3.5^{+1.5}_{-1.3}\,\mathrm{g\,cm^{-3}}$, respectively) are too low for either planet to have a purely terrestrial (iron plus rock) composition. Therefore, the planets must possess an envelope with some amount of H$_2$O and/or H/He, in order to explain their masses and radii. The right panel of Fig.~\ref{fig:massradius} shows limiting cases for each planet in which the envelope composition is either purely H$_2$O or purely H/He. The mass and radius of TOI-776~b can be explained to within $1\sigma$ ($\chi^2 \leq 1$) with a pure H$_2$O envelope of 12--73\% by mass or a pure H/He envelope with mass fraction $1.1 \times 10^{-4}$--$5.2 \times 10^{-3}$. Best-fit solutions (those which minimise $\chi^2$) for pure envelopes are found at $x_{\rm H_2O}=0.3$ and $x_{\rm H/He}=1.1 \times 10^{-3}$. TOI-776~c might have larger envelopes; within $1\sigma$, it is consistent with a pure H$_2$O layer of $\geq$18\% or a pure H/He envelope with mass fraction $5.4 \times 10^{-4}$--$1.2 \times 10^{-2}$. The best-fit pure-envelope solutions for TOI-776~c are $x_{\rm H_2O}=0.58$ and $x_{\rm H/He}=3.6 \times 10^{-3}$. Each of the best-fit models, shown in the right panel of  Fig.~\ref{fig:massradius}, have a radiative-convective boundary at $P_{\rm rc}=10\,\mathrm{bar}$.

It is also possible that the planets in this system have both H$_2$O and H/He components, as well as an iron/rock core. For the three components, we explored the full range of plausible values ($x_{\rm core}$, $x_{\rm H_2O}$ and $x_{\rm H/He}$) that could explain the interior compositions of each planet. We considered two different temperature profiles for each planet, with $P_{\rm rc}=1$ and 100\,bar. Figure~\ref{fig:h2o_hhe} shows themass fractions of water and H/He compatible to within 1$\sigma$ ($\chi^2 \leq 1$) with the masses and radii of TOI-776~b and c. We obtained upper limits on the total H$_2$O and H/He mass fractions for TOI-776~b: $x_{\rm H_2O} \leq 73\%$ and $x_{\rm H/He} \leq 0.52\%$. These correspond to cases with pure H$_2$O or H/He envelopes as previously discussed. For TOI-776~c, we find that $x_{\rm H/He} \leq 1.2\%$. A $100\%$ H$_2$O planet would theoretically be consistent with the mass and radius of TOI-776~c, but this would be unrealistic from a planet formation perspective, as some rocky material is needed for further accretion of ice and gas \citep{Lee2016ApJ...817...90L}. Figure~$\ref{fig:h2o_hhe}$ shows as well a significant overlap between the best-fitting shaded regions for the two planets, meaning that the planets could also share the same composition.

The masses and radii of TOI-776~b and c allow for a wide range of possible solutions, from water worlds with steam atmospheres to mostly rocky planets with hydrogen-rich envelopes, however they are inconsistent with bare rocks without atmospheres. Our models assume a surface pressure of 0.1\,bar, meaning a water-world solution for either planet yields a steam atmosphere. On the other hand, a higher surface pressure could result in liquid H$_2$O at the surface. A rocky planet with an outgassed secondary atmosphere which includes carbon compounds is unlikely: \citet{ElkinsTanton2008} placed an upper limit on the mass fraction for this type of atmosphere at 5\%. The lower mass limits in the case of pure H$_2$O envelopes are 8\% and 18\% for TOI-776~b and c respectively. On the other hand, in a carbon-rich atmosphere, the dominant species, CO$_2$, has a higher mean molecular weight than H$_2$O, leading to a lower atmospheric scale height. All things considered, we can infer that a 5\% carbon-rich atmosphere is less than what would be needed to explain the planet radii. However, determining whether the two planets have H$_2$O- or H/He-rich atmospheres is impossible with the present data. Atmospheric observations of the planets would be required in order to break this degeneracy.

\subsection{Radius gap in M dwarfs}

\begin{figure*}
    \centering
    \includegraphics[width=\hsize]{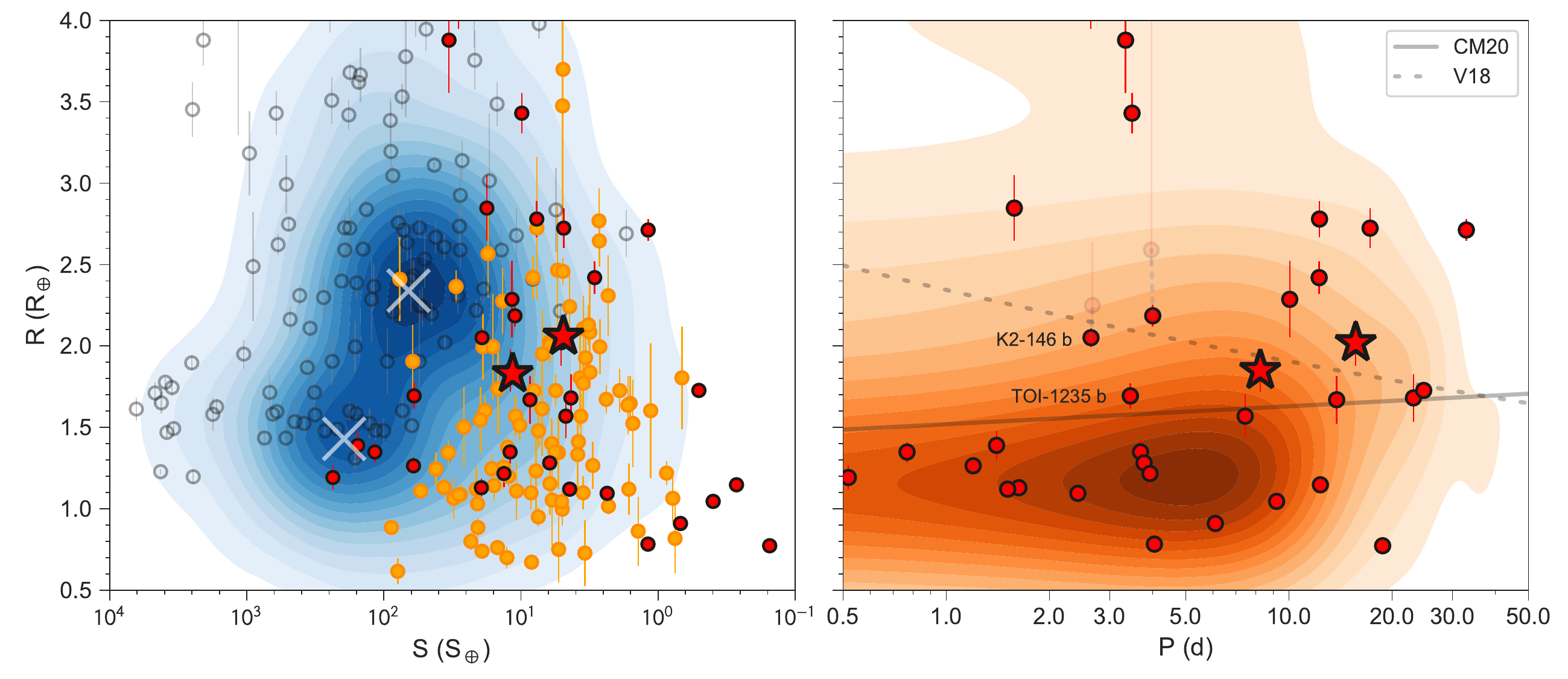}
    \caption{Insolation-radius ({\em left}) and period-radius ({\em right}) diagrams in Earth units. In both panels the different circles represent the same planets as in Fig.~\ref{fig:massradius} from the TEPCat database \citep{Southworth2011MNRAS.417.2166S}. In the left panel, we plot in blue the $R$-$S$ point density of all the known confirmed transiting planets with contours, and sub-Neptunes and super-Earths density maxima with white crosses. In the right panel, the orange contours represent the density of planets around M dwarfs with mass determinations worse than 30\% or without mass constraints at all (orange circles in the {\em left} panel). The dashed line represents the location of the radius valley for F-, G-, and K-type stars from \citet{VanEylen2018MNRAS.479.4786V}, consistent with the predictions from photo-evaporation and core-powered mass loss models, while the solid line represents the location of the radius valley for mid-K and mid-M dwarfs from \citet{CloutierMenou2020AJ....159..211C}, consistent with gas-poor formation scenarios. Together with TOI-776~b, the other two systems that are within both lines are K2-146 \citep[solid,][translucent]{Hamann2019AJ....158..133H,Lam2020AJ....159..120L} and TOI-1235 \citep{Bluhm2020arXiv200406218B,Cloutier2020AJ....160...22C}. }
    \label{fig:radiusgap}
\end{figure*}

The occurrence rate distribution of close-in planets exhibits a paucity of planets between $1.7-2.0\,R_\oplus$ \citep{Fulton17,Fulton18,Hardegree-Ullman2020ApJS..247...28H} around FGK stars ($T_{\rm eff}>4700\,\mathrm{K}$) and between $1.4-1.7\,R_\oplus$ \citep{Hirano2018AJ....155..127H,CloutierMenou2020AJ....159..211C} around mid-K to mid-M dwarfs ($T_{\rm eff}<4700\,\mathrm{K}$). This feature is pointed out as the result of the transition from small rocky planets to larger non-rocky planets with volatile-rich envelopes \citep{2014ApJ...783L...6W,Dress15}. Recent studies showed that the location of the radius gap depends on the orbital period or, alternatively, on the planet's insolation \citep{VanEylen2018MNRAS.479.4786V,Martinez2019ApJ...875...29M,CloutierMenou2020AJ....159..211C}. Additionally, the width and center of the radius gap also depends on whether the host star is single or part of a multiple star system \citep{Teske2018AJ....156..292T}.

According to the above discussion, if we consider the radius axis in Fig.~\ref{fig:massradius}, TOI-776~b and c belong, within the uncertainties, to the radius gap in the case of FGK stars. On the other hand, they are well above the radius gap if we account for mid-K to mid-M dwarfs. Similarly, when looking at the distribution of transiting planets in a radius-insulation diagram (Fig.~\ref{fig:radiusgap}, left panel), the TOI-776 planets lie above the radius valley --- the two-dimensional view of the radius gap --- that separates rocky super-Earths from volatile-rich sub-Neptunes around FGK stars.

The right panel of Fig.~\ref{fig:radiusgap} shows the period-radius diagram of all known exoplanets with precise bulk density measurements that orbit M dwarfs. The dashed line marks the empirical location of the radius valley for FGK stars, following \citet{VanEylen2018MNRAS.479.4786V}, while the solid line indicates the location for mid-K to mid-M dwarfs as in \citet{CloutierMenou2020AJ....159..211C}. The change in slope as a function of stellar type is the result of a change in the dominant mechanism responsible for sculpting the radius valley. For instance, the thermally driven mass loss, caused by photo-evaporation or core-powered mechanisms, becomes less efficient toward low-mass stars. The measured slope for mid-K to mid-M dwarfs suggests that gas-poor formation \citep{Lee2014ApJ...797...95L,Lee2016ApJ...817...90L,Lopez2018MNRAS.479.5303L} might be the main process from which small planets form. However, thousands of small planets around low-mass stars with precise radii are needed in order to robustly state if the radius valley is the result of the erosion or the gas-poor formation scenarios \citep{Cloutier2020AJ....160...22C}. Although enriching the sample of exoplanet systems orbiting M dwarfs is nowadays possible thanks to \textit{TESS} and future space-based missions such as \textit{PLATO} an alternative is to obtain precise bulk density measurements of exoplanets lying in the region of discrepancy between models. 

TOI-776~b joins TOI-1235~b \citep{Bluhm2020arXiv200406218B,Cloutier2020AJ....160...22C} and K2-146~b \citep{Lam2020AJ....159..120L,Hamann2019AJ....158..133H} inside the period-radius region where thermally driven mass loss models disagree with the predictions from gas-poor formation. However, K2-146~b belongs to this region if we refer to the parameters reported in \citet{Hamann2019AJ....158..133H}, because the radius estimated by \citet{Lam2020AJ....159..120L} (see translucent points in Fig.
~\ref{fig:radiusgap}) is more than 2$\sigma$ higher, causing the planet to be placed outside the radius valley. Our previous analyses show that both TOI-776~b and c are likely to have retained a significant atmosphere, with slightly different envelope mass fractions. This result, given their period and radius, would be consistent with the predictions from gas-poor formation models. 

On the other hand, the system's composition may be reconciled with thermally driven mass loss because the inner, most irradiated planet, has a smaller envelope mass fraction compared to its outer companion. Unlike other known systems whose planets straddle both sides of the radius gap \citep[e.g.,][]{Dumusque2014ApJ...789..154D,2017AJ....154..266N,Nowak2020A&A...642A.173N}, TOI-776 is an interesting case where photo-evaporation could have stopped or become inefficient early in the planet's history. But, it is possible that the planets are currently undergoing mass loss under the core-powered mechanism, which erodes sub-Neptune planets into rocky super-Earths in Gyr timescales \citep{Ginzburg2018MNRAS.476..759G}, contrary to the few Myr timescale when photo-evaporation is effective \citep{SanzForcada2011A&A...532A...6S}. As reported in Table~\ref{tab:star}, the age of TOI-776 is between 2 and 10\,Gyr. However, the current data precision and limited number of known planets in this specific regime hamper any further investigation in favor of one or the other mechanism of formation. New studies on the dependence of the radius valley with other stellar parameters such as the age or metallicity, together with a larger sample of well-characterized planets in or near the radius valley, will help discerning between them in a demographic sense \citep{Hardegree-Ullman2020ApJS..247...28H,Berger2020AJ....160..108B,Gupta2020MNRAS.493..792G}.

However, for the first time, we can compare between planets which belong to this region of the parameter space where formation models make opposing predictions. TOI-1235~b has a rocky composition with a 90\% confidence upper limit in the envelope mass fraction of 0.5\%, thus incompatible with a gas-poor formation scenario. We reach the opposite conclusion for TOI-776~b and c, whose bulk densities imply the presence of a volatile envelope making them compatible with the predictions from gas poor formation mechanisms, given their periods and radii. Therefore, although other stellar parameters might need to be taken into account, we can tentatively predict that the stellar mass below which thermally-driven mass loss is no longer the main formation pathway for sculpting the radius valley is probably between 0.63 and 0.54\,$M_\odot$, which correspond to the host stellar masses of TOI-1235 and TOI-776, respectively. More planets in this interesting region of the parameter space with precise bulk density measurements are key to reveal the mechanisms responsible of the radius valley emergence around low-mass stars with respect to solar-like stars.

\subsection{Atmospheric characterization}

\subsubsection{Transmission Spectroscopy Metric (TSM)}

\begin{figure*}[ht]
    \centering
    \includegraphics[width=\hsize]{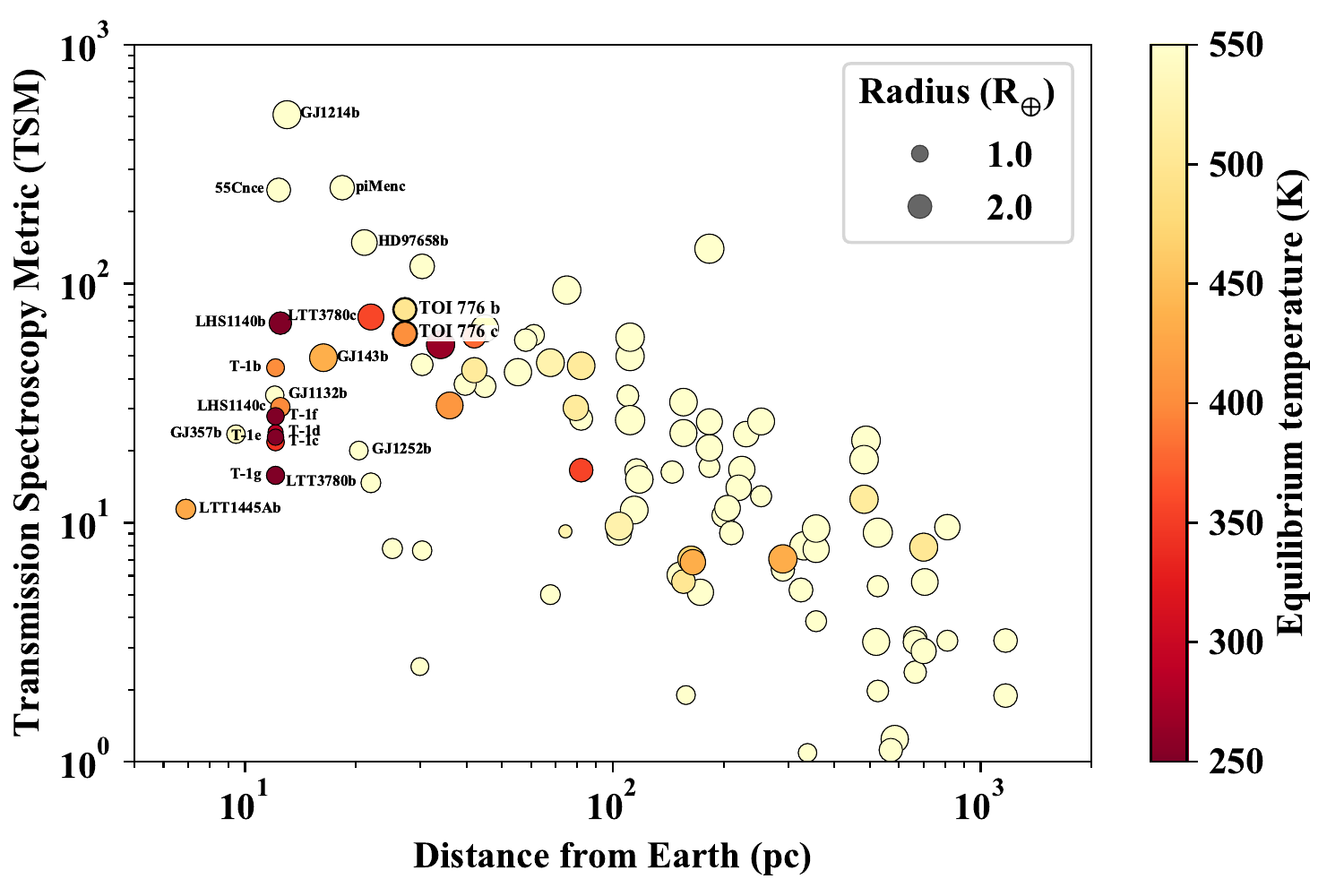}
    \caption{The transmission spectroscopy metric (TSM) for exoplanets from the Exoplanet Encyclopedia with a radius less than 3\,$R_\oplus$ and a mass determination by RVs or TTVs. TOI-776~b and c are labeled and marked with thicker black borderlines.}
    \label{fig:TSM}
\end{figure*}

We used the proposed metric by \cite{kempton2018framework} to evaluate the suitability of the TOI-776 planets for atmospheric characterization studies. Figure~\ref{fig:TSM} shows the transmission spectroscopy metric (TSM) for all exoplanets in the Exoplanet Encyclopedia\footnote{\url{www.exoplanet.eu}} with a radius less than $3\,R_\oplus$. We used the scale factors listed in Table~1 from \citet{kempton2018framework} as opposed to the suggested value for temperate planets, 0.167, to compute the TSM values in Fig.~\ref{fig:TSM}. The estimated TSM of TOI-776~b and c are 77.9 and 61.8 respectively, which places them among the top priority targets for atmospheric follow-ups of small planets around nearby stars. This is not surprising, because TOI-776 is one of the brightest M dwarfs with known transiting planets. However, most of the planets shown in Fig.~\ref{fig:TSM} are well below the radius gap which makes the TOI-776 system a valuable target for atmospheric characterization in order to trace the formation and evolution of multi-planetary systems orbiting low-mass stars and break the degeneracy of internal composition models.

\subsubsection{Synthetic spectra}

\begin{figure*}
    \centering
    \includegraphics[width=\textwidth]{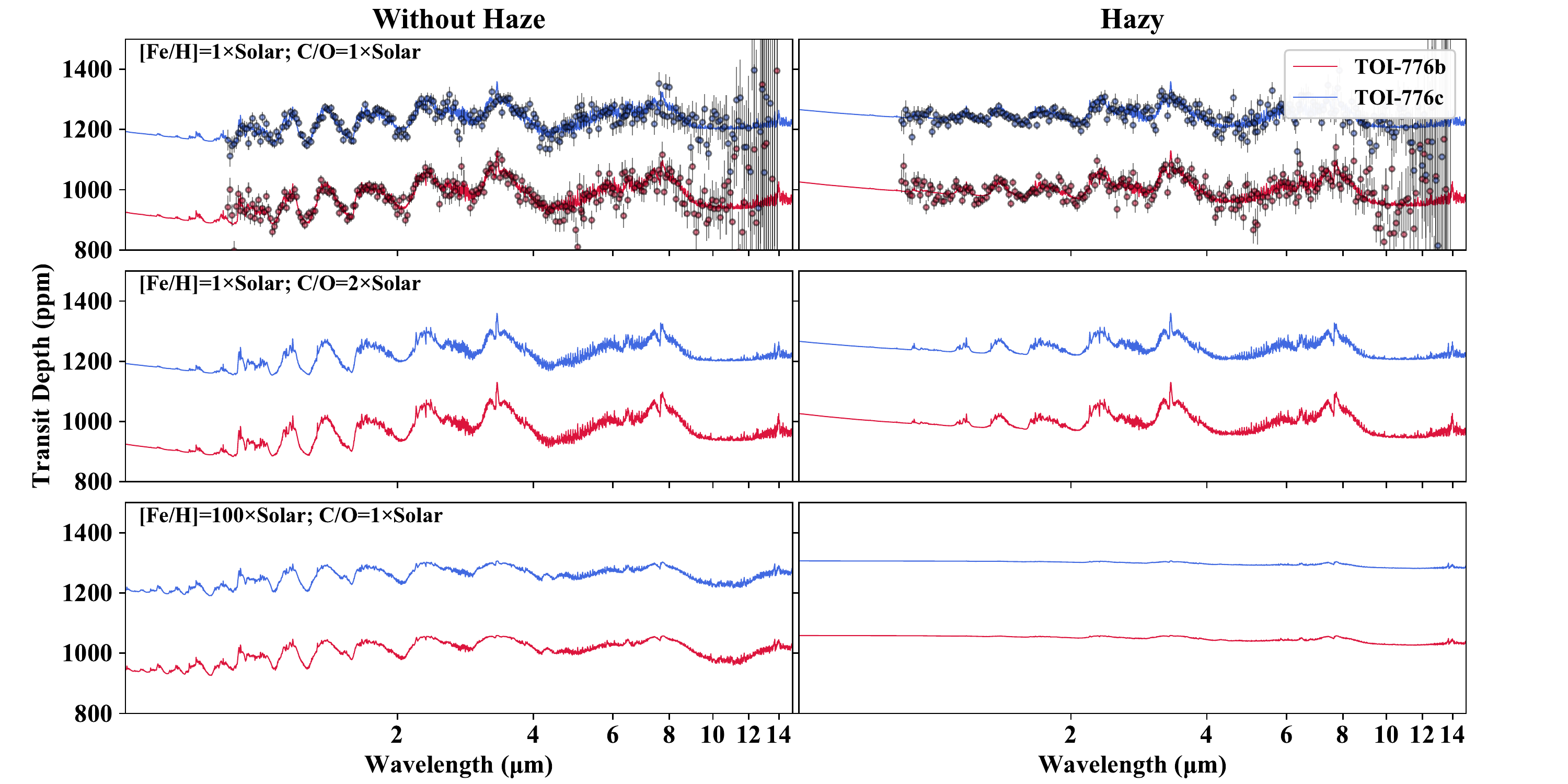}
    \caption{Synthetic atmospheric spectra of TOI-776~b (red) and c (blue). {\it Top}: Fiducial models with solar abundance (solid lines). Estimated uncertainties are shown for \textit{JWST} NIRISS-SOSS, NIRSpec-G395M, and MIRI-LRS configurations, assuming two transits and binned for $R=50$. {\it Middle}: Enhanced carbon-to-oxygen ratio by a factor of two. {\it Bottom}: Enhanced metallicity by a factor of 100. The left column represents spectra without haze opacity and the right column with haze opacity.}
    \label{fig:atm_spectra}
\end{figure*}

In order to quantitatively assess the possibility of TOI-776~b and c's atmospheric characterization with the \textit{James Webb Space Telescope} (\textit{JWST}), we investigated a suite of atmospheric scenarios and calculated their \textit{JWST} synthetic spectra using the photo-chemical model \texttt{ChemKM} \citep{molaverdikhani2019coldb} and petitRADTRANS \citep{molliere2019petitradtrans}. We based the temperature structure of these planets on modern Earth’s temperature structure, and we increased the surface temperature for it to be consistent with the equilibrium temperature of TOI-776~b and c \citep{kawashima2019detectable}. We followed a similar approach as in \citet{Luque2019A&A...628A..39L}: we estimated TOI-776’s ($T_{\rm eff}$ = 3709\,K) flux in the range between X-rays and optical wavelengths, using as reference GJ~832 geometric mean spectra ($T_{\rm eff} = 3816\,\mathrm{K}$). The stellar data were obtained from the MUSCLES database \citep{france2016muscles}. To set up the models, we used \citet{hebrard2012neutral} chemical network with 135 species and 788 reactions, and an updated version of \citet{hebrard2012neutral}’s UV absorption cross sections and branching yields.

Figure~\ref{fig:atm_spectra} shows the synthetic transmission spectra of TOI-776~b and c assuming different metallicities, carbon-to-oxygen ratios and haze opacities. For our fiducial model (top left panel of Fig.~\ref{fig:atm_spectra}), we assume solar abundances. Such spectra are predominantly consisting of water and methane features, as expected for this type of planets \citep{molaverdikhani2019colda}. The significance of these features are on the order of 100\,ppm, well above conservative \textit{JWST} expected noise floor \citep[20\,ppm for NIRISS and 50\,ppm for MIRI,][]{greene2016characterizing}. We calculated the NIRISS-SOSS, NIRSpec-G395M, and MIRI-LRS uncertainties with PandExo \citep{batalha2017pandexo}, assuming two transits and binned for $R=50$, supporting the previous statement. In this scenario, the contribution from haze opacity partially obscures molecular features below \SI{2}{\micro\metre}, but it is almost ineffective at longer wavelengths (see left and right upper panels of Fig.~\ref{fig:atm_spectra}). We note, however, that the radiative feedback of haze particles might significantly affect the temperature structure and the composition of atmosphere \citep{molaverdikhani2020role}. We did not take this effect into account in this work in order to keep the temperature profiles consistent with the Earth’s profile.

Smaller planets are expected to have enhanced metallicities \citep[e.g.][]{wakeford2017hat}. Therefore, we investigated two deviations from our solar abundance fiducial model: 1) an enhanced carbon-to-oxygen ratio (C/O) two-times the solar value, and 2) an enhanced metallicity of hundred times higher than solar. C/O enhancement alone does not affect the composition and spectral features substantially, as seen in Fig.~\ref{fig:atm_spectra} middle panels. On the other hand, one might expect a higher metallicity to result in more pronounced spectral features, due to higher species abundances. However, the bottom panels of Fig.~\ref{fig:atm_spectra} discard this possibility. On the contrary, an enhanced metallicity causes a higher mean molecular weight, which in turn shrinks the spectral significance (bottom-left panel of Fig.~\ref{fig:atm_spectra}), and, simultaneously, it results in a higher haze production, which also obscures the spectra significantly (bottom-right panel of Fig.~\ref{fig:atm_spectra}). Therefore, a flat transmission spectrum may indicate a hazy atmosphere with a high metallicity \citep{kreidberg2014clouds} as opposed to a non-existing atmosphere \citep{kreidberg2019absence}. Complementary observations, such as ground-based high-resolution spectroscopy or spectroscopy of the reflected light, are required to reveal the true nature of these flat spectra.

\section{Summary} \label{sec:summary}

We present the discovery and characterization of the two-planet system transiting the bright ($V=11.54\,\mathrm{mag}, J=8.48\,\mathrm{mag}$) M1\,V star TOI-776. Both planets were detected by the \textit{TESS} mission, confirmed from ground-based transit follow-up observations and have their dynamical masses determined with precise RV measurements using HARPS. In addition, fifteen years of ground-based photometric monitoring by ASAS-SN, ASAS, NSVS, Catalina, and SuperWASP help us to measure a rotational period between 30 to 40\,d, typical of inactive early-type M dwarfs. Our findings are summarized below:
\begin{itemize}

    
    \item A joint fit of all the available transit photometry from \textit{TESS}, MEarth, and LCOGT and the precise RVs from HARPS reveals that the TOI-776 system consists of two transiting planets, namely TOI-776~b, which has a period of 8.25\,d, 
    a radius of $1.85\pm0.13\,R_\oplus$, a mass of $4.0\pm0.9\,M_\oplus$, a bulk density of $3.4^{+1.1}_{-0.9}\,\mathrm{g\,cm^{-3}}$, and an equilibrium temperature of $514\pm17\,\mathrm{K}$; and TOI-776~c, which has a period of 15.66\,d, 
    a radius of $2.02\pm0.14\,R_\oplus$, a mass of $5.3\pm1.8\,M_\oplus$, a bulk density of $3.5^{+1.4}_{-1.3}\,\mathrm{g\,cm^{-3}}$, and an equilibrium temperature of $415\pm14\,\mathrm{K}$. The RV data show one additional signal, with a period of 34\,d, associated with the star's rotation, in agreement from our analyses of the photometry and spectral line indicators.
    
    \item The bulk densities of TOI-776~b and c allow for a wide range of possible interior compositions, from water worlds to rocky planets with H/He-rich atmospheres, but they are too low for either planet to have a purely terrestrial (iron plus rock) composition. Thus, an atmosphere is expected for both planets.
    
    \item From its location in a period-radius diagram, TOI-776~b lies in the transition region where formation and evolution models make different predictions for planetary systems orbiting M dwarfs. For the TOI-776 system, the planets lie above the radius valley carved by gas-poor formation mechanisms, in agreement with their bulk densities being incompatible with the absence of an atmosphere. Still, it is possible that the planets are still undergoing slow thermally driven mass loss under the core-powered scenario.
    
    \item The TOI-776 system is an excellent target for the \textit{JWST}. It is the only known multi-planetary system with planets inside and near the radius valley for which all planets: 1) have a bulk density determination with at least 30\% relative uncertainties, and 2) are extremely suitable for atmospheric characterization. Thanks to the brightness of its host star, it is a remarkable laboratory to break the degeneracy in planetary interior models and to test formation and evolution theories of small planets around low-mass stars.
    
\end{itemize}

\begin{acknowledgements}
This work was supported by the KESPRINT\footnote{\url{www.kesprint.science}.} collaboration, an international consortium devoted to the characterization and research of exoplanets discovered with space-based missions. 
We are very grateful to the ESO staff members for their precious support during the observations. We warmly thank Xavier Dumusque and Fran\c{c}ois Bouchy for coordinating the shared observations with HARPS and Jaime Alvarado Montes, Xavier Delfosse, Guillaume Gaisn\'e, Melissa Hobson, and Felipe Murgas who helped collecting the data. 
This paper includes data collected by the \textit{TESS} mission. Funding for the \textit{TESS} mission is provided by the NASA Explorer Program. We acknowledge the use of \textit{TESS} Alert data, which is currently in a beta test phase, from pipelines at the \textit{TESS} Science Office and at the \textit{TESS} Science Processing Operations Center. Resources supporting this work were provided by the NASA High-End Computing (HEC) Program through the NASA Advanced Supercomputing (NAS) Division at Ames Research Center for the production of the SPOC data products. This research has made use of the Exoplanet Follow-up Observation Program website, which is operated by the California Institute of Technology, under contract with the National Aeronautics and Space Administration under the Exoplanet Exploration Program. 
This work has made use of data from the European Space Agency (ESA) mission {\it Gaia} (\url{https://www.cosmos.esa.int/gaia}), processed by the {\it Gaia} Data Processing and Analysis Consortium (DPAC, \url{https://www.cosmos.esa.int/web/gaia/dpac/consortium}). Funding for the DPAC has been provided by national institutions, in particular the institutions participating in the {\it Gaia} Multilateral Agreement.
The MEarth Team gratefully acknowledges funding from the David and Lucile Packard Fellowship for Science and Engineering (awarded to D.C.). This material is based upon work supported by the National Science Foundation under grants AST-0807690, AST-1109468, AST-1004488 (Alan T. Waterman Award), and AST-1616624, and upon work supported by the National Aeronautics and Space Administration under Grant No. 80NSSC18K0476 issued through the XRP Program. This work is made possible by a grant from the John Templeton Foundation. The opinions expressed in this publication are those of the authors and do not necessarily reflect the views of the John Templeton Foundation.
This work makes use of observations from the LCOGT network.
Some of the Observations in the paper made use of the High-Resolution Imaging instrument Zorro. Zorro was funded by the NASA Exoplanet Exploration Program and built at the NASA Ames Research Center by Steve B. Howell, Nic Scott, Elliott P. Horch, and Emmett Quigley. Data were reduced using a software pipeline originally written by Elliott Horch and Mark Everett. Zorro was mounted on the Gemini South telescope, and NIRI was mounted on the Gemini North telescope, of the international Gemini Observatory, a program of NSF’s OIR Lab, which is managed by the Association of Universities for Research in Astronomy (AURA) under a cooperative agreement with the National Science Foundation on behalf of the Gemini partnership: the National Science Foundation (United States), National Research Council (Canada), Agencia Nacional de Investigación y Desarrollo (Chile), Ministerio de Ciencia, Tecnología e Innovación (Argentina), Ministério da Ciência, Tecnologia, Inovações e Comunicações (Brazil), and Korea Astronomy and Space Science Institute (Republic of Korea). Data collected under program GN-2019A-LP-101.
Based in part on observations obtained at the Southern Astrophysical Research (SOAR) telescope, which is a joint project of the Minist\'{e}rio da Ci\^{e}ncia, Tecnologia e Inova\c{c}\~{o}es (MCTI/LNA) do Brasil, the US National Science Foundation’s NOIRLab, the University of North Carolina at Chapel Hill (UNC), and Michigan State University (MSU).
This work was enabled by observations made from the Gemini North telescope, located within the Maunakea Science Reserve and adjacent to the summit of Maunakea. We are grateful for the privilege of observing the Universe from a place that is unique in both its astronomical quality and its cultural significance.
Based on observations collected at the European Organization for Astronomical Research in the Southern Hemisphere under ESO programme 0103.C-0449(A).
R.\,L. has received funding from the European Union’s Horizon 2020 research and innovation program under the Marie Skłodowska-Curie grant agreement No.~713673 and financial support through the “la Caixa” INPhINIT Fellowship Grant LCF/BQ/IN17/11620033 for Doctoral studies at Spanish Research Centers of Excellence from “la Caixa” Banking Foundation, Barcelona, Spain. 
This work is partly financed by the Spanish Ministry of Economics and Competitiveness through projects ESP2016-80435-C2-2-R and ESP2016-80435-C2-1-R. 
L.\,M.\,S. and D.\,G. gratefully acknowledge financial support from the CRT foundation under Grant No. 2018.2323 ``Gaseous or rocky? Unveiling the nature of small worlds". 
This work is supported by JSPS KAKENHI Grant Numbers JP19K14783, JP18H01265 and JP18H05439, and JST PRESTO Grant Number JPMJPR1775. 
I.\,J.\,M.\,C. acknowledges support from the NSF through grant AST-1824644. 
J.\,K., Sz.\,Cs., M.\,E., A.\,P.\,H., K.\,W.\,F.\,L., S.\,G. acknowledge support by DFG grants PA525/ 18-1, PA525/ 19-1, PA525/ 20-1, HA3279/ 12-1 and RA714/ 14-1 within the DFG Schwerpunkt SPP 1992, Exploring the Diversity of Extrasolar Planets. 
M.\,F, I.\,G. and C.\,M.\,P. gratefully acknowledge the support of the Swedish National Space Agency (DNR 65/19 and 174/18). 
P.\,K. and J.\,S. acknowledge the grant INTER-TRANSFER number LTT20015. 
S.\,A. knowledges support from the Danish Council for Independent Research, through a DFF Sapere Aude Starting Grant no. 4181-00487B. Funding for the Stellar Astrophysics Centre is provided by The Danish National Research Foundation (Grant agreement no.: DNRF106). 
H.\,J.\,D. acknowledges support by grant ESP2017-87676-C5-4-R of the Spanish Secretary of State for R\&D\&i (MINECO).
D.\,G. warmly thanks Javier Alarcón, Pablo Arias, Duncan Castex, Mónica Castillo, Cecilia Farias, Mario Herrera, Francisco Labraña, Angélica León, and Ariel Sánchez (ESO La Silla) for the inspiring conversations.
 
\end{acknowledgements}

\bibliographystyle{aa} 
\bibliography{biblio} 


\begin{appendix} 

\section{Joint fit priors.}

\begin{table*}
    \centering
    \caption{Priors used for the models presented in Sect.~\ref{sec:fit} using \texttt{juliet}. The prior labels of $\mathcal{N}$, $\mathcal{U}$, $\mathcal{B}$ and $\mathcal{J}$ represent normal, uniform, Beta and Jeffrey's distributions. The parameterization for $(p,b)$ using $(r_1,r_2)$ \citep{Espinoza18} and the linear $(q_1)$ and quadratic $(q_1,q_2)$ limb darkening parameterization \citep{Kipping13} are both described in Sect.~\ref{subsubsec:photonly}.}
    \label{tab:priors}
    \begin{tabular}{lllr} 
        \hline
        \hline
        \noalign{\smallskip}
        Parameter name & Prior & Units & Description \\
        \noalign{\smallskip}
        \hline
        \noalign{\smallskip}
        \multicolumn{4}{c}{\it Stellar parameters} \\
        \noalign{\smallskip}
        $\rho_\star$ & $\mathcal{N}(5300,1500^2)$ & $\mathrm{kg\,m\,^{-3}}$ & Stellar density. \\
        \noalign{\smallskip}
        \multicolumn{4}{c}{\it Planet parameters} \\
        \noalign{\smallskip}
        $P_{b}$                 & $\mathcal{N}(8.24,0.05^2)$        & d                 & Period of planet b. \\
        $P_{c}$                 & $\mathcal{N}(15.65,0.05^2)$       & d                 & Period of planet c. \\
        $t_{0,b} - 2450000$     & $\mathcal{N}(8571.41,0.01^2)$     & d                 & Time of transit-center of planet b. \\
        $t_{0,c} - 2450000$     & $\mathcal{N}(8572.60,0.01^2)$     & d                 & Time of transit-center of planet c. \\
        $r_{1,b}$               & $\mathcal{U}(0,1)$                & \dots             & Parametrization for $p$ and $b$ of planet b. \\
        $r_{2,b}$               & $\mathcal{U}(0,1)$                & \dots             & Parametrization for $p$ and $b$ of planet b. \\
        $r_{1,c}$               & $\mathcal{U}(0,1)$                & \dots             & Parametrization for $p$ and $b$ of planet c. \\
        $r_{2,c}$               & $\mathcal{U}(0,1)$                & \dots             & Parametrization for $p$ and $b$ of planet c. \\
        $K_{b}$                 & $\mathcal{U}(0,20)$               & $\mathrm{m\,s^{-1}}$      & RV semi-amplitude of planet b. \\
        $K_{c}$                 & $\mathcal{U}(0,20)$               & $\mathrm{m\,s^{-1}}$      & RV semi-amplitude of planet c. \\
        $e_{b}$                 & $\mathcal{B}(1.52,29)$            & \dots             & Eccentricity of planet b. \\
        $e_{c}$                 & $\mathcal{B}(1.52,29)$            & \dots             & Eccentricity of planet c. \\
        $\omega_{b}$            & $\mathcal{U}(-180,180)$           & deg               & Argument of periastron of planet b. \\
        $\omega_{c}$            & $\mathcal{U}(-180,180)$           & deg               & Argument of periastron of planet c. \\
        \noalign{\smallskip}
        \multicolumn{4}{c}{\it Photometry parameters} \\
        \noalign{\smallskip}
        $\sigma_{\textnormal{TESS}}$            & $\mathcal{J}(1,1000)$     & ppm       & Extra jitter term for {\it TESS}. \\
        $D_{\textnormal{TESS}}$                 & 1.0 (fixed)               & \dots     & Dilution factor for {\it TESS}. \\
        $M_{\textnormal{TESS}}$                 & 0.0 (fixed)               & ppm       & Relative flux offset for {\it TESS}. \\
        $q_{1,\textnormal{TESS}}$               & $\mathcal{U}(0,1)$        & \dots     & Quadratic limb darkening parametrization for {\it TESS}. \\
        $q_{2,\textnormal{TESS}}$               & $\mathcal{U}(0,1)$        & \dots     & Quadratic limb darkening parametrization for {\it TESS}. \\
        $\sigma_{\textnormal{LCO-CTIO}}$        & $\mathcal{J}(10,10^5)$    & ppm       & Extra jitter term for LCO-CTIO. \\
        $M_{\textnormal{LCO-CTIO}}$             & $\mathcal{N}(0,0.01^2)$   & ppm       & Relative flux offset for LCO-CTIO. \\
        $\theta_{\textnormal{LCO-CTIO}}$        & $\mathcal{U}(-1.0,1.0)$   & \dots     & Airmass regression coefficients for LCO-CTIO. \\
        $q_{1,\textnormal{LCO-CTIO}}$           & $\mathcal{U}(0,1)$        & \dots     & Linear limb darkening parametrization for LCO-CTIO. \\
        $\sigma_{\textnormal{LCO-SAAO}}$        & $\mathcal{J}(10,10^5)$    & ppm       & Extra jitter term for LCO-SAAO. \\
        $M_{\textnormal{LCO-SAAO}}$             & $\mathcal{N}(0,0.01^2)$   & ppm       & Relative flux offset for LCO-SAAO. \\
        $\theta_{\textnormal{LCO-SAAO}}$        & $\mathcal{U}(-1.0,1.0)$   & \dots     & Airmass regression coefficients for LCO-SAAO. \\
        $q_{1,\textnormal{LCO-SAAO}}$           & $\mathcal{U}(0,1)$        & \dots     & Linear limb darkening parametrization for LCO-SAAO. \\
        $\sigma_{\textnormal{LCO-SSO}}$         & $\mathcal{J}(10,10^5)$    & ppm       & Extra jitter term for LCO-SSO. \\
        $M_{\textnormal{LCO-SSO}}$              & $\mathcal{N}(0,0.01^2)$   & ppm       & Relative flux offset for LCO-SSO. \\
        $\theta_{\textnormal{LCO-SSO}}$         & $\mathcal{U}(-1.0,1.0)$   & \dots     & Airmass regression coefficients for LCO-SSO. \\
        $q_{1,\textnormal{LCO-SSO}}$            & $\mathcal{U}(0,1)$        & \dots     & Linear limb darkening parametrization for LCO-SSO. \\
        $\sigma_{\textnormal{MEarth}}$          & $\mathcal{J}(10,10^5)$    & ppm       & Extra jitter term for MEarth. \\
        $M_{\textnormal{MEarth}}$               & $\mathcal{N}(0,0.01^2)$   & ppm       & Relative flux offset for MEarth. \\
        $q_{1,\textnormal{MEarth}}$             & $\mathcal{U}(0,1)$        & \dots     & Linear limb darkening parametrization for MEarth. \\
        \noalign{\smallskip}
        \multicolumn{4}{c}{\it RV parameters} \\
        \noalign{\smallskip}
        $\mu_{\textnormal{HARPS}}$          & $\mathcal{U}(-100,100)$       & $\mathrm{m\,s^{-1}}$ & Systemic velocity for HARPS. \\
        $\sigma_{\textnormal{HARPS}}$       & $\mathcal{J}(0.1,100)$        & $\mathrm{m\,s^{-1}}$ & Extra jitter term for HARPS. \\
        \noalign{\smallskip}
        \multicolumn{4}{c}{\it GP hyperparameters and additional sinusoid} \\
        \noalign{\smallskip}
        $\sigma_\mathrm{GP,TESS}$       & $\mathcal{J}(10^{-2},10^6)$       & ppm                       & Amplitude of GP component for TESS. \\
        $T_\mathrm{GP,TESS}$            & $\mathcal{J}(10^{-6},10^4)$       & d                         & Length scale of GP component for TESS. \\
        $K$                             & $\mathcal{U}(0,20)$               & $\mathrm{m\,s^{-1}}$      & RV semi-amplitude of the additional sinusoid. \\
        $t_{0} - 2450000$               & $\mathcal{U}(8575.0,8655.0)$      & d                         & Time of transit-center the additional sinusoid. \\
        $P$                             & $\mathcal{N}(35.0,10.0^2)$        & d                         & Period of the additional sinusoid. \\
        \noalign{\smallskip}
        \hline
    \end{tabular}
\end{table*}

\section{HARPS RV measurements and spectral line indicators.}

\begin{table*}
    \centering
    \caption{\texttt{serval} extraction.}
    \label{tab:rvs-serval}
    \begin{tabular}{crcrcrc}
        \hline
        \hline
        \noalign{\smallskip}
  \multicolumn{1}{c}{BJD$_\mathrm{TBD} - 2457000$} &
  \multicolumn{1}{c}{RV (m\,s$^{-1}$)} &
  \multicolumn{1}{c}{$\sigma_\mathrm{RV}$ (m\,s$^{-1}$)} &
  \multicolumn{1}{c}{CRX (m\,s$^{-1}$\,Np$^{-1}$)} &
  \multicolumn{1}{c}{$\sigma_\mathrm{CRX}$ (m\,s$^{-1}$\,Np$^{-1}$)} &
  \multicolumn{1}{c}{dLW (m$^2$\,s$^{-2}$)} &
  \multicolumn{1}{c}{$\sigma_\mathrm{dLW}$ (m$^2$\,s$^{-2}$)} \\
        \noalign{\smallskip}
        \hline
        \noalign{\smallskip}
  1884.75667 & 4.3 & 3.1 & $-$4.0 & 26.4 & $-$3.4 & 4.0\\
  1886.88043 & $-$0.2 & 1.6 & $-$4.7 & 13.3 & $-$15.1 & 2.0\\
  1887.79526 & 2.2 & 1.9 & 10.9 & 15.8 & $-$14.7 & 2.1\\
  1888.83087 & 1.2 & 1.3 & $-$12.3 & 10.7 & $-$13.6 & 1.9\\
  1889.79811 & 1.2 & 1.6 & 5.1 & 12.9 & $-$12.1 & 2.5\\
  1890.80821 & 7.3 & 2.4 & $-$2.6 & 19.2 & $-$15.0 & 3.0\\
  1894.81627 & $-$0.2 & 2.1 & 25.4 & 16.8 & $-$23.2 & 2.9\\
  1898.85272 & 4.5 & 1.4 & $-$3.7 & 11.2 & $-$26.8 & 1.5\\
  1899.86392 & 9.3 & 1.5 & 14.7 & 12.0 & $-$25.8 & 1.5\\
  1900.84340 & 6.0 & 1.4 & $-$12.9 & 11.0 & $-$24.8 & 2.0\\
  1902.80747 & 0.5 & 1.5 & 14.9 & 12.2 & $-$24.4 & 1.9\\
  1903.81551 & 0.2 & 2.0 & $-$4.2 & 16.2 & $-$27.4 & 1.7\\
  1910.81439 & 5.0 & 1.3 & $-$1.8 & 10.4 & $-$15.2 & 1.8\\
  1911.72654 & 8.3 & 1.5 & 8.6 & 12.4 & $-$19.5 & 1.8\\
  1912.77545 & 10.0 & 1.5 & $-$7.8 & 11.8 & $-$23.4 & 2.3\\
  1914.81139 & 10.9 & 1.5 & 13.4 & 12.2 & $-$20.5 & 2.0\\
  1915.74379 & 7.6 & 1.2 & 0.3 & ~9.7 & $-$23.1 & 1.8\\
  1916.69654 & 6.0 & 1.3 & 8.3 & 10.8 & $-$25.2 & 1.4\\
  1918.78064 & $-$0.2 & 1.3 & 13.0 & 10.4 & $-$19.5 & 1.6\\
  1919.64336 & 3.2 & 1.4 & $-$0.8 & 11.0 & $-$7.3 & 1.8\\
  1924.80573 & 0.4 & 1.4 & 5.0 & 11.2 & $-$16.8 & 1.8\\
  1925.68888 & 2.7 & 1.3 & $-$13.1 & 10.6 & $-$16.3 & 1.8\\
  1925.83889 & $-$0.5 & 1.5 & $-$19.7 & 12.0 & $-$20.7 & 1.8\\
  1926.79127 & 0.4 & 1.3 & 3.8 & 10.7 & $-$20.7 & 1.5\\
  1927.83074 & 1.8 & 1.6 & $-$14.7 & 12.8 & $-$18.8 & 2.2\\
  1928.78734 & 2.9 & 1.3 & $-$2.8 & 10.3 & $-$16.6 & 1.9\\
  1929.76211 & 5.2 & 1.7 & $-$18.0 & 13.5 & $-$7.4 & 2.4\\
  1930.83118 & 6.0 & 1.6 & $-$10.0 & 13.2 & $-$7.2 & 2.2\\
  1931.76389 & 7.3 & 1.4 & $-$17.5 & 11.4 & $-$7.1 & 2.4\\
        \noalign{\smallskip}
        \hline
\end{tabular}
\end{table*}

\begin{table*}
    \centering
    \caption{\texttt{TERRA} extraction.}
    \label{tab:rvs-terra}
    \begin{tabular}{crccrccc}
        \hline
        \hline
        \noalign{\smallskip}
  \multicolumn{1}{c}{BJD$_\mathrm{TBD} - 2457000$} &
  \multicolumn{1}{c}{RV (m\,s$^{-1}$)} &
  \multicolumn{1}{c}{$\sigma_\mathrm{RV}$ (m\,s$^{-1}$)} &
  \multicolumn{1}{c}{H$_\alpha$} &
  \multicolumn{1}{c}{S-index} &
  \multicolumn{1}{c}{$\sigma_\mathrm{S-index}$} &
  \multicolumn{1}{c}{NaD$_1$} &
  \multicolumn{1}{c}{NaD$_2$} \\
        \noalign{\smallskip}
        \hline
        \noalign{\smallskip}
  1884.75667 & 0.7 & 3.2 & 0.353 & 1.501 & 0.025 & 1.050 & 0.783\\
  1886.88043 & $-$3.1 & 1.8 & 0.345 & 1.555 & 0.018 & 1.056 & 0.800\\
  1887.79526 & $-$2.4 & 1.6 & 0.358 & 1.380 & 0.012 & 1.046 & 0.792\\
  1888.83087 & $-$3.5 & 1.4 & 0.357 & 1.391 & 0.012 & 1.044 & 0.795\\
  1889.79811 & $-$2.0 & 1.4 & 0.315 & 1.496 & 0.013 & 1.041 & 0.789\\
  1890.80821 & 1.5 & 2.1 & 0.336 & 1.508 & 0.015 & 1.047 & 0.797\\
  1894.81627 & $-$5.6 & 1.8 & 0.346 & 1.407 & 0.016 & 1.057 & 0.787\\
  1898.85272 & 1.3 & 1.4 & 0.361 & 1.356 & 0.014 & 1.055 & 0.797\\
  1899.86392 & 5.7 & 1.5 & 0.350 & 1.399 & 0.016 & 1.060 & 0.798\\
  1900.84340 & 2.6 & 1.6 & 0.364 & 1.336 & 0.015 & 1.057 & 0.799\\
  1902.80747 & $-$3.2 & 1.4 & 0.357 & 1.340 & 0.013 & 1.063 & 0.796\\
  1903.81551 & $-$4.0 & 1.2 & 0.366 & 1.342 & 0.018 & 1.060 & 0.799\\
  1910.81439 & 3.6 & 1.2 & 0.339 & 1.445 & 0.014 & 1.053 & 0.799\\
  1911.72654 & 5.3 & 1.5 & 0.349 & 1.394 & 0.013 & 1.049 & 0.800\\
  1912.77545 & 5.6 & 1.4 & 0.315 & 1.559 & 0.015 & 1.054 & 0.790\\
  1914.81139 & 6.5 & 1.6 & 0.341 & 1.460 & 0.018 & 1.056 & 0.796\\
  1915.74379 & 4.6 & 1.0 & 0.324 & 1.462 & 0.013 & 1.056 & 0.792\\
  1916.69654 & 2.8 & 1.1 & 0.348 & 1.368 & 0.011 & 1.059 & 0.801\\
  1918.78064 & $-$2.5 & 1.3 & 0.359 & 1.298 & 0.013 & 1.066 & 0.804\\
  1919.64336 & $-$0.3 & 1.2 & 0.356 & 1.338 & 0.012 & 1.061 & 0.794\\
  1924.80573 & $-$3.3 & 1.3 & 0.361 & 1.246 & 0.015 & 1.054 & 0.798\\
  1925.68888 & $-$2.1 & 1.2 & 0.375 & 1.286 & 0.012 & 1.059 & 0.805\\
  1925.83889 & $-$4.3 & 1.5 & 0.369 & 1.267 & 0.014 & 1.054 & 0.789\\
  1926.79127 & $-$2.5 & 1.2 & 0.365 & 1.231 & 0.014 & 1.054 & 0.797\\
  1927.83074 & 0.0 & 1.8 & 0.347 & 1.275 & 0.018 & 1.046 & 0.798\\
  1928.78734 & $-$1.6 & 1.3 & 0.331 & 1.391 & 0.017 & 1.047 & 0.797\\
  1929.76211 & 2.2 & 1.7 & 0.349 & 1.324 & 0.018 & 1.044 & 0.802\\
  1930.83118 & 3.3 & 1.5 & 0.356 & 1.386 & 0.020 & 1.045 & 0.782\\
  1931.76389 & 4.0 & 1.4 & 0.353 & 1.415 & 0.020 & 1.039 & 0.777\\
        \noalign{\smallskip}
        \hline
\end{tabular}
\end{table*}

\section{Corner plots.}

\begin{figure*}
    \centering
    \includegraphics[width=\textwidth]{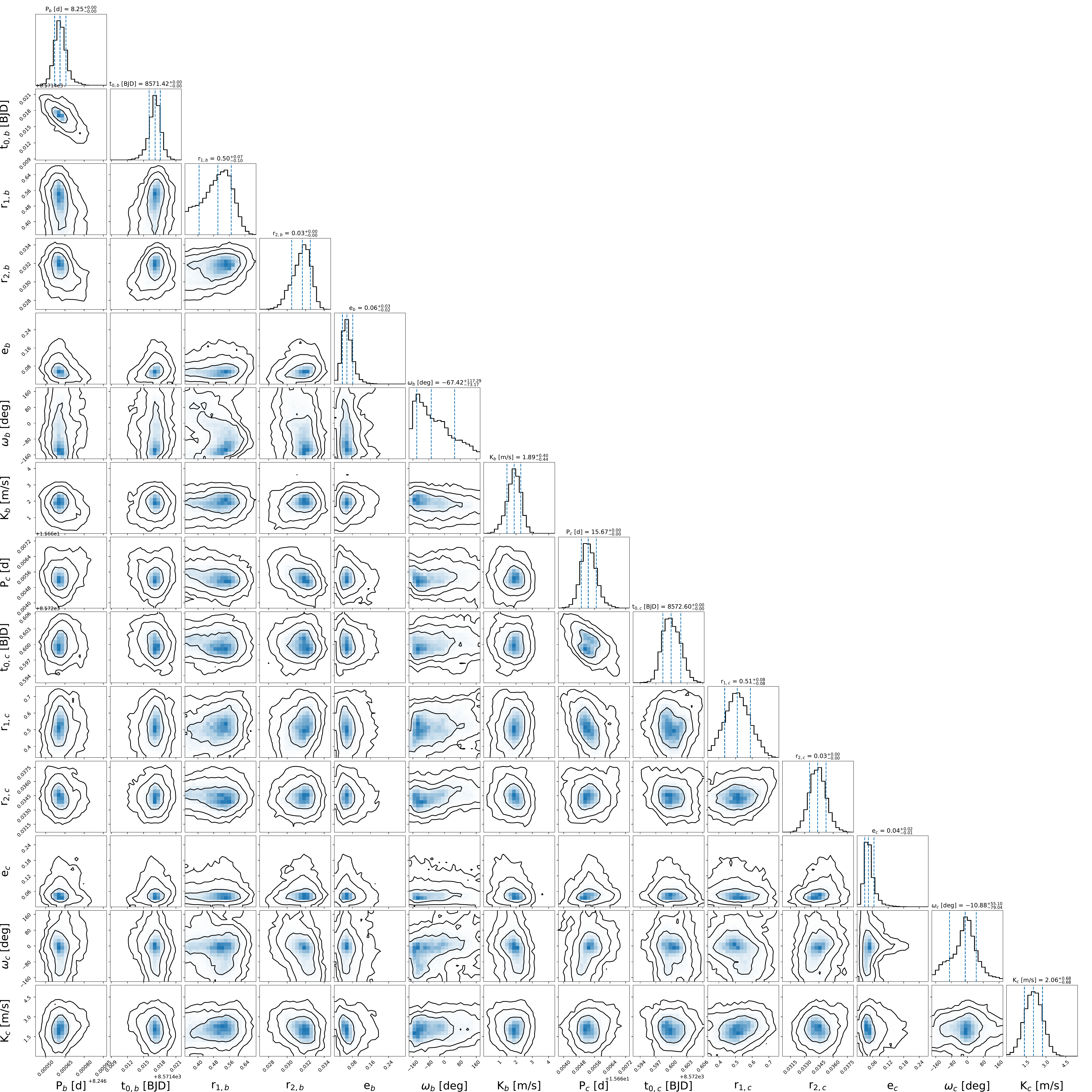}
    \caption{Posterior distributions of the orbital parameters of the TOI-776 system. Each panel contains $\sim 220\,000$ samples. The top panels of the corner plot show the probability density distributions of each orbital parameter. The vertical dashed lines indicate the 16th, 50th, and the 84th percentiles of the samples. Contours are drawn to improve the visualization of the 2D histograms and indicate the 68.3\%, 95.5\%, and 99.7\% confidence interval levels (i.e., 1$\sigma$, 2$\sigma,$ and 3$\sigma$).}
    \label{fig:corner_main}
\end{figure*}

\end{appendix}

\end{document}